\documentclass[acmsmall]{acmart}

\AtBeginDocument{%
  \providecommand\BibTeX{{%
    \normalfont B\kern-0.5em{\scshape i\kern-0.25em b}\kern-0.8em\TeX}}}

\setcopyright{acmlicensed}
\acmJournal{PACMHCI}
\acmYear{2024} \acmVolume{8} \acmNumber{MHCI} \acmArticle{262} \acmMonth{9}\acmDOI{10.1145/3676507}

\usepackage[utf8]{inputenc}
\usepackage[T1]{fontenc}
\usepackage{tabularx}
\usepackage{soul}
\usepackage{booktabs, makecell}
\usepackage{graphicx}
\usepackage{xcolor}
\usepackage{xspace}
\usepackage{colortbl}
\usepackage{subcaption}
\usepackage{xtab}
\usepackage{enumitem}
\usepackage{longtable}
\usepackage{lscape}
\usepackage{multirow}
\usepackage{eurosym}
\usepackage{balance}

\begin{document}

\title[Risks and Benefits of AI in Mobile and Wearable Uses]{Good Intentions, Risky Inventions: A Method for Assessing the Risks and Benefits of AI in Mobile and Wearable Uses}

\author{Marios Constantinides}
\affiliation{%
  \institution{Nokia Bell Labs}
  \city{Cambridge}
  \country{United Kingdom}
}
\email{marios.constantinides@nokia-bell-labs.com}

\author{Edyta Bogucka}
\affiliation{%
  \institution{Nokia Bell Labs}
  \city{Cambridge}
  \country{United Kingdom}
}
\email{edyta.bogucka@nokia-bell-labs.com}

\author{Sanja Scepanovic}
\affiliation{%
  \institution{Nokia Bell Labs}
  \city{Cambridge}
  \country{United Kingdom}
}
\email{sanja.scepanovic@nokia-bell-labs.com}

\author{Daniele Quercia}
\affiliation{%
  \institution{Nokia Bell Labs}
  \city{Cambridge}
  \country{United Kingdom}
}
\email{daniele.quercia@nokia-bell-labs.com}

\renewcommand{\shortauthors}{Marios Constantinides, Edyta Bogucka, Sanja Scepanovic, \& Daniele Quercia}

\begin{abstract}
Integrating Artificial Intelligence (AI) into mobile and wearables offers numerous benefits at individual, societal, and environmental levels. Yet, it also spotlights concerns over emerging risks. Traditional assessments of risks and benefits have been sporadic, and often require costly expert analysis. We developed a semi-automatic method that leverages Large Language Models (LLMs) to identify AI uses in mobile and wearables, classify their risks based on the EU AI Act, and determine their benefits that align with globally recognized long-term sustainable development goals; a manual validation of our method by two experts in mobile and wearable technologies, a legal and compliance expert, and a cohort of nine individuals with legal backgrounds who were recruited from Prolific, confirmed its accuracy to be over 85\%. We uncovered that specific applications of mobile computing hold significant potential in improving well-being, safety, and social equality. However, these promising uses are linked to risks involving sensitive data, vulnerable groups, and automated decision-making. To avoid rejecting these risky yet impactful mobile and wearable uses, we propose a risk assessment checklist for the Mobile HCI community. 
\end{abstract}



\begin{teaserfigure}
 \centering
 \includegraphics[width=\textwidth]{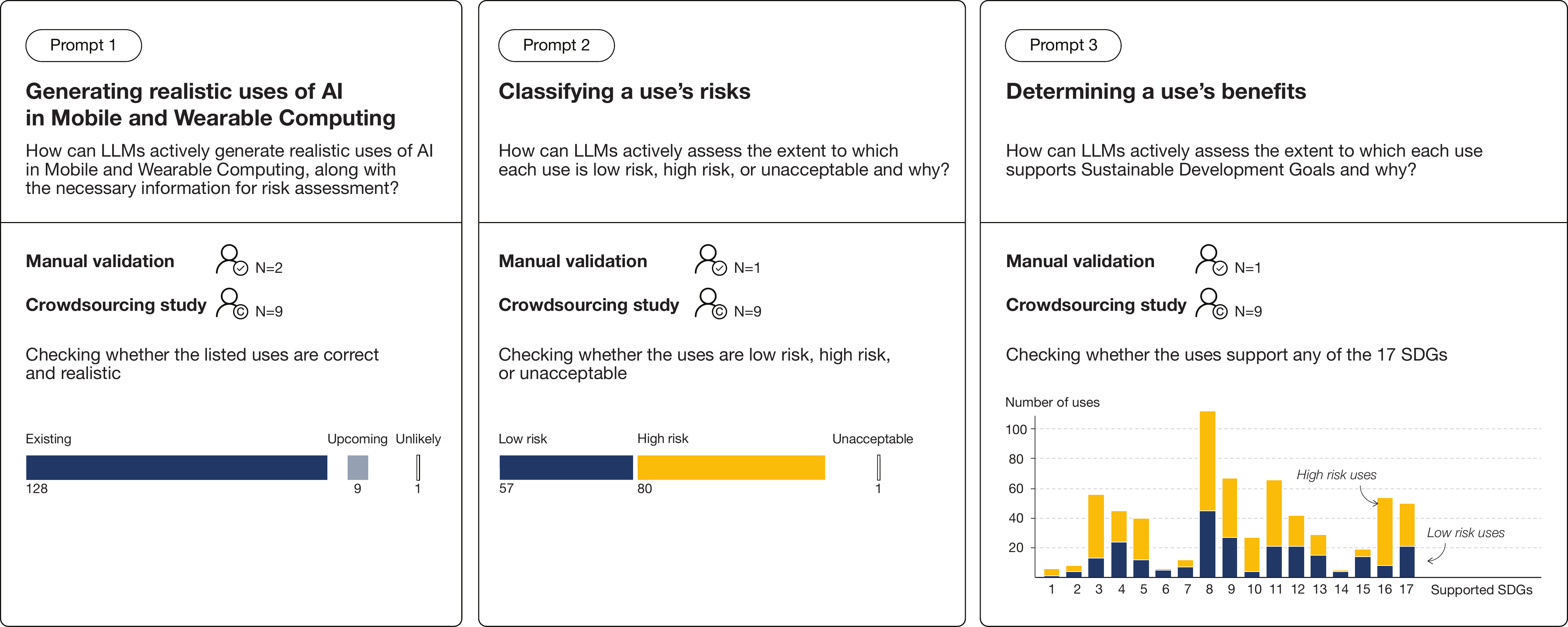}
    \caption{Overview of our method. A three LLM-prompt pipeline for generating mobile and wearable uses of AI (prompt \#1), classifying each use's risks according to the EU AI Act~\cite{EUACT2021} (prompt \#2), and determining whether each generated use is beneficial according to the UN's Sustainability Development Goals~\cite{sdgs} (prompt \#3). Out of 138 generated uses, as many as 80 were considered high risk according to the EU AI Act, primarily aligning with Sustainable Development Goals 3
  (good health and well-being), 10 (reduced inequalities), and 16 (peace and justice). Our method was validated by two experts in mobile and wearable technologies, a legal and compliance expert, and a cohort of nine individuals with legal backgrounds who were recruited from Prolific, confirming its accuracy to be over 85\%.}
  \Description{Overview of our method.}
 \label{fig:teaser}
\end{teaserfigure}

{\def\thefootnote{}\footnotetext{Project website: \url{https://social-dynamics.net/risky-mobile}}

\maketitle

\section{Introduction}
\label{sec:intro}

The integration of Artificial Intelligence (AI) into mobile and wearable devices has unlocked new capabilities, from improving workplace productivity to enhancing healthcare and education. Mobile assistants and wearables, equipped with various sensors such as accelerometers and heart rate monitors, have significantly increased workplace productivity~\cite{mirjafari2019differentiating, aseniero2020meetcues, choi2021kairos, constantinides2020comfeel}. Smart wearable devices provide low-cost, objective measures of physical activity, mental health, and sleep quality assessments~\cite{zhou2023circadian, park2023social, constantinides2018personalized}, contributing significantly to scalable behavior monitoring in large populations~\cite{perez2021wearables}. In sports medicine, wearable technology can detect patterns in physiological variables, assisting athletes in improving performance and preventing injuries~\cite{chidambaram2022using}. In education, smartwatches can be used to collect real-time data during students' learning activities, improving the analytics and insights into learning processes~\cite{ciolacu2019education}. 

Despite the potential advantages, any technology comes with risks. Mobile and wearable technology uses are no exception. Privacy concerns are particularly prominent, as AI integration can lead to data leakage and unauthorized surveillance~\cite{haris2014privacy, constantinides2022good, das2023algorithmic}. However, recent advances in federated learning~\cite{mcmahan2017communication} and differential privacy~\cite{dwork2008differential} may enhance the data privacy of AI-based mobile and wearable uses, potentially transforming high-risk uses into low-risk ones. As these uses often involve automated decision-making, it is crucial to provide explanations, especially in high-stakes domains. For example, there is an increasing demand for explainable AI in healthcare to ensure trustworthy AI-based decisions~\cite{saraswat2022explainable}. Security issues are also critical. In addition to well-known challenges such as the potential to compromise sensitive data or use it to maliciously infer private information~\cite{crager2017information}, wearable devices may also cause physically harm to the wearer~\cite{mills2016wearing}. For example, there have been reports of skin irritation and allergic reactions caused by the materials used (e.g., nickel in the metal components), or due to the accumulation of moisture under the device, which can lead to skin irritation or even infections~\cite{fitbit_irritations}.

Traditional assessments of risks and benefits have been sporadic, and often require costly expert analysis. In fact, risk identification requires comprehensive domain expertise~\cite{hassel2023governing}. According to the OECD, legal advisory costs required to ensure compliance with regulatory systems of different countries is estimated to be nearly 780 billions dollars per year worldwide~\cite{ifac, moraes2021smile}. That is why traditional assessments of risks and benefits tend to be fragmented, focusing on specific aspects of mobile and wearable use (e.g., efficiency in risk assessment for work-related activities~\cite{yang2018towards}, or adoption based on privacy risk versus perceived benefits~\cite{li2016examining}), and come with technical and ethical challenges. On the technical side, mobile and wearables pose significant privacy risks~\cite{cecchinato2015smartwatches}, while, on the ethical side, these devices can potentially augment (or replace) human intelligence in ways that might not always be desirable~\cite{xu2023regulation}. This dichotomy underscores the need for a more systematic approach to evaluate the benefits and risks associated with these technologies. Issues such as data privacy, the algorithms accuracy, and the potential for social and individual opportunities (outlined, for example, by the United Nation's Sustainability Development Goals report~\cite{sdgs}) must be considered to ensure that the integration of AI truly benefits users without compromising their safety and privacy~\cite{puntoni2021consumers}. To address these challenges, we implemented a semi-automatic way of conducting a risks and benefits assessment of AI for mobile and wearable uses and, in so doing, we made two main contributions:

\begin{enumerate}
\item We engineered and validated three Large Language Model (LLM) prompts capable of (\S\ref{sec:methods}): \emph{(a)} generating a diverse range of realistic uses of AI for mobile and wearable computing; \emph{(b)} classifying each generated use risks (whether it is unacceptable, high risk, or low risk) according to the EU AI Act~\cite{EUACT2021}; and \emph{(c)} determining the benefits associated with each generated use. For classifying risks, we chose the EU AI Act because it is the most advanced legal framework available at the time of writing, which specifically adopts a risk-based approach at a use level~\cite{EUACT2021}. This means that stakeholders involved in the development of an AI technology are required to assess its risks for specific uses, rather than assessing the technology itself for broad risks. For determining the benefits, we chose the Sustainable Development Goals (SDGs). The SDGs are globally recognized and allow for the evaluation of the AI's long-term impacts related to five priority areas that concern people, the planet, prosperity, peace, and partnerships~\cite{sdgs}. To achieve these tasks, we used a set of novel prompt elements, including a comprehensive list of \emph{domains}, five \emph{risk concepts} for describing uses, and detailed manual \emph{evaluations} of prompts.

\item Our method generated 138 uses, which we reviewed and thematically grouped them in two categories (\S\ref{sec:results}): uses promoting SDGs and being low risk; and uses promoting SDGs yet being high risk. We found that specific applications of mobile computing hold significant potential in improving well-being, safety, and social equality. However, these promising uses are linked to risks involving sensitive data, vulnerable groups, and automated decision-making. 
\end{enumerate}

In light of these results, we discuss challenges in conducting semi-automatic assessments, and present a Risk Assessment Checklist for Mobile Computing uses as a  practical solution for balancing risks and benefits trade-offs in the use of AI for mobile and wearable technologies (\S\ref{sec:discussion}).
\section{Related Work and Background}
\label{sec:related}

Next, we surveyed previous literature that our work draws upon, and grouped it into three main
areas: \emph{i)}  AI regulations and impact assessments (\S\ref{subsec:ai_regulations}), \emph{ii)} risk and benefits assessments of AI for mobile and wearable uses (\S\ref{sec:related-risk}), and \emph{iii)} the use of LLMs to transition from manual to semi-automatic assessment through prompt engineering (\S\ref{sec:related-prompt}).

\subsection{AI Regulations and Impact Assessments}
\label{subsec:ai_regulations}
The increasing deployment of AI has prompted calls for regulatory oversight~\cite{tahaei2023human, borenstein2021emerging, Floridi2021}. The US Office of Science and Technology Policy has released a non-binding AI Bill of Rights Blueprint, highlighting principles such as safety, non-discrimination, data privacy, and AI transparency. Similarly, the European Commission's AI Act, which is legally binding, aims to balance innovation with safeguarding societal values and rights. The Act classifies AI applications based on risk levels, from low to unacceptable, and prohibits AI applications that are harmful or manipulative~\cite{EUACT2021, amendments2023}. It emphasizes a risk-based regulatory framework, requiring comprehensive evaluations for high risk AI systems. 

Typically, an impact assessment is conducted and a report is produced that documents the potential impacts of AI. CredoAI's report template includes sections on formal system evaluations and compliance with laws, regulations, and standards \cite{sherman2023riskProfiles}. Similarly, the report developed by the National Institute of Standards and Technology (NIST) includes listings of potential biases, evaluations of the impact's magnitude and likelihood, technical specifications, third-party technologies, and legal and compliance issues \cite{nist2023aiRisk}. More recently, a study has identified the five necessary components (i.e., domain, purpose, capability, AI user, and AI subject) to conduct regulatory risk assessment according to the EU AI Act~\cite{Golpayegani2023Risk}. The \emph{domain} specifies the industry or sector, such as health. The \emph{purpose} explains the goal, such as improving well-being. The \emph{capability} describes the technology behind it, such as mood and stress level tracking. The \emph{AI user} is the one using the system, such as mental health professionals. The \emph{AI subject} is the one affected by the system, such as patients. Our study operationalizes these five components in a semi-automatic way using LLMs. 

\subsection{Risk and Benefit Assessments of AI for Mobile and Wearable Uses}
\label{sec:related-risk}
With the upcoming enforcement of the EU AI Act~\cite{EUACT2021}, AI-based technologies will be subject to legal compliance. This raises particular concerns for the use of AI in mobile and wearables, as the Act highlights factors that could classify certain uses as high risk or even unacceptable. For example, mobile and wearable devices, often placed near or worn on the body, operate in the background unnoticed, and collect personal information about individuals. This characteristic may categorize some uses as high risk due to the sensitive nature of the data handling. However, recent advances in federated learning and differential privacy can ensure data privacy and security for such applications, potentially transforming high-risk uses into low-risk ones. Federated learning, for example, allows AI models to be trained across multiple decentralized devices holding local data samples, and never exchanging them~\cite{mcmahan2017communication}. Additionally, differential privacy introduces mathematical guarantees to ensure that the output of AI models does not reveal sensitive information, and allows data analysis without compromising individual privacy~\cite{dwork2008differential}. Similarly, homomorphic encryption methods offer the potential for performing computations on encrypted data, providing robust security assurances~\cite{gentry2009fully}. Additionally, mobile and wearable uses are often non-generalizable due to limited training datasets (e.g., skewed towards predominantly white cohorts~\cite{sjoding2020racial}), introducing biases and limiting their applicability across diverse populations. Therefore, it is important to understand the risks these uses entail. Assessing both risks and benefits allows for a well-informed perspective, puts risks into context, and helps determine which risks are worth taking or which mitigation strategies should be prioritized to preserve the benefits. 

The traditional approach to assess risks and benefits in mobile and wearable uses has often been sporadic, required expert knowledge, and primarily confined in research papers. \citet{heidel2020potential} conducted a scoping review to identify benefits and risks arising from the introduction of health apps and wearables into the German statutory health care system. They noted that health apps and wearables lead to improved treatment quality through enhanced patient monitoring, disease management, personalized therapy, and better health education. However, they also expressed concerns about data privacy and security. The primary concern regarding privacy and human rights with wearables lies in the quantity and type of data that individuals provide. Zuboff, in ``The Age of Surveillance Capitalism,''~\cite{zuboff2023age} wrote that ``every time we encounter a digital interface we make our experience available to `datafication', thus ‘rendering unto surveillance capitalism’ its continuous tithe of raw-material supplies". Experts typically test mobile and wearable apps for privacy and security vulnerabilities using methods such as man-in-the-middle attacks, eavesdropping, and packet injection~\cite{cusack2017assessment, fuster2023analysis} to prevent attacks and data misuse.

To sum up, risks and benefits assessments for AI in mobile and wearable uses have been hitherto hindered by two primary limitations. First, these analyses often focus on individual AI use cases or only a selection of few, potentially overlooking the broader spectrum of uses. Second, assessments are predominantly conducted manually by mobile and wearable experts or researchers, identifying potential benefits and risks through scoping reviews. Overcoming these limitations requires the development of new approaches that automate or semi-automate part of the assessment process. Next, we turn to LLMs, which present a promising ground for exploring this approach.

\subsection{LLMs and Prompt Engineering}
\label{sec:related-prompt}

LLMs such as OpenAI's ChatGPT~\cite{OpenAI}, which are trained on a large corpora of text using cutting-edge deep learning techniques~\cite{Wei2023AnOO}, have the capability to produce text that closely resembles human writing in response to various prompts. These models are extremely efficient for tasks such as answering queries and generating content, including text annotation~\cite{gilardi2023chatgpt,chi_llm_coauthor, chimLLM_stories} and powering chatbots that assist with mental health concerns~\cite{chen2023llm, sharma2023cognitive}. LLMs are known for providing insights that often exceed general public knowledge~\cite{gilardi2023chatgpt}, playing a significant role in collaborative processes between humans and AI~\cite{chimLLM_stories, chi_human_ai_colab, CHI_AI_CHAINS, chi_llm_coauthor}, and in some instances, their performance can match that of expert opinions~\cite{aydin2022openai, dowling2023chatgpt, byun2023dispensing}. To achieve the desired output from LLMs, a set of best practices for prompt engineering has emerged~\cite{Saravia_Prompt_Engineering_Guide_2022, OpenAIGuide}. Typically, these practices can be grouped into five areas. 

The first area concerns the choice of the prompt's elements. This includes the system's role, which is key to initiating interactions with the language model as it primes the model to behave in a manner that aligns with the role's expectations~\cite{microsoftazure2023}. Whether the model is instructed to act as a mobile developer, a creative assistant, or a project manager, this initial framing is a critical step in guiding the model’s responses and ensuring they are contextually relevant and appropriately styled. Additionally, the role can be extended with additional cues to direct the model to the desired output (e.g., initial words for the sentence the model is expected to complete~\cite{microsoftazure2023, OpenAIGuide}).

The second area is about the prompt's elements arrangement. For example, the order in which elements are placed greatly affects the outputs due to a phenomenon called recency bias. It is recommended to place the most important elements of the prompt at the beginning and end~\cite{liu2023lost}.

The third area is about the instructions to the model. 
Successful instruction requires clear and precise communication of tasks. It involves choosing the right action verbs (e.g., describe, summarize, explain), specifying the preferred format of the output (e.g., JSON or a single sentence), incorporating knowledge that is required for the model to learn. Part of the instructions are also different learning methods that can boost the model's performance. Beyond zero-shot learning, where models depend only on their pre-existing knowledge, few-shot learning provides task-specific examples of input-output pairs, enhancing performance on specialized tasks. Chain-of-thought (CoT) reasoning, which prompts the model to provide step-by-step answers, is especially beneficial for tasks that involve counting and mathematical problems~\cite{wei2021finetuned, brown2020language, NEURIPS2022_9d560961}.

The fourth area is about adjustment of model parameters. For example, temperature modulates the balance between unpredictability and predictability in the produced text. Higher temperature values encourage more varied responses, while lower values result in outputs that are more deterministic.

The fifth area is about the evaluation of the output. LLMs may generate biased or incorrect content, a phenomenon known as hallucinations \cite{dziri2022origin}. Therefore, for tasks that require certain levels of quality or creativity, the model's output should be evaluated. Evaluations may include human reviewers to guarantee that the content adheres to quality benchmarks and remains authentic~\cite{eldan2023tinystories, dziri2022origin}.

As highlighted, traditional risks and benefits assessments of AI for mobile and wearable uses have been sporadic, fragmented, and resource-intensive. LLMs incorporating these best practices have the potential to uncover benefits and risks in an automated manner.
\section{Author Positionality Statement}
\label{sec:positionality}

Recognizing the importance of author positionality is essential for transparently examining our perspectives on methodology and analysis~\cite{olteanu2023responsible,frluckaj2022gender}. In this paper, we situate ourselves in a Western country (United Kingdom) during the 21\textsuperscript{st} century, writing as authors primarily engaged in academic and industry research. Our team comprises two males and two females from Southern, Eastern, and Central Europe with diverse ethnic and religious backgrounds. Our combined expertise covers a range of areas, including human-computer interaction (HCI), ubiquitous computing, software engineering, artificial intelligence, natural language processing, data visualization, and digital humanities.

It is also important to acknowledge the use of a large language model as part of our methodology. As researchers from a primarily Western institution, we understand the importance of broadening the perspectives on methodology presented in this paper, and retain responsibility for the content and interpretation of the findings. Consequently, our positionality may have influenced the subjectivity inherent in selecting our methodology, designing our study, and interpreting and analyzing our data.

\section{Methods}
\label{sec:methods}

We developed three LLM prompts (Figure~\ref{fig:teaser}). The first prompt generated a wide range of AI uses for mobile and wearables, the second classified each use (whether it is unacceptable, high risk, or low risk) according to the EU AI Act~\cite{EUACT2021}, and the third determined whether each use is helpful for reaching Sustainability Development Goals~\cite{sdgs}. We chose the EU AI Act for classifying a use's risks because it is the most advanced risk-based legal framework for AI systems, and the SDGs for determining a use's benefits because it is globally recognized and allows for evaluation of long-term impacts related to people, the planet, prosperity, peace, and partnerships. Each prompt was manually validated by two authors to verify the correctness of its outputs. We also consulted a legal and compliance expert to further validate the prompt's outputs as well as conducted a crowdsourcing study with a cohort of nine individuals with legal backgrounds who were recruited from Prolific, confirming the accuracy of our approach to be over 85\%. As our LLM model, we chose OpenAI's GPT-4\footnote{\url{https://cdn.openai.com/papers/gpt-4.pdf}} due to its best performance across benchmarks,\footnote{\url{https://lmsys.org/blog/2023-06-22-leaderboard/}} and its demonstrated capability in interpreting legal documents~\cite{zheng2023llm}. To allow for reproducibility, in line with open-source language models, we made our prompts and code publicly available.\footnote{Available at the project's page: \url{https://social-dynamics.net/risky-mobile}}

\subsection{Prompt \#1: Generating Uses}\label{sec:generation}
This prompt generates a list of AI uses for mobile and wearable technologies. It achieves this by introducing two innovative prompt elements---a cue for different \emph{domains} where these technologies could be applied, and a cue for five \emph{components} that guide the specific description of each use and enable risk assessment as per the EU AI Act.

Prompt \#1 combines three elements: \emph{system role}, \emph{instructions}, and \emph{output format}. 

First, the \emph{system role} is a \emph{Senior Mobile and Wearable Systems Specialist, with experience in sensing technologies and their applications in multiple domains;} this choice guides the model to generate content aligned with advanced expertise required of this position~\cite{giray2023prompt}. We selected this role because it brings a comprehensive understanding of both the technical aspects and practical deployment of wearable technologies, ensuring our study addresses real-world applicability alongside theoretical advancements.  Compared to other potential roles (e.g., a Health Informatics Specialist, Consumer Electronics Product Manager, or IoT Architect), a more broader expertise is particularly suited to guide the development and deployment of mobile and wearable technologies. Additionally, we included a cue to further contextualize the role. This cue is a list of mobile and wearable sensors (e.g., accelerometers, gyroscopes, photoplethysmography, cameras, and global positioning systems) along with their possible placement (e.g., placed in various objects or parts of the body, such as the torso, wrist, and pocket) and their technical capabilities (e.g., monitoring human behavior, inferring physical, mental, emotional, and social status).

Second, the \emph{instructions} include: (1) the definitions of five components~\cite{Golpayegani2023Risk} that are required for risk assessment as per the EU AI Act~\cite{EUACT2021}. The five components are: the \emph{domain}, that specifies the industry or sector (e.g., health); the \emph{purpose}, that explains the goal (e.g., monitoring fitness); the \emph{capability}, that describes the technology behind the use (e.g., mood and stress level tracking); the \emph{AI user}, who operates the system (e.g., doctors); and the \emph{AI subject}, who is affected by the system (e.g., patients); (2) a list of application 46 domains (e.g., health, family, law enforcement), systematically developed based on the EU AI Act by selecting eight high risk domains listed in Annex III and 32 additional domains inferred from the Act's text and amendments~\cite{amendments2023}. This list guides the LLM to generate specialized and nuanced uses across individual, interpersonal, institutional, community, and public policy contexts; (3) the task of generating three AI uses for each of the 46 application domains.

Third, the \emph{output format} specifies how the LLM's output should be displayed in the five-component format (e.g., for monitoring fitness using wearables [Domain, Purpose, Capability, AI user, AI subject] becomes [``Health'', ``Monitoring fitness'', ``Mood and stress level tracking'', ``Doctors'', ``Patients'']).
\smallskip

\noindent\textbf{Manual validation.} After generating the uses, two authors with experience in mobile and wearable computing reviewed each one of them. The authors independently assessed each generated use to determine whether it is correct and whether it is existing or upcoming. They considered all 138 generated uses to be correct, with 128 (93\%) already existing (practically viable and in use), 9 (6\%) being upcoming (practically viable but still in early prototype stage), and 1 being unlikely (practically viable but only discussed as possibilities in research papers).

\subsection{Prompt \#2: Classifying Use Risks}\label{sec:safety}
This prompt classifies the risks of each generated use according to the EU AI Act~\cite{EUACT2021}. It combines four elements: \emph{system role}, \emph{input}, \emph{instructions}, and \emph{output format}. First, the \emph{system role} is a \emph{Senior AI Technology Expert, specializing in compliance with the EU AI Act}. Second, the input includes articles from the EU AI Act and its Annex III regarding high risk AI systems. Third, the instructions use a Chain-of-Thought approach (i.e., dividing the task into a series of smaller, intermediate reasoning steps that lead to the final output~\cite{NEURIPS2022_9d560961}). This approach involves three steps: (1) drafting a concise description of the system's use through the five component-format; (2) producing a risk classification as per the EU AI Act (whether the system's use is unacceptable, high risk, or low risk); and (3) providing justification for the classification. Fourth, the output format of risk assessment as per the EU AI Act contains three fields: ``[Use Risk Classification], [Classification Justification], [Relevant EU AI Act Article].'' 
\smallskip

\noindent\textbf{Manual validation.}
Two authors familiarized themselves with the EU AI Act articles, and manually classified each of the 138 generated uses as either unacceptable, high risk, or low risk. The validation criteria aligned with the specific requirements and definitions outlined in the EU AI Act concerning risk levels associated with AI uses. We defined the following criteria: potential impact on privacy and data protection, and the likelihood of adverse outcomes affecting individuals or groups. Additionally, we considered the operational contexts in which a mobile or wearable use is deployed such as the type of data collected and the functionality of the device. 80 uses (58\%) were identified as high risk, 57 (41.3\%) as low risk, and 1 (0.7\%) use as unacceptable risk. The agreement between the two authors and the LLM's classification was 85\%. In the rest 15\% of cases in which LLM was wrong, the disagreements were about uses related to vulnerable groups, personalization, government and democracy, and critical infrastructure (e.g., energy), all of which the LLM classified as low risk instead of high risk. The only use categorized as unacceptable risk (where both the two authors and the LLM agreed) was related to enhanced surveillance capabilities using cameras. These disagreements were resolved in two steps. First, the two authors held an initial review meeting in which they discussed their assessments against those done by the LLM, and consulted the rest of the research team. Second, we consulted a compliance expert specialized with the EU AI Act to inspect the 15\% of the uses that were wrongly classified by the LLM. The expert is an academic affiliated with TU Delft in Netherlands and works at the intersection of technology, policy and management. The expert manually classified each use's risk (i.e., unacceptable, high risk, or low risk), confirming that indeed these uses were wrongly classified by the LLM. Upon the consultations with the team and the expert, we reached the final decision.

\subsection{Prompt \#3: Determining Use Benefits}\label{sec:SDGs}
This prompt determines whether each generated use is helpful for reaching  Sustainability Development Goals~\cite{sdgs}. It combines four elements: \emph{role}, \emph{input}, \emph{instructions}, and \emph{output format}. First, the role is a ``Senior Specialist in the field of Mobile and Wearable Technologies with a dedicated focus on understanding, promoting, and implementing the SDGs''. Second, the input includes the 17 definitions of the SDGs, covering a broad range of potential areas for beneficial AI applications across the economy, society, and the environment. Third, the instructions (using Chain-of-Thought as in the previous prompt) involve: (1) summarizing the system's use based on the five-component description; (2) listing the benefits, reflecting how the use promotes each SDG. Fourth, the output format contains two fields: ``[SDG Goal], [Reasoning for Promoting the Goal]''. 
\smallskip

\noindent\textbf{Manual validation.} Two authors familiarized themselves with the 17 definitions and descriptions of the UN Sustainable Development Goals, and manually reviewed the feasibility of benefits associated with each goals~\cite{sdgs}. To ensure a robust manual validation, the authors defined the following criteria: \emph{a)} the applicability of the use to an SDG goal is evident; and \emph{b)} the feasibility of benefits is evaluated based on current technological capabilities. The agreement between the two authors and the LLM's output was 61\%. Disagreements were mainly related to goals 5 (gender equality), 16 (peace, justice and strong institutions) and 17 (partnerships for the goals), all of which were manually corrected. Disagreements were resolved in a similar way to classifying uses risks. First, the two authors discussed disagreements among them, and then consulted the rest of the research team to reach a final decision.

\subsection{Crowdsourcing Study}

\noindent\textbf{Setup and Procedure} We developed a web-based survey and administered it on Prolific\footnote{\url{https://www.prolific.com/}}, which included the LLM classifications for each use (Figure~\ref{fig:crowdsourcing}). The survey consisted of six pages. The first page outlined the study's description and the two tasks that the crowdworkers had to execute. Initially, they were asked to read the definitions of risky' and beneficial' uses, and then they were presented with assessment cards for 46 uses to assess. The second and third pages included the definitions of risky and beneficial uses according to the EU AI Act and the SDGs, respectively. The fourth and fifth pages presented 46 uses (23 each in page), where trap questions were interspersed to ensure data quality. We added two types of trap questions: the first type included two attention checks: \emph{When asked for your favorite color, you must select `Green'}; and \emph{When asked for your favorite city, you must select `Rome'.} Participants had to correctly respond to these checks after completing each task. The second type included an unacceptable use (i.e., a social scoring system based on facial recognition in public spaces) according to the EU AI Act, which we deliberately classified as low risk. Participants who failed either of the attention checks or to correct the classification of this use were disregarded and not paid. Additionally, we disabled pasting from external sources and editing previous responses to ensure original and thoughtful answers. The final page included a thank-you note and redirected participants on Prolific. To ensure reproducibility and assist researchers who aim to use our methodology, we have made the code for our survey publicly available.\footnote{Available at the project's page: \url{https://social-dynamics.net/risky-mobile}}

\noindent\textbf{Participants.} To recruit individuals with a background in legal and compliance, we relied on Prolific's screening criteria. We searched for participants likely involved in revising AI systems as part of their legal roles, using AI at least 1-6 times a week. In total, we recruited 9 participants with ``legal'' as the ``function in organization'', suggesting that these individuals had prior legal expertise (Table~\ref{tab:userstudy-demographics}). Six self-identified as males and three as females, with their ages ranging between 21 and 49 years old ($\mu = 37$ and $\sigma = 8.73$). Their current countries of residence were within the European Union, including participants from Italy, Poland, Germany, the Netherlands, Portugal, and Slovenia. On average, participants took 40 minutes to 1 hour to complete a batch of 46 uses, and were compensated \euro{9.29}.
\smallskip

\begin{figure*}
    \includegraphics[width=\textwidth]{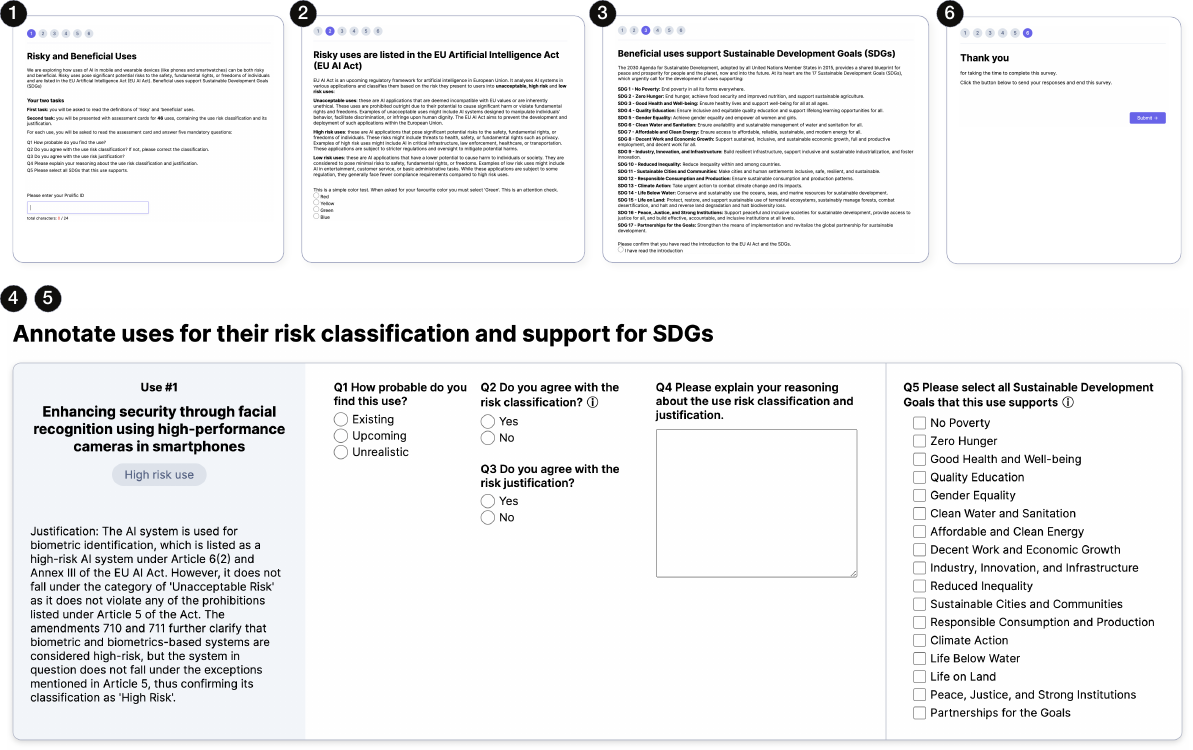}
    \caption{Crowdsourcing study survey. Participants were given instructions about the study (1) and were provided the definitions of risk classifications as per the EU AI Act (2) and the definitions of the 17 Sustainable Development Goals (3). 
    Then participants were presented with 46 mobile HCI uses (4,5) and asked to assess how probable is the use (Q1), whether they agree with the LLM-generated risk classification (Q2) and its justification (Q3), explain their reasoning about the classification and justification (Q4), and select the SDGs that they believe the use supports. After annotating all uses, they were redirected to the Prolific confirmation page (6).}
    \label{fig:crowdsourcing}
\end{figure*}

\begin{table*}[t!]
    \centering
    \caption{Demographics of participants in the crowdsourcing study.}
    \label{tab:userstudy-demographics}
    \begin{tabular}{p{2.2cm} p{0.7cm} p{1cm} p{1cm} p{2cm} p{4cm}}
    \toprule
    \textbf{Batch} & \textbf{ID} & \textbf{Gender} & \textbf{Age}  & \textbf{Country} & \textbf{Weekly AI use} \\ \midrule
    
    \multirow{3}{*}{} & P1 & Male & 33 & Italy & Every day  \\
    B1 & P2 & Male & 41 & Poland & 2-6 times a week \\
     & P3 & Male & 38 & Germany & 2-6 times a week \\
     \midrule            
    \multirow{3}{*}{} & P4 & Male & 40 & Germany & 2-6 times a week  \\
    B2 & P5 & Female & 47 & Italy & 2-6 times a week \\
     & P6 & Male & 25 & Poland & Multiple times every day \\
     \midrule      
     \multirow{3}{*}{} & P7 & Female & 49 & Portugal & Every day  \\
    B3 & P8 & Female & 36 & Slovenia & Every day \\
     & P9 & Male & 21 & Germany & Multiple times every day \\
     
    \bottomrule
    \end{tabular}
\end{table*}

\noindent\textbf{Analysis and Results.} Each use was annotated by three crowdworkers, and each of them annotated a total of 46 uses (i.e., all 138 uses were divided into three batches).

When they classified the three risk levels based on the  EU AI Act classifications, there was a full agreement in 48\% of uses. Full agreement was about uses that have clear benefits and utility (e.g., tracking health status); are widely-adopted (e.g., facilitating mobile payments); or are in already regulated domains (e.g., enhancing food safety). Conversely, disagreements were about uses that present privacy, security, and ethical concerns (e.g., tracking vulnerable groups like children and the elderly); or are deemed emerging (e.g., mobile VR for educational content).  We then used majority voting to establish the final classifications by the crowdworkers and compared them against the LLM classification. The agreement  was 92.75\%. By manually inspecting the disagreements, crowdworkers corrected the LLM classification and cited similar reasons  to those reported during the two authors'
manual validation (e.g., cases involving uses about democracy and personalization).

When the crowdworkers then determined the 17 SDGs applicable to the uses at hand (one use could have multiple SDGs assigned), there was a full agreement in 11\% of uses (i.e., crowdworkers selected the exact same SDG(s) for each use). Full agreement was about uses that had a clear positive impact on daily life (e.g., monitoring health status with wearables); aligned with broader societal goals (e.g., improving election transparency); or  improved safety and security (e.g., enhancing surveillance capabilities for public safety). Conversely, disagreements were about uses with varied perceptions of usefulness (e.g., tracking emotions for well-being); or had privacy, security, and ethical concerns (e.g., involving facial recognition). For each use, we then took the union of the three crowdworkers' annotations to establish the final classification and compared this list against the LLM classification. The agreement was 37.89\%. Compared to the two authors' validation (61\% agreement with the LLM classification), crowdworkers had lower agreement. This greater discrepancy could be attributed to the difficulty in determining the applicability of SDGs, which may require expertise beyond the legal domain. Additionally, it requires complex cross-referencing of the use with 17 different goals and their numerous targets.

\section{Results}
\label{sec:results}
Having the set of 138 generated uses and their associated risks and benefits (detailed in Table~\ref{tab:all_uses} in the Appendix; the notation \# indicates the use's ID), we report the results in two steps. First, for each SDG, we computed the proportion of beneficial uses categorized as low risk (\S\ref{subsec:low_risk}) and high risk (\S\ref{subsec:high_risk}), and then discuss example uses from the goals with the highest proportions of each risk category. Second, we conducted a thematic analysis to understand the reasons why a use is risky or not, and why it is beneficial (\S\ref{subsec:thematic}). Note that among the 138 uses, only one was classified as unacceptable according to the EU AI Act. This use involves \emph{enhancing surveillance capabilities using mobile cameras in public spaces} (use \#64). In our analysis, we did not focus on unacceptable uses because they are banned. Instead, high-risk AI uses, which are allowed to be deployed, must be subject to an impact assessment for compliance purposes. Therefore, our analysis focuses on distinguishing between high-risk and low-risk uses.

\subsection{Uses Promoting SDGs and Being Low Risk}
\label{subsec:low_risk}

Uses that pose low risk primarily promote goals related to environmental protection, such as life below the water (goal 14), life on land (goal 15), and clean water and sanitation (goal 6) (navy bars in Figure \ref{fig:goals_safety}). These uses are classified as low risk by the EU AI Act. Next, we discuss examples of these uses.

\begin{description}
\item \emph{Monitoring crop health and growth, use \#49.} IoT sensors are used to continuously monitor environmental parameters such as temperature, humidity, and soil moisture, alerting farmers via mobile apps if conditions are unsuitable for their crops~\cite{tyagi2020crop}. Such uses promote sustainable agricultural practices and environmental conservation (goal 15). Similar uses can involve in vivo plant sensors (e.g., organic electrochemical transistor-based biosensors~\cite{vurro2023vivo}) that detect nutrient deficiencies or drought stress and allow for the precise application of agrichemicals, fertilizers, and water, thereby maximizing yields~\cite{roper2021emerging}. 

\item \emph{Tracking livestock health and location, use \#50.}  GPS collars have been effectively used to track livestock, including their location, movement patterns, and grazing behavior (e.g., by creating ``virtual geofences'' which alert farmers if an animal strays from a designated area~\cite{ilyas2020smart}); uses that enables precise livestock management (goal 15). These collars can also be equipped with more specialized sensors to monitor the behavioral and physiological parameters of livestock, allowing farmers to assess the animals' health and welfare over time. Examples include pedometers and microelectromechanical (MEMS) activity sensors, which are easily attached to animals~\cite{helwatkar2014sensor}. Additionally, tracking livestock movements and interactions can help in understanding and managing disease outbreaks, such as those caused by the influenza virus.~\cite{animal-health}. 

\item \emph{Optimizing irrigation systems, use \#51.} 
IoT sensors have been used to optimize water usage (goal 6) and predict harvest times (goal 15). \citet{mohamed2021smart} demonstrated a smart irrigation system equipped with sensors for monitoring water levels, irrigation efficiency, and climate conditions. Wearable soil moisture sensors, strategically placed throughout a field, can now transmit data to a central system, accessible via mobile devices. Farmers can adjust irrigation schedules directly from their phones, optimizing water usage based on real-time soil moisture data. This approach not only preserves (and saves) water but also ensures that crops receive the appropriate amount of water at the right time, which is essential for crop health and yield~\cite{obaideen2022overview}. 

\item \emph{Monitoring environmental conditions, use \#112.} IoT sensors are instrumental in monitoring a wide range of environmental parameters (e.g., heat, sound, pressure, temperature, air quality, water quality), promoting sustainable management of terrestial ecosystems (goal 15). Integrating these sensors into smart environmental monitoring systems allows for large-scale environmental assessments~\cite{andrachuk2019smartphone}. Battery-free, wireless cameras and sensors can now be deployed in underwater environments to monitor water quality, marine life, and underwater ecosystems~\cite{afzal2022battery} (goal 14). These sensors are not only limited to monitoring air and water quality but also extend to waste management~\cite{laha2022advancement}. For example, in domestic environments, they can effectively monitor conditions such as temperature, humidity, and gas levels, with the data being easily accessible through mobile apps or web pages~\cite{chaudhari2018effective}. 

\item \emph{Tracking wildlife movements, use \#113.} GPS collars are used to study animal behavior, and understand their habitat preferences and migration behaviors~\cite{allan2013cost} (goal 15). These technologies also enable real-time monitoring of wildlife movements, including proximity, geofencing, movement rate, and immobility~\cite{wall2014novel} and can enable modeling epidemics in wildlife hosts, such as outbreaks of dolphin morbillivirus~\cite{morris2015partially}.
Similar uses can involve multi-sensory wearable devices that employ neural networks for classifying animal behavior in national parks~\cite{rios2016sensor} or for predicting animal personality~\cite{meegahapola2023quantified}. Wireless sensor networks track small turtles, monitoring their micro-climate and hibernation periods~\cite{joshi2008gps}, while GPS-powered systems can effectively track wild animals straying from sanctuaries, aiding in poaching prevention~\cite{gor2017gata}.
\end{description}

\begin{figure}
    \centering
     \includegraphics[width=0.95\columnwidth]{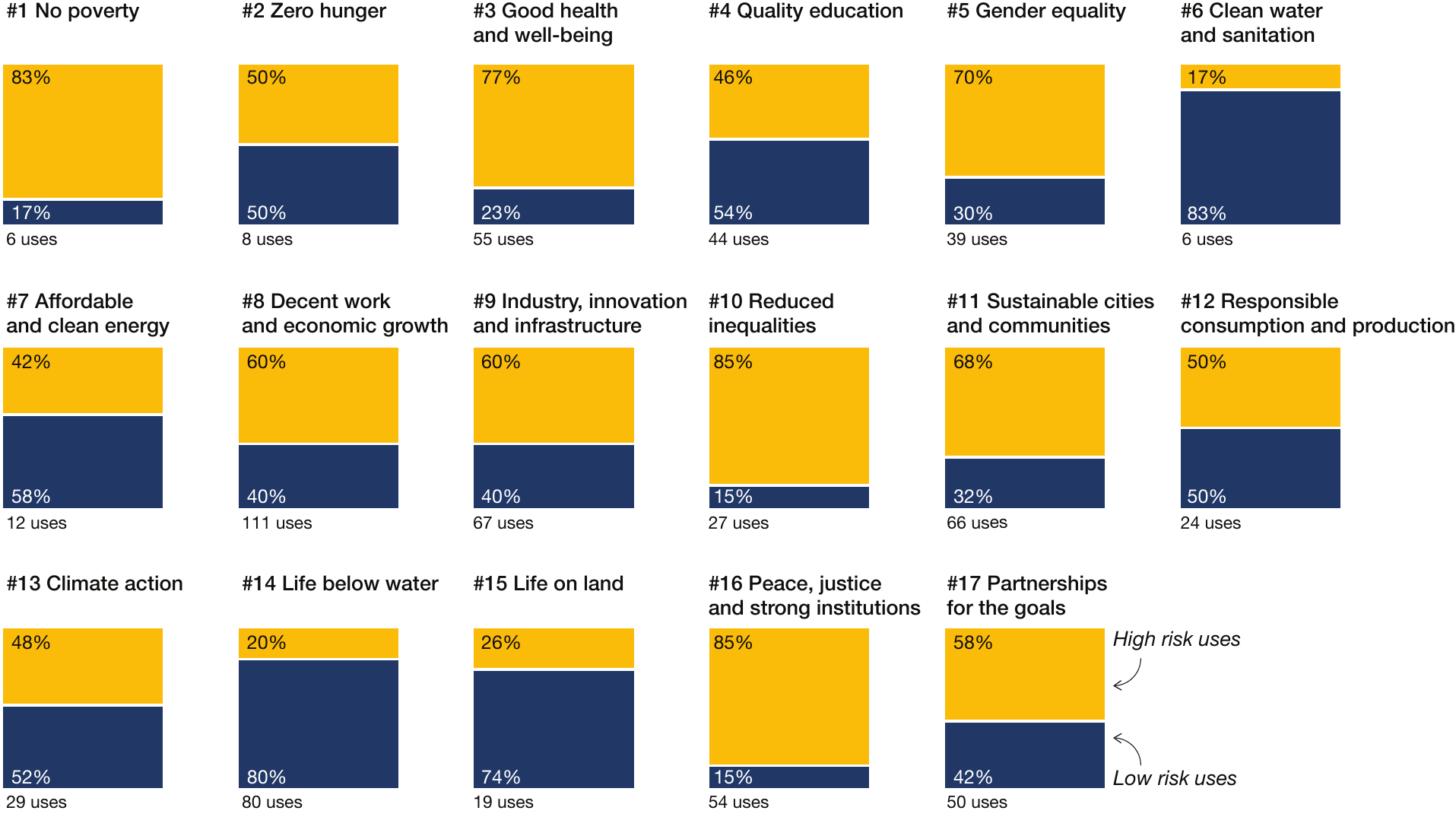}
    \caption{Percentages of uses categorized as low risk and high risk for each goal. The total number of uses per goal is in parentheses below the goal's name.}
    \label{fig:goals_safety}
 \end{figure}

\subsection{Uses Promoting SDGs yet Being High Risk}
\label{subsec:high_risk}

Uses that promote certain goals but pose high risks primarily relate to peace, justice, and strong institution (goal 16), good health and well-being (goal 3), and reduced inequalities (goal 10) (gold bars in Figure \ref{fig:goals_safety}). These uses entail high risks, either due to the use of risky technology (e.g., cameras for facial recognition, or motion sensors paired with physiological sensors), or because they operate in high risk domains such as migration management, military and defense, or healthcare involving individuals from vulnerable groups. These uses include:

\begin{description}
    \item \emph{Enhancing security through facial recognition, use \#1.} High-performance cameras in smartphones have been typically used for various authentication purposes such as phone unlocking, banking, and access control, by analyzing face images~\cite{corsetti2019face, buriro2016hold}. Such uses ensure safety and security by preventing unauthorized access (goal 16). However, as per EU AI Act's Article 6 ``biometric identification and categorisation of natural persons'' is under the high risk category. However, the very same technology (i.e., cameras) may well be used for crowd surveillance; a use that has unacceptable risks. 
    \item \emph{Identifying individuals in criminal investigations, use \#3.} Smartphones and wearable watches can provide crucial information in criminal investigations (e.g., documenting an individual's activities), which can be used to determine the veracity of witness testimony~\cite{dorai2020data}. For example, a smartphone synced with a wearable watch can offer evidence related to the execution of a violent act, aiding in pinpointing the time and nature of the incident~\cite{mcnary2018using} and ensuring safety and security (goal 16). Suspicious pattern detection techniques from mobile device data can also help identify individuals involved in activities such as cyberbullying or low-level drug dealing~\cite{barmpatsalou2018mobile}. According to EU AI Act's Article 6, the amendment 711 classifies such a use as risky because it can used by ``law enforcement authorities for making individual risk assessments of natural persons in order to assess the risk of a natural person for offending or reoffending or the risk for potential victims of criminal offences''.    
    \item \emph{Monitoring children's location for safety, use \#4.} Multi-sensor wearable such as wristbands and IoT-enabled devices have been designed to track children's location, body temperature, humidity, and heartbeat, with the capability to alert parents through SMS text messages in case of emergency~\cite{chowdhury2019multi, sunehra2020raspberry}. Such uses strive for improving children's well-being and safety (goal 3). However, as per the EU AI Act's article 69 and amendment 634, the use is risky because it ``may affect vulnerable persons or groups of persons, including children, the elderly, migrants and persons with disabilities''.
    \item \emph{Monitoring elderly health and activity, use \#5.} Wearable devices, such as wristbands equipped with sensors like accelerometers, gyroscopes, and heart rate monitors, are used for activity detection and monitoring in elderly care and rehabilitation~\cite{nweke2019data}, improving their well-being and safety (goal 3). More broadly, IoT-based smart healthcare systems provide real-time monitoring of the elderly's location, activity patterns, and health status, enabling early detection of health risks and improved response to emergencies~\cite{lin2022iot}. As this use may affect vulnerable groups is classified as risky under the EU AI Act's article 69 and amendment 634. However, the same on-body sensors could be used for surveillance, a concept Yuval Noah Harari termed ``under-the-skin surveillance''~\cite{harari}, introducing unacceptable risks. This type of surveillance goes beyond traditional monitoring by collecting intimate physiological data directly from a person's body, potentially invading privacy and autonomy on a deeply personal level.
    \item \emph{Enhancing mental health treatment,  use \#12.} Wearable sensors for electrodermal activity and photoplethysmography provide real-time physiological data crucial for mental health monitoring and treatment~\cite{fletcher2010wearable}, ensuring good (mental) health and well-being to individuals (goal 3). A comprehensive study highlighted 46 systems focusing on continuous monitoring, diagnosis, and care of various mental disorders through mobile and wearable technologies~\cite{bardram2020decade}. Music streaming services, when repurposed as therapies for affective disorders, exemplify the potential of these technologies in mental health treatment~\cite{schriewer2016music}. Additionally, smartphones and wearable devices support mental health research by facilitating the collection of novel naturalistic and longitudinal data relevant to psychiatry~\cite{torous2017new}. However, similarly to uses affecting children or elderly, this use is again classified as risky the EU AI Act's article 69 and amendment 634 as it may affect vulnerable groups who suffer from mental health problems.
    \item \emph{Preventing unauthorized access, use \#43.}
    Wearable devices are used in access control systems to protect information and define restrictions on information handling~\cite{diez2019lightweight} in military facilities, prevent unauthorized data leakage, and safeguarding sensitive information~\cite{yang2018backscatter} (goal 16). These devices can also be used in user authentication, improving the security of mobile payment systems and other applications where sensitive data is involved~\cite{wong2016enhanced}. However, As per the EU AI Act's Article 6 and amendment 711, the use is risky because it can be used for ``biometric identification''.
    \item \emph{Predicting infrastructure failures,  use \#62.} Wearable sensors have been applied to predict health-related issues, such as COVID-19 incidence, by tracking the spread of the virus, demonstrating their potential in predictive health infrastructure~\cite{seshadri2020wearable}. Analyzing mobility patterns in IoT networks (e.g., georeferenced traces that citizens leave when they navigate a city) are used to predict infrastructure failures and optimize road traffic~\cite{fazio2021innovative} such as road traffic, ensuring inclusive access to society (goal 16). However, the EU AI Act's amendments 713 and 714 classify the use as risky because it is intended to be used as ``safety component in the management and operation of road traffic and the supply of water, gas, heating and electricity''.
    \item \emph{Enhancing surveillance capabilities, use \#64.}
     Mobile cameras, accelerometers, and gyroscopes are typically employed for continuous monitoring and activity detection in surveillance applications~\cite{cornacchia2016survey}, ensuring safety by maintaining public order and general welfare in a democratic society. The concept of ``sousveillance,'' which refers to surveillance from below as opposed to from above, has been facilitated by wearable devices, generating new types of information in social surveillance situations~\cite{mann2003sousveillance}. Mobile sensors, like those in Android smartphones, are used in real-time surveillance systems for tasks such as human face tracking~\cite{saraceni2012active}, even in public spaces. High-tech gear like textile-based wearable microphones, miniature cameras, and wireless personal displays can also enhance surveillance capabilities while appearing natural to the user~\cite{venkateswarlu2006wearable}. However, as per the EU AI Act's article 5.1, the ``use of real-time remote biometric identification systems in publicly accessible spaces, unless and in as far as such use is strictly necessary''.    
\end{description}

\subsection{Understanding the Reasons Why a Use Is Low Risk, High Risk, and Beneficial}
\label{subsec:thematic}
Two authors conducted an inductive thematic analysis (bottom-up) to understand the reasons why a use is risky or not, and why it is beneficial (i.e., as per Prompt 2 and 3 outputs), following established coding methodologies~\cite{saldana2015coding, miles1994qualitative, mcdonald2019reliability}. The authors used sticky notes on the Miro platform~\cite{miro2022} to capture the reasons and inductively construct the themes. They held four meetings, totaling 8 hours, to discuss the emerging themes that arose during the analysis process. Next, we discuss the resulting themes and grouped them into three categories, that is, reasons for uses being: \emph{a)} low risk; \emph{b)} high risk; and \emph{c)} beneficial.

\smallskip
\noindent\textbf{Low Risk:} 
Low risk uses were about the \emph{environment and sustainability}, and \emph{logistics}. The reasons why these uses are low risk are because they handle data about non-human subjects and the environment, and operate in domains that are not considered high risk. For environment and sustainability, mobile and wearables enable environmental monitoring (e.g., air and water quality) that allow for timely interventions to mitigate pollution; and monitor energy consumption and provide feedback to users for encouraging energy-saving behaviors and contributing to carbon footprint reduction. For logistics, wearables devices equipped with RFID or GPS technology enable real-time tracking of goods that can significantly reducing the time needed for inventory management and supply chain operations; sensor-equipped wearables (e.g., soil moisture sensors) provide precise information on crop needs to help minimize the use of water and fertilizers. 
\smallskip

\noindent\textbf{High Risk:} 
High risk uses were about the \emph{sensitive data handling}, \emph{vulnerable user groups}, and \emph{automated decision making}. The reasons why these uses are high risk are because they generally collect sensitive data, even about vulnerable individuals at time or engage in automatic decision making aimed at behavioural change. For sensitive data handling, the uses involved collecting data with different levels of sensitivity: biometric data such as facial images or footprints, voice prints or gait data, and other personal data which can reveal information about physical, physiological or behavioural traits of people such as fitness data or financial records. For vulnerable user groups, the uses were affecting vulnerable persons or groups of persons, including children, the elderly, migrants, and persons with disabilities. For automated decision making, the uses were designed in a way to influence habits (e.g., recommender systems), leading to potential negative impacts on mental health and social interactions.

\smallskip
\noindent\textbf{Beneficial:} Independent to whether uses are risky or not, they were beneficial mainly because they \emph{foster participation and collaboration} in society, \emph{improve well-being}, and \emph{improve safety and security}. For fostering participation and collaboration in society, mobile and wearables enable seamless communication and connectivity, breaking down geographical and temporal barriers; allow users through apps to share their experiences, achievements, and challenges, fostering a sense of community and mutual support. For improving well-being, mobile and wearables allow for physical activity and sleep patterns monitoring; used to detect early signs of health issues and provide recommendations for stress management (e.g., meditation and psychological support). For improving safety and security, emergency response apps allow users to quickly alert emergency services; GPS tracking features on mobile and wearables are beneficial for monitoring the safety of children, the elderly, or individual in high risk environments. 
\section{Discussion}\label{sec:discussion}
Our method offers a semi-automatic way to assess the risks and benefits of AI uses in mobile and wearables using three LLM prompts. We found that low risk uses primarily benefit the environment and logistics. Conversely, high risk uses, while promising improved well-being, safety, and equality, come with significant concerns over sensitive data, vulnerable groups, and automated decision-making. These uses often collect biometric data from cameras or operate in high risk domains such as healthcare, particularly when involving vulnerable groups. Next, we discuss challenges in conducting such an assessment highlighting our lessons learned, and present practical solutions (e.g., a checklist) for balancing the trade-offs between risks and benefits in mobile and wearable technologies. 

\subsection{Challenges in Conducting Risks and Benefits Assessment Using LLMs}
The use of LLMs to assess the risks and benefits of AI (in our case, on mobile and wearable technologies) presented unique challenges, which are likely to be applicable to other domains as well. Although LLMs may be able to provide a comprehensive analysis across various use cases, risks, and benefits, their outcomes should be treated as preliminary steps rather than definitive conclusions. We specifically highlight three areas that require careful attention. 

First, a key challenge is ensuring the accuracy and relevance of the LLM' output, as these models heavily rely on the quality and scope of their training datasets. We overcome this challenge by employing a number of best prompt engineering practices~\cite{Saravia_Prompt_Engineering_Guide_2022, OpenAIGuide} such as specifying the model's role (e.g., senior mobile and wearable specialist), detailing its task(s), providing additional knowledge (e.g., excerpts of the EU AI Act), and defining the expected output format. 

Second, the complexity and often opaque nature of LLMs make their output's interpretation challenging. To ensure that the language model produced realistic uses, and, in turn, classified each use's risks and determined its benefits, two authors manually inspected the output in every step of the process. A human-in-the-loop approach helps to alleviate model's hallucinations. Additionally, by employing a Chain-of-Though approach~\cite{NEURIPS2022_9d560961}, we ensured that the model explained its reasoning to arrive at the final output.

Finally, LLMs might not fully capture all ethical and societal dimensions linked to mobile and wearables. More broadly, identifying risks associated with AI systems requires deep organizational, technical, and regulatory expertise~\cite{bucinca2023harmGeneration}, which can be achieved by integrating experts' knowledge in the process. The use of AI for evaluating AI systems (and their use) introduces a number of ethical considerations that require careful examination. One prominent concern is the recursive nature of such assessments, wherein AI systems are employed to evaluate outputs generated by similar AI systems, potentially leading to feedback loops that amplify biases or errors without sufficient human oversight~\cite{hagendorff2020ethics}. Another concern is about transparency. Documenting the decision-making processes, the criteria used for evaluations, and the limitations of the AI tools employed is essential for allowing independent verification and fostering trust~\cite{mittelstadt2016ethics}. For example, to allow for reproducibility, we made our code publicly available.

Despite the advancements in LLMs capabilities, there are inherent limitations in their ability to make judgments against complex ethical standards. Current LLMs may struggle to comprehend and apply nuanced ethical principles, leading to potential gaps and inconsistencies in such assessments~\cite{jobin2019global}. Addressing these limitations requires ongoing research aimed at enhancing LLMs ethical reasoning capabilities and ensuring alignment with universally accepted ethical norms and standards. Additionally, we foresee that semi-automated ways for conducting risk and benefits assessments should be complemented with human-in-the-loop validations and/or non-AI approaches (e.g., crowdsourcing studies), which we discuss next.

\subsection{Challenges in Conducting AI-Based Risk and Benefit Assessments Versus Human-Based Assessments}
Next, we turn into discussing challenges for conducting risks and benefits assessments using AI-based approaches (e.g., LLMs) versus human-based ones (e.g., crowdsourcing). While AI-based assessments offer speed, scalability, and consistency, our findings reveal several challenges that call for a combined approach incorporating both methodologies. 

The first challenge is about consistency and contextual judgment. Our AI-based assessment using LLMs demonstrated, to a great extent, high agreement rate with human judgments when assessing the risk classification level of mobile and wearable uses. However, this consistency may overlook nuanced interpretations that human experts can provide. Human assessors, especially those with specialized knowledge, can offer contextual judgments that AI might miss, as evidenced by the adjustments made by crowdworkers based on deeper understandings of democracy and personalization issues. In other words, this  illustrates the limitations of AI in areas requiring ethical consideration and complex contextual analysis, suggesting that while AI can guide assessments, it should not replace human judgment.

The second challenge is about the depth of understanding. This was particularly evident in the lower agreement rates on SDGs, where human expertise was crucial for comprehensive evaluation. This difference suggests that AI should operate as an exploratory tool, with final decisions deferred to human experts who can assess the broader ethical and societal implications of a use.

The third challenge is about reliability. The human-based approach introduced diverse judgment perspectives, but at the same time demonstrated potential inconsistencies. This variability, while introducing diverse viewpoints, also necessitates rigorous control measures to ensure reliability (e.g., using trap questions and attention checks). One possibility would be to use AI-based assessments as a baseline for consistency and comparison with human-based ones.

Given these challenges, we believe that a promising approach should combine both AI-based assessment and human-based ones. Such a hybrid approach would use AI for initial assessments due to its efficiency and scalability, and human experts for providing the final judgments. To aid researchers interested in adopting a similar approach in different contexts, we recommend following these steps: \emph{(1)} describe a use case using the five-component format suggested by \citet{Golpayegani2023Risk} (i.e., Domain, Purpose, Capability, AI User, AI Subject) as required for risk assessment under the EU AI Act~\cite{EUACT2021}; \emph{(2)} incorporate expert human validation with clear intervention protocols throughout the process; and \emph{(3)} implement continuous feedback loops between the AI-based assessments and the human-based ones.

\subsection{Balancing the Trade-Offs Between Risks and Benefits of AI in Mobile and Wearable Uses}
The use of mobile and wearables in sensitive or high-stakes domains underscores the need to balance risks and benefits. While these technologies contribute significantly to Sustainable Development Goals such as good health and well-being, reduced inequalities, and peace and justice, they also pose inherent risks due to their sensitive nature. Challenges include privacy concerns, potential misuse for surveillance, and the risk of exacerbating existing inequalities or societal issues. For example, sensitive personal data (e.g., health information) could be misused if not adequately protected, leading to unauthorized access, surveillance, or infringement on individual freedoms and rights. Additionally, it is hard to obtain large representative samples across diverse populations when training models for mobile and wearable uses, making such uses more susceptible to biases~\cite{sjoding2020racial}.

\begin{figure}[t!]
    \centering
    \includegraphics[width=0.9\columnwidth]{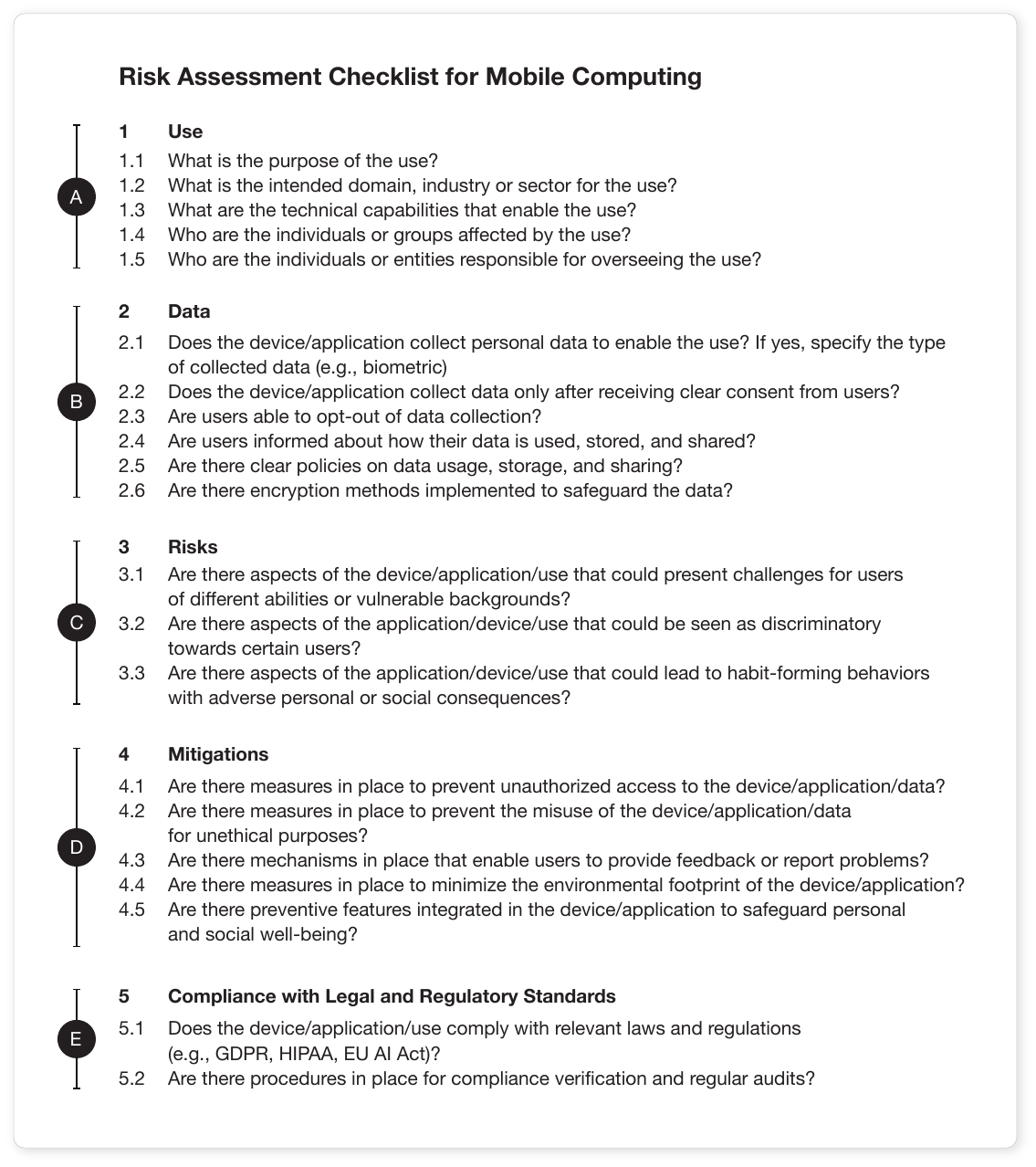}
    \caption{Risk Assessment Checklist for Mobile Computing. It helps to systematically consider a mobile or wearable system's: (A) use, (B) data, (C) risks, (D) mitigations and (D) compliance with legal and regulatory standards.}
    \label{fig:checklist}
 \end{figure}

To balance risks and benefits requires active research, policy updates, and open dialogue as to what is considered as an acceptable technology use~\cite{constantinides2022good}. For example, it is essential to enhance privacy protections beyond those outlined in the EU AI Act as mobile and wearables capabilities continue to evolve. This can be achieved not only by understanding the direct impact of these technologies (e.g., producing impact assessment cards similar to model cards~\cite{mitchell2019model}) but also by considering their broader societal implications. It is of prime importance to ensure that the benefits of these technologies do not exacerbate existing vulnerabilities or inequalities. Another solution would be for the mobile computing community to adopt practices for reporting harms and risks associated with the use of specific mobile or wearable sensors and technologies. Conferences such as the Neural Information Processing Systems (NeurIPS)~\cite{ashurst2020guide} and the International Conference on Machine Learning (ICML) have already started mandating statements that include ``risks associated with the proposed methods, methodology, application or data collection, and data usage''. More recently, \citet{olteanu2023responsible} argued that even responsible AI research needs impact statements, too; such statements aim at disclosing any possible negative consequences, contributing to more inclusive research. Additionally, researchers have proposed algorithmic impact assessments (AIAs) as a form of accountability for organizations that build and deploy automated decision-support systems~\cite{metcalf2021algorithmic}, and frameworks for bridging the gap between technical and ethical aspects of AI systems~\cite{kasirzadeh2021reasons}.
\smallskip

\noindent\textbf{Risk Assessment Checklist for Mobile Computing:} 
Drawing from similar initiatives in AI and Responsible AI, we developed a risk assessment checklist for mobile computing based on: the learnings from developing our semi-automatic method (\S\ref{sec:methods}) and the findings about what makes a use risky or not (\S\ref{sec:results}). These two translated into a five-section checklist, which aligns with recently agreed ways of reporting an AI system's use risks (including the indented use, risks, mitigations, compliance)~\cite{sherman2023riskProfiles, microsoft2022Assessment, ada_lovelace, equalAI2023nist}. However, we do not argue that the checklist is exhaustive, but rather aims to serve as a starting point for the Mobile HCI community to facilitate and foster transparency and accountability in the development and use of mobile and wearable uses.

The first section covers the intended use (Figure~\ref{fig:checklist}A). This is informed by the EU AI Act's requirements for risk assessment, and the fact that prior work emphasized the importance of documenting a system's use in a structured five-component format~\cite{Golpayegani2023Risk}. This format is not only suitable for risk assessment but also comprehensive enough to be usable by both technical and non-technical stakeholders (i.e., due to its simple yet compact structure).

The second section is about the data (Figure~\ref{fig:checklist}B). From the thematic analysis (\S\ref{subsec:thematic}), it was evident that the type of data (e.g., sensitive vs. non-sensitive) that a use handles plays a critical role in determining its risk level. For example, uses that handle data primarily about non-human subjects (e.g., animals, as per use \#50) and the environment (e.g., heat, sound, and air quality, as per use \#112), they are considered low risk. Conversely, uses that handle primarily sensitive data (e.g., facial images, as per use \#1), they are considered high risk.

The third section is about the risks (Figure~\ref{fig:checklist}C). From both the inspection of high risk uses (\S\ref{subsec:high_risk}) and the thematic analysis (\S\ref{subsec:thematic}), these uses tend to involve individuals from vulnerable groups or engage in automatic decision making aimed at behavioural change. Therefore, by formulating questions to identify and assess these risks, this section aims to preemptively address and mitigate potential negative impacts on individuals and society.

The fourth section is about the mitigations of risks associated with the specific mobile and wearable uses as well as the data they handle (Figure~\ref{fig:checklist}D). 

Finally, the fifth section is about the regulatory compliance (Figure~\ref{fig:checklist}E). Given that our method is directly influenced by the upcoming enforcement of the EU AI Act, this section aims to prompt for adherence with existing and upcoming regulations as well as procedures for verification and regular audits. Such procedures will be a legal necessity to monitor a use and determine its risk level. 

We foresee that this checklist can be applied across a range of stakeholders and use cases: \emph{AI practitioners} can use it to identify and assess risks at various stages of the product development cycle; \emph{AI teams} can use it to raise awareness and facilitate discussions about potential risks, leading to more comprehensive risk mitigation strategies; \emph{researchers} can use it to navigate the publication process and to help the community avoid rejecting risky technologies that may have positive impact; and \emph{end users} can use it to understand the various aspects (and AI's use) they should consider before using mobile or wearables.

\subsection{Limitations and Future Work}
We acknowledge three main limitations of our work that call for future research efforts. First, our method incorporates domain experts into the assessment process. The two authors who manually validated each prompt's output have experience in the design, development, and deployment of mobile and wearable technologies. While not achieving full automation, it expedites the experts' tasks. It systematically compiles a list of uses, presenting them in an organized manner for rigorous risk assessment. Consequently, experts can optimize their time, reducing the need for extensive literature reviews to identify potential uses, risks, and benefits. Similarly, our crowdsourcing study experienced relatively low participation because, in reality, only a handful of experts can provide the insights required to validate our method. At the same time, we chose the majority voting to establish the final classification among crowdworkers. However, this may present a situation wherein a majority's decisions overshadow those of a minority (a phenomenon also known as the `tyranny of the majority''~\cite{read2009majority, sumrall2001global}). Future studies could explore ways to form cohorts of experts that are broadly available and accessible to the research community, offering diverse perspectives in evaluating LLM-generated classifications. 

Second, the significant level of disagreement among the experts themselves highlights the subjective nature of such assessments. This suggests that even expert evaluations can be heavily biased by individual experiences, domain expertise, and cultural values~\cite{schaekermann2019understanding, hoth2016uncovering}. The diversity in the cultural and professional backgrounds of our experts, while enriching the study with varied perspectives, also introduced challenges in reaching consensus. This observation highlights the need for developing more robust methods to use expert disagreement constructively.

Third, LLMs may not cover all potential impact areas, particularly emerging ethical issues (e.g., existential risks) or unforeseen societal consequences, leading to an oversimplification of risks. This may lead to overly simplistic approaches in dealing with complex socio-technical AI systems, especially those that deal with applications domains flagged by the EU AI Act as high risk (e.g., concerning vulnerable groups, or law enforcement). The output of such an assessment is highly advisable to be vetted by lawyers and regulatory experts.

\section{Conclusion}
We proposed a method for assessing the risks and benefits of AI in mobile and wearable uses. Using our method, we generated 138 such uses of AI, while also evaluating their associated risks as defined by the EU AI Act and benefits in line with the UN Sustainable Development Goals; upon manual validation of our method, we confirmed its accuracy to be over 85\%. We found that a specific set of mobile computing uses (low risk) primarily benefit the environment and logistics. Interestingly, another set of uses (high risk) that hold significant potential in improving well-being, safety, and social equality are linked to risks involving sensitive data, vulnerable groups, and automated decision-making. To not dismiss the potential of these uses, we proposed a checklist for assessing the risks of mobile and wearable uses.


\balance
\bibliographystyle{ACM-Reference-Format}
\bibliography{main}


\begin{thebibliography}{128}


\ifx \showCODEN    \undefined \def \showCODEN     #1{\unskip}     \fi
\ifx \showDOI      \undefined \def \showDOI       #1{#1}\fi
\ifx \showISBNx    \undefined \def \showISBNx     #1{\unskip}     \fi
\ifx \showISBNxiii \undefined \def \showISBNxiii  #1{\unskip}     \fi
\ifx \showISSN     \undefined \def \showISSN      #1{\unskip}     \fi
\ifx \showLCCN     \undefined \def \showLCCN      #1{\unskip}     \fi
\ifx \shownote     \undefined \def \shownote      #1{#1}          \fi
\ifx \showarticletitle \undefined \def \showarticletitle #1{#1}   \fi
\ifx \showURL      \undefined \def \showURL       {\relax}        \fi
\providecommand\bibfield[2]{#2}
\providecommand\bibinfo[2]{#2}
\providecommand\natexlab[1]{#1}
\providecommand\showeprint[2][]{arXiv:#2}

\bibitem[{Ada Lovelace Institute}(2022)]%
        {ada_lovelace}
\bibfield{author}{\bibinfo{person}{{Ada Lovelace Institute}}.} \bibinfo{year}{2022}\natexlab{}.
\newblock \bibinfo{booktitle}{\emph{{Algorithmic Impact Assessment: AIA Template}}}.
\newblock
\urldef\tempurl%
\url{https://www.adalovelaceinstitute.org/resource/aia-template/}
\showURL{%
Retrieved January 22, 2024 from \tempurl}


\bibitem[Afzal et~al\mbox{.}(2022)]%
        {afzal2022battery}
\bibfield{author}{\bibinfo{person}{Sayed~Saad Afzal}, \bibinfo{person}{Waleed Akbar}, \bibinfo{person}{Osvy Rodriguez}, \bibinfo{person}{Mario Doumet}, \bibinfo{person}{Unsoo Ha}, \bibinfo{person}{Reza Ghaffarivardavagh}, {and} \bibinfo{person}{Fadel Adib}.} \bibinfo{year}{2022}\natexlab{}.
\newblock \showarticletitle{Battery-free wireless imaging of underwater environments}.
\newblock \bibinfo{journal}{\emph{Nature communications}} \bibinfo{volume}{13}, \bibinfo{number}{1} (\bibinfo{year}{2022}), \bibinfo{pages}{5546}.
\newblock


\bibitem[Allan et~al\mbox{.}(2013)]%
        {allan2013cost}
\bibfield{author}{\bibinfo{person}{Blake~M Allan}, \bibinfo{person}{John~PY Arnould}, \bibinfo{person}{Jennifer~K Martin}, {and} \bibinfo{person}{Euan~G Ritchie}.} \bibinfo{year}{2013}\natexlab{}.
\newblock \showarticletitle{A cost-effective and informative method of GPS tracking wildlife}.
\newblock \bibinfo{journal}{\emph{Wildlife Research}} \bibinfo{volume}{40}, \bibinfo{number}{5} (\bibinfo{year}{2013}), \bibinfo{pages}{345--348}.
\newblock


\bibitem[Andrachuk et~al\mbox{.}(2019)]%
        {andrachuk2019smartphone}
\bibfield{author}{\bibinfo{person}{Mark Andrachuk}, \bibinfo{person}{Melissa Marschke}, \bibinfo{person}{Charlotte Hings}, {and} \bibinfo{person}{Derek Armitage}.} \bibinfo{year}{2019}\natexlab{}.
\newblock \showarticletitle{Smartphone technologies supporting community-based environmental monitoring and implementation: a systematic scoping review}.
\newblock \bibinfo{journal}{\emph{Biological Conservation}}  \bibinfo{volume}{237} (\bibinfo{year}{2019}), \bibinfo{pages}{430--442}.
\newblock


\bibitem[Aseniero et~al\mbox{.}(2020)]%
        {aseniero2020meetcues}
\bibfield{author}{\bibinfo{person}{Bon~Adriel Aseniero}, \bibinfo{person}{Marios Constantinides}, \bibinfo{person}{Sagar Joglekar}, \bibinfo{person}{Ke Zhou}, {and} \bibinfo{person}{Daniele Quercia}.} \bibinfo{year}{2020}\natexlab{}.
\newblock \showarticletitle{MeetCues: Supporting online meetings experience}. In \bibinfo{booktitle}{\emph{2020 IEEE Visualization Conference (VIS)}}. IEEE, \bibinfo{pages}{236--240}.
\newblock


\bibitem[Ashurst et~al\mbox{.}(2020)]%
        {ashurst2020guide}
\bibfield{author}{\bibinfo{person}{Carolyn Ashurst}, \bibinfo{person}{Markus Anderljung}, \bibinfo{person}{Carina Prunkl}, \bibinfo{person}{Jan Leike}, \bibinfo{person}{Yarin Gal}, \bibinfo{person}{Toby Shevlane}, {and} \bibinfo{person}{Allan Dafoe}.} \bibinfo{year}{2020}\natexlab{}.
\newblock \showarticletitle{A guide to writing the NeurIPS impact statement}.
\newblock \bibinfo{journal}{\emph{Centre for the Governance of AI. URL: https://perma. cc/B5R8-2B9V}} (\bibinfo{year}{2020}).
\newblock


\bibitem[Azure(2023)]%
        {microsoftazure2023}
\bibfield{author}{\bibinfo{person}{Microsoft Azure}.} \bibinfo{year}{2023}\natexlab{}.
\newblock \bibinfo{booktitle}{\emph{Introduction to prompt engineering}}.
\newblock
\urldef\tempurl%
\url{https://learn.microsoft.com/en-us/azure/ai-services/openai/concepts/prompt-engineering#best-practices}
\showURL{%
Retrieved September 8, 2023 from \tempurl}


\bibitem[Bardram and Matic(2020)]%
        {bardram2020decade}
\bibfield{author}{\bibinfo{person}{Jakob~E Bardram} {and} \bibinfo{person}{Aleksandar Matic}.} \bibinfo{year}{2020}\natexlab{}.
\newblock \showarticletitle{A decade of ubiquitous computing research in mental health}.
\newblock \bibinfo{journal}{\emph{IEEE Pervasive Computing}} \bibinfo{volume}{19}, \bibinfo{number}{1} (\bibinfo{year}{2020}), \bibinfo{pages}{62--72}.
\newblock


\bibitem[Barmpatsalou et~al\mbox{.}(2018)]%
        {barmpatsalou2018mobile}
\bibfield{author}{\bibinfo{person}{Konstantia Barmpatsalou}, \bibinfo{person}{Tiago Cruz}, \bibinfo{person}{Edmundo Monteiro}, {and} \bibinfo{person}{Paulo Simoes}.} \bibinfo{year}{2018}\natexlab{}.
\newblock \showarticletitle{Mobile forensic data analysis: Suspicious pattern detection in mobile evidence}.
\newblock \bibinfo{journal}{\emph{IEEE Access}}  \bibinfo{volume}{6} (\bibinfo{year}{2018}), \bibinfo{pages}{59705--59727}.
\newblock


\bibitem[Borenstein and Howard(2021)]%
        {borenstein2021emerging}
\bibfield{author}{\bibinfo{person}{Jason Borenstein} {and} \bibinfo{person}{Ayanna Howard}.} \bibinfo{year}{2021}\natexlab{}.
\newblock \showarticletitle{Emerging challenges in AI and the need for AI ethics education}.
\newblock \bibinfo{journal}{\emph{AI and Ethics}}  \bibinfo{volume}{1} (\bibinfo{year}{2021}), \bibinfo{pages}{61--65}.
\newblock


\bibitem[Brown et~al\mbox{.}(2020)]%
        {brown2020language}
\bibfield{author}{\bibinfo{person}{Tom~B. Brown} {et~al\mbox{.}}} \bibinfo{year}{2020}\natexlab{}.
\newblock \showarticletitle{Language Models Are Few-Shot Learners}. In \bibinfo{booktitle}{\emph{Proceedings of the 34th International Conference on Neural Information Processing Systems}} (Vancouver, BC, Canada) \emph{(\bibinfo{series}{NIPS'20})}. \bibinfo{publisher}{Curran Associates Inc.}, \bibinfo{address}{Red Hook, NY, USA}, Article \bibinfo{articleno}{159}, \bibinfo{numpages}{25}~pages.
\newblock
\showISBNx{9781713829546}


\bibitem[Buriro et~al\mbox{.}(2016)]%
        {buriro2016hold}
\bibfield{author}{\bibinfo{person}{Attaullah Buriro}, \bibinfo{person}{Bruno Crispo}, \bibinfo{person}{Filippo Delfrari}, {and} \bibinfo{person}{Konrad Wrona}.} \bibinfo{year}{2016}\natexlab{}.
\newblock \showarticletitle{Hold and sign: A novel behavioral biometrics for smartphone user authentication}. In \bibinfo{booktitle}{\emph{2016 IEEE security and privacy workshops (SPW)}}. IEEE, \bibinfo{pages}{276--285}.
\newblock


\bibitem[Buçinca et~al\mbox{.}(2023)]%
        {bucinca2023harmGeneration}
\bibfield{author}{\bibinfo{person}{Zana Buçinca}, \bibinfo{person}{Chau~Minh Pham}, \bibinfo{person}{Maurice Jakesch}, \bibinfo{person}{Marco~Tulio Ribeiro}, \bibinfo{person}{Alexandra Olteanu}, {and} \bibinfo{person}{Saleema Amershi}.} \bibinfo{year}{2023}\natexlab{}.
\newblock \bibinfo{title}{AHA!: Facilitating AI Impact Assessment by Generating Examples of Harms}.
\newblock
\newblock
\showeprint[arxiv]{2306.03280}~[cs.HC]


\bibitem[Byun et~al\mbox{.}(2023)]%
        {byun2023dispensing}
\bibfield{author}{\bibinfo{person}{Courtni Byun}, \bibinfo{person}{Piper Vasicek}, {and} \bibinfo{person}{Kevin Seppi}.} \bibinfo{year}{2023}\natexlab{}.
\newblock \showarticletitle{Dispensing with Humans in Human-Computer Interaction Research}. In \bibinfo{booktitle}{\emph{Extended Abstracts of the 2023 CHI Conference on Human Factors in Computing Systems}} (Hamburg, Germany) \emph{(\bibinfo{series}{CHI EA '23})}. \bibinfo{publisher}{Association for Computing Machinery}, \bibinfo{address}{New York, NY, USA}, Article \bibinfo{articleno}{413}, \bibinfo{numpages}{26}~pages.
\newblock
\showISBNx{9781450394222}
\urldef\tempurl%
\url{https://doi.org/10.1145/3544549.3582749}
\showDOI{\tempurl}


\bibitem[Cecchinato et~al\mbox{.}(2015)]%
        {cecchinato2015smartwatches}
\bibfield{author}{\bibinfo{person}{Marta~E Cecchinato}, \bibinfo{person}{Anna~L Cox}, {and} \bibinfo{person}{Jon Bird}.} \bibinfo{year}{2015}\natexlab{}.
\newblock \showarticletitle{Smartwatches: the good, the bad and the ugly?}. In \bibinfo{booktitle}{\emph{Proceedings of the 33rd Annual ACM Conference extended abstracts on human factors in computing systems}}. \bibinfo{pages}{2133--2138}.
\newblock


\bibitem[Chaudhari et~al\mbox{.}(2018)]%
        {chaudhari2018effective}
\bibfield{author}{\bibinfo{person}{Apurva Chaudhari}, \bibinfo{person}{Bhushan Mapari}, {and} \bibinfo{person}{Shreenivas Jog}.} \bibinfo{year}{2018}\natexlab{}.
\newblock \showarticletitle{Effective environmental monitoring \& domestic home conditions by implementation of IoT}. In \bibinfo{booktitle}{\emph{2018 Fourth International Conference on Computing Communication Control and Automation (ICCUBEA)}}. IEEE, \bibinfo{pages}{1--5}.
\newblock


\bibitem[Chen et~al\mbox{.}(2023)]%
        {chen2023llm}
\bibfield{author}{\bibinfo{person}{Siyuan Chen}, \bibinfo{person}{Mengyue Wu}, \bibinfo{person}{Kenny~Q Zhu}, \bibinfo{person}{Kunyao Lan}, \bibinfo{person}{Zhiling Zhang}, {and} \bibinfo{person}{Lyuchun Cui}.} \bibinfo{year}{2023}\natexlab{}.
\newblock \showarticletitle{LLM-empowered Chatbots for Psychiatrist and Patient Simulation: Application and Evaluation}.
\newblock \bibinfo{journal}{\emph{arXiv preprint arXiv:2305.13614}} (\bibinfo{year}{2023}).
\newblock
\urldef\tempurl%
\url{https://arxiv.org/abs/2305.13614}
\showURL{%
\tempurl}


\bibitem[Chidambaram et~al\mbox{.}(2022)]%
        {chidambaram2022using}
\bibfield{author}{\bibinfo{person}{Swathikan Chidambaram}, \bibinfo{person}{Yathukulan Maheswaran}, \bibinfo{person}{Kian Patel}, \bibinfo{person}{Viknesh Sounderajah}, \bibinfo{person}{Daniel~A Hashimoto}, \bibinfo{person}{Kenneth~Patrick Seastedt}, \bibinfo{person}{Alison~H McGregor}, \bibinfo{person}{Sheraz~R Markar}, {and} \bibinfo{person}{Ara Darzi}.} \bibinfo{year}{2022}\natexlab{}.
\newblock \showarticletitle{Using artificial intelligence-enhanced sensing and wearable technology in sports medicine and performance optimisation}.
\newblock \bibinfo{journal}{\emph{Sensors}} \bibinfo{volume}{22}, \bibinfo{number}{18} (\bibinfo{year}{2022}), \bibinfo{pages}{6920}.
\newblock


\bibitem[Choi et~al\mbox{.}(2021)]%
        {choi2021kairos}
\bibfield{author}{\bibinfo{person}{Jun-Ho Choi}, \bibinfo{person}{Marios Constantinides}, \bibinfo{person}{Sagar Joglekar}, {and} \bibinfo{person}{Daniele Quercia}.} \bibinfo{year}{2021}\natexlab{}.
\newblock \showarticletitle{KAIROS: Talking heads and moving bodies for successful meetings}. In \bibinfo{booktitle}{\emph{Proceedings of the 22nd International Workshop on Mobile Computing Systems and Applications}}. \bibinfo{pages}{30--36}.
\newblock


\bibitem[Chowdhury et~al\mbox{.}(2019)]%
        {chowdhury2019multi}
\bibfield{author}{\bibinfo{person}{Ushashi Chowdhury}, \bibinfo{person}{Pranjal Chowdhury}, \bibinfo{person}{Sourav Paul}, \bibinfo{person}{Anwesha Sen}, \bibinfo{person}{Partho~Protim Sarkar}, \bibinfo{person}{Shubhankur Basak}, {and} \bibinfo{person}{Abari Bhattacharya}.} \bibinfo{year}{2019}\natexlab{}.
\newblock \showarticletitle{Multi-sensor wearable for child safety}. In \bibinfo{booktitle}{\emph{2019 IEEE 10th Annual Ubiquitous Computing, Electronics \& Mobile Communication Conference (UEMCON)}}. IEEE, \bibinfo{pages}{0968--0972}.
\newblock


\bibitem[Chung et~al\mbox{.}(2022)]%
        {chimLLM_stories}
\bibfield{author}{\bibinfo{person}{John Joon~Young Chung}, \bibinfo{person}{Wooseok Kim}, \bibinfo{person}{Kang~Min Yoo}, \bibinfo{person}{Hwaran Lee}, \bibinfo{person}{Eytan Adar}, {and} \bibinfo{person}{Minsuk Chang}.} \bibinfo{year}{2022}\natexlab{}.
\newblock \showarticletitle{TaleBrush: Sketching Stories with Generative Pretrained Language Models}. In \bibinfo{booktitle}{\emph{Proceedings of the 2022 CHI Conference on Human Factors in Computing Systems}} (New Orleans, LA, USA) \emph{(\bibinfo{series}{CHI '22})}. \bibinfo{publisher}{Association for Computing Machinery}, \bibinfo{address}{New York, NY, USA}, Article \bibinfo{articleno}{209}, \bibinfo{numpages}{19}~pages.
\newblock
\showISBNx{9781450391573}
\urldef\tempurl%
\url{https://doi.org/10.1145/3491102.3501819}
\showDOI{\tempurl}


\bibitem[Ciolacu et~al\mbox{.}(2019)]%
        {ciolacu2019education}
\bibfield{author}{\bibinfo{person}{Monica~Ionita Ciolacu}, \bibinfo{person}{Leon Binder}, \bibinfo{person}{Paul Svasta}, \bibinfo{person}{Ioan Tache}, {and} \bibinfo{person}{Dan Stoichescu}.} \bibinfo{year}{2019}\natexlab{}.
\newblock \showarticletitle{Education 4.0--jump to innovation with IoT in higher education}. In \bibinfo{booktitle}{\emph{2019 IEEE 25th International Symposium for Design and Technology in Electronic Packaging (SIITME)}}. IEEE, \bibinfo{pages}{135--141}.
\newblock


\bibitem[Comission(2021)]%
        {EUACT2021}
\bibfield{author}{\bibinfo{person}{European Comission}.} \bibinfo{year}{2021}\natexlab{}.
\newblock \bibinfo{booktitle}{\emph{Proposal for a Regulation of the European Parliament and of the Council laying down harmonised rules on Artificial Intelligence (Artificial Intelligence Act) and amending certain Union legislative acts}}.
\newblock European Comission.
\newblock
\urldef\tempurl%
\url{https://eur-lex.europa.eu/legal-content/EN/TXT/?uri=CELEX%3A52021PC0206}
\showURL{%
Retrieved July 13, 2023 from \tempurl}


\bibitem[Comission(2023)]%
        {amendments2023}
\bibfield{author}{\bibinfo{person}{European Comission}.} \bibinfo{year}{2023}\natexlab{}.
\newblock \bibinfo{booktitle}{\emph{Amendments adopted by the European Parliament on 14 June 2023 on the proposal for a regulation of the European Parliament and of the Council on laying down harmonised rules on artificial intelligence (Artificial Intelligence Act) and amending certain Union legislative acts}}.
\newblock European Comission.
\newblock
\urldef\tempurl%
\url{https://www.europarl.europa.eu/doceo/document/TA-9-2023-0236_EN.html}
\showURL{%
Retrieved July 13, 2023 from \tempurl}


\bibitem[Constantinides et~al\mbox{.}(2018)]%
        {constantinides2018personalized}
\bibfield{author}{\bibinfo{person}{Marios Constantinides}, \bibinfo{person}{Jonas Busk}, \bibinfo{person}{Aleksandar Matic}, \bibinfo{person}{Maria Faurholt-Jepsen}, \bibinfo{person}{Lars~Vedel Kessing}, {and} \bibinfo{person}{Jakob~E Bardram}.} \bibinfo{year}{2018}\natexlab{}.
\newblock \showarticletitle{Personalized versus generic mood prediction models in bipolar disorder}. In \bibinfo{booktitle}{\emph{Proceedings of the 2018 ACM International Joint Conference and 2018 International Symposium on Pervasive and Ubiquitous Computing and Wearable Computers}}. \bibinfo{pages}{1700--1707}.
\newblock


\bibitem[Constantinides and Quercia(2022)]%
        {constantinides2022good}
\bibfield{author}{\bibinfo{person}{Marios Constantinides} {and} \bibinfo{person}{Daniele Quercia}.} \bibinfo{year}{2022}\natexlab{}.
\newblock \showarticletitle{Good Intentions, Bad Inventions: How Employees Judge Pervasive Technologies in the Workplace}.
\newblock \bibinfo{journal}{\emph{IEEE Pervasive Computing}} \bibinfo{volume}{22}, \bibinfo{number}{1} (\bibinfo{year}{2022}), \bibinfo{pages}{69--76}.
\newblock


\bibitem[Constantinides et~al\mbox{.}(2020)]%
        {constantinides2020comfeel}
\bibfield{author}{\bibinfo{person}{Marios Constantinides}, \bibinfo{person}{Sanja {\v{S}}{\'c}epanovi{\'c}}, \bibinfo{person}{Daniele Quercia}, \bibinfo{person}{Hongwei Li}, \bibinfo{person}{Ugo Sassi}, {and} \bibinfo{person}{Michael Eggleston}.} \bibinfo{year}{2020}\natexlab{}.
\newblock \showarticletitle{ComFeel: Productivity is a matter of the senses too}.
\newblock \bibinfo{journal}{\emph{Proceedings of the ACM on Interactive, Mobile, Wearable and Ubiquitous Technologies}} \bibinfo{volume}{4}, \bibinfo{number}{4} (\bibinfo{year}{2020}), \bibinfo{pages}{1--21}.
\newblock


\bibitem[Cornacchia et~al\mbox{.}(2016)]%
        {cornacchia2016survey}
\bibfield{author}{\bibinfo{person}{Maria Cornacchia}, \bibinfo{person}{Koray Ozcan}, \bibinfo{person}{Yu Zheng}, {and} \bibinfo{person}{Senem Velipasalar}.} \bibinfo{year}{2016}\natexlab{}.
\newblock \showarticletitle{A survey on activity detection and classification using wearable sensors}.
\newblock \bibinfo{journal}{\emph{IEEE Sensors Journal}} \bibinfo{volume}{17}, \bibinfo{number}{2} (\bibinfo{year}{2016}), \bibinfo{pages}{386--403}.
\newblock


\bibitem[Corsetti et~al\mbox{.}(2019)]%
        {corsetti2019face}
\bibfield{author}{\bibinfo{person}{Barbara Corsetti}, \bibinfo{person}{Raul Sanchez-Reillo}, \bibinfo{person}{Richard~M Guest}, {and} \bibinfo{person}{Marco Santopietro}.} \bibinfo{year}{2019}\natexlab{}.
\newblock \showarticletitle{Face image analysis in mobile biometric accessibility evaluations}. In \bibinfo{booktitle}{\emph{2019 International Carnahan Conference on Security Technology (ICCST)}}. IEEE, \bibinfo{pages}{1--5}.
\newblock


\bibitem[Crager and Maiti(2017)]%
        {crager2017information}
\bibfield{author}{\bibinfo{person}{Kirsten Crager} {and} \bibinfo{person}{Anindya Maiti}.} \bibinfo{year}{2017}\natexlab{}.
\newblock \showarticletitle{Information leakage through mobile motion sensors: User awareness and concerns}. In \bibinfo{booktitle}{\emph{Proceedings of the European Workshop on Usable Security (EuroUSEC)}}.
\newblock


\bibitem[Cusack et~al\mbox{.}(2017)]%
        {cusack2017assessment}
\bibfield{author}{\bibinfo{person}{Brian Cusack}, \bibinfo{person}{Bryce Antony}, \bibinfo{person}{Gerard Ward}, {and} \bibinfo{person}{Shaunak Mody}.} \bibinfo{year}{2017}\natexlab{}.
\newblock \showarticletitle{Assessment of security vulnerabilities in wearable devices}.
\newblock  (\bibinfo{year}{2017}).
\newblock


\bibitem[Das~Swain et~al\mbox{.}(2023)]%
        {das2023algorithmic}
\bibfield{author}{\bibinfo{person}{Vedant Das~Swain}, \bibinfo{person}{Lan Gao}, \bibinfo{person}{William~A Wood}, \bibinfo{person}{Srikruthi~C Matli}, \bibinfo{person}{Gregory~D Abowd}, {and} \bibinfo{person}{Munmun De~Choudhury}.} \bibinfo{year}{2023}\natexlab{}.
\newblock \showarticletitle{Algorithmic Power or Punishment: Information Worker Perspectives on Passive Sensing Enabled AI Phenotyping of Performance and Wellbeing}. In \bibinfo{booktitle}{\emph{Proceedings of the 2023 CHI Conference on Human Factors in Computing Systems}}. \bibinfo{pages}{1--17}.
\newblock


\bibitem[Diez et~al\mbox{.}(2019)]%
        {diez2019lightweight}
\bibfield{author}{\bibinfo{person}{Fidel~Paniagua Diez}, \bibinfo{person}{Diego~Su{\'a}rez Touceda}, \bibinfo{person}{Jos{\'e} Mar{\'\i}a~Sierra C{\'a}mara}, {and} \bibinfo{person}{Sherali Zeadally}.} \bibinfo{year}{2019}\natexlab{}.
\newblock \showarticletitle{Lightweight access control system for wearable devices}.
\newblock \bibinfo{journal}{\emph{IT professional}} \bibinfo{volume}{21}, \bibinfo{number}{1} (\bibinfo{year}{2019}), \bibinfo{pages}{50--58}.
\newblock


\bibitem[Dorai et~al\mbox{.}(2020)]%
        {dorai2020data}
\bibfield{author}{\bibinfo{person}{Gokila Dorai}, \bibinfo{person}{Shiva Houshmand}, {and} \bibinfo{person}{Sudhir Aggarwal}.} \bibinfo{year}{2020}\natexlab{}.
\newblock \showarticletitle{Data Extraction and Forensic Analysis for Smartphone Paired Wearables and IoT Devices.}. In \bibinfo{booktitle}{\emph{HICSS}}. \bibinfo{pages}{1--10}.
\newblock


\bibitem[Dowling and Lucey(2023)]%
        {dowling2023chatgpt}
\bibfield{author}{\bibinfo{person}{Michael Dowling} {and} \bibinfo{person}{Brian Lucey}.} \bibinfo{year}{2023}\natexlab{}.
\newblock \showarticletitle{{ChatGPT} for (Finance) research: The Bananarama Conjecture}.
\newblock \bibinfo{journal}{\emph{Finance Research Letters}}  \bibinfo{volume}{53} (\bibinfo{year}{2023}), \bibinfo{pages}{103662}.
\newblock
\urldef\tempurl%
\url{https://doi.org/10.1016/j.frl.2023.103662}
\showDOI{\tempurl}


\bibitem[Dwork(2008)]%
        {dwork2008differential}
\bibfield{author}{\bibinfo{person}{Cynthia Dwork}.} \bibinfo{year}{2008}\natexlab{}.
\newblock \showarticletitle{Differential privacy: A survey of results}. In \bibinfo{booktitle}{\emph{International conference on theory and applications of models of computation}}. Springer, \bibinfo{pages}{1--19}.
\newblock
\urldef\tempurl%
\url{https://doi.org/10.1007/978-3-540-79228-4_1}
\showDOI{\tempurl}


\bibitem[Dziri et~al\mbox{.}(2022)]%
        {dziri2022origin}
\bibfield{author}{\bibinfo{person}{Nouha Dziri}, \bibinfo{person}{Sivan Milton}, \bibinfo{person}{Mo Yu}, \bibinfo{person}{Osmar Zaiane}, {and} \bibinfo{person}{Siva Reddy}.} \bibinfo{year}{2022}\natexlab{}.
\newblock \showarticletitle{On the origin of hallucinations in conversational models: Is it the datasets or the models?}
\newblock \bibinfo{journal}{\emph{arXiv preprint arXiv:2204.07931}} (\bibinfo{year}{2022}).
\newblock
\urldef\tempurl%
\url{https://arxiv.org/abs/2204.07931}
\showURL{%
\tempurl}


\bibitem[Eldan and Li(2023)]%
        {eldan2023tinystories}
\bibfield{author}{\bibinfo{person}{Ronen Eldan} {and} \bibinfo{person}{Yuanzhi Li}.} \bibinfo{year}{2023}\natexlab{}.
\newblock \showarticletitle{TinyStories: How Small Can Language Models Be and Still Speak Coherent English?}
\newblock \bibinfo{journal}{\emph{arXiv preprint arXiv:2305.07759}} (\bibinfo{year}{2023}).
\newblock
\urldef\tempurl%
\url{https://arxiv.org/abs/2305.07759}
\showURL{%
\tempurl}


\bibitem[Fazio et~al\mbox{.}(2021)]%
        {fazio2021innovative}
\bibfield{author}{\bibinfo{person}{Peppino Fazio}, \bibinfo{person}{Miralem Mehic}, {and} \bibinfo{person}{Miroslav Voznak}.} \bibinfo{year}{2021}\natexlab{}.
\newblock \showarticletitle{An innovative dynamic mobility sampling scheme based on multiresolution wavelet analysis in IoT networks}.
\newblock \bibinfo{journal}{\emph{IEEE Internet of Things Journal}} \bibinfo{volume}{9}, \bibinfo{number}{13} (\bibinfo{year}{2021}), \bibinfo{pages}{11336--11350}.
\newblock


\bibitem[Fletcher et~al\mbox{.}(2010)]%
        {fletcher2010wearable}
\bibfield{author}{\bibinfo{person}{Richard~R Fletcher}, \bibinfo{person}{Ming-Zher Poh}, {and} \bibinfo{person}{Hoda Eydgahi}.} \bibinfo{year}{2010}\natexlab{}.
\newblock \showarticletitle{Wearable sensors: Opportunities and challenges for low-cost health care}. In \bibinfo{booktitle}{\emph{2010 Annual International Conference of the IEEE Engineering in Medicine and Biology}}. IEEE, \bibinfo{pages}{1763--1766}.
\newblock


\bibitem[Floridi et~al\mbox{.}(2021)]%
        {Floridi2021}
\bibfield{author}{\bibinfo{person}{Luciano Floridi}, \bibinfo{person}{Josh Cowls}, \bibinfo{person}{Monica Beltrametti}, \bibinfo{person}{Raja Chatila}, \bibinfo{person}{Patrice Chazerand}, \bibinfo{person}{Virginia Dignum}, \bibinfo{person}{Christoph Luetge}, \bibinfo{person}{Robert Madelin}, \bibinfo{person}{Ugo Pagallo}, \bibinfo{person}{Francesca Rossi}, \bibinfo{person}{Burkhard Schafer}, \bibinfo{person}{Peggy Valcke}, {and} \bibinfo{person}{Effy Vayena}.} \bibinfo{year}{2021}\natexlab{}.
\newblock \bibinfo{booktitle}{\emph{An Ethical Framework for a Good AI Society: Opportunities, Risks, Principles, and Recommendations}}.
\newblock \bibinfo{publisher}{Springer International Publishing}, \bibinfo{address}{Cham}, \bibinfo{pages}{19--39}.
\newblock
\urldef\tempurl%
\url{https://doi.org/10.1007/978-3-030-81907-1_3}
\showDOI{\tempurl}


\bibitem[Frluckaj et~al\mbox{.}(2022)]%
        {frluckaj2022gender}
\bibfield{author}{\bibinfo{person}{Hana Frluckaj}, \bibinfo{person}{Laura Dabbish}, \bibinfo{person}{David~G. Widder}, \bibinfo{person}{Huilian~Sophie Qiu}, {and} \bibinfo{person}{James Herbsleb}.} \bibinfo{year}{2022}\natexlab{}.
\newblock \showarticletitle{Gender and Participation in Open Source Software Development}.
\newblock \bibinfo{journal}{\emph{Proceedings of the ACM on Human-Computer Interaction}} \bibinfo{volume}{6}, \bibinfo{number}{CSCW2} (\bibinfo{year}{2022}), \bibinfo{pages}{1--31}.
\newblock
\urldef\tempurl%
\url{https://doi.org/10.1145/3555190}
\showDOI{\tempurl}


\bibitem[Fuster et~al\mbox{.}(2023)]%
        {fuster2023analysis}
\bibfield{author}{\bibinfo{person}{Jaime Fuster}, \bibinfo{person}{Sonia Solera-Cotanilla}, \bibinfo{person}{Jaime P{\'e}rez}, \bibinfo{person}{Mario Vega-Barbas}, \bibinfo{person}{Rafael Palacios}, \bibinfo{person}{Manuel Alvarez-Campana}, {and} \bibinfo{person}{Gregorio Lopez}.} \bibinfo{year}{2023}\natexlab{}.
\newblock \showarticletitle{Analysis of security and privacy issues in wearables for minors}.
\newblock \bibinfo{journal}{\emph{Wireless Networks}} (\bibinfo{year}{2023}), \bibinfo{pages}{1--17}.
\newblock


\bibitem[Gentry(2009)]%
        {gentry2009fully}
\bibfield{author}{\bibinfo{person}{Craig Gentry}.} \bibinfo{year}{2009}\natexlab{}.
\newblock \bibinfo{booktitle}{\emph{A fully homomorphic encryption scheme}}.
\newblock \bibinfo{publisher}{Stanford university}.
\newblock


\bibitem[Gilardi et~al\mbox{.}(2023)]%
        {gilardi2023chatgpt}
\bibfield{author}{\bibinfo{person}{Fabrizio Gilardi}, \bibinfo{person}{Meysam Alizadeh}, {and} \bibinfo{person}{Ma{\"e}l Kubli}.} \bibinfo{year}{2023}\natexlab{}.
\newblock \showarticletitle{{ChatGPT} outperforms crowd workers for text-annotation tasks}.
\newblock \bibinfo{journal}{\emph{Proceedings of the National Academy of Sciences 120 (30)}} (\bibinfo{year}{2023}).
\newblock
\urldef\tempurl%
\url{https://doi.org/10.1073/pnas.2305016120}
\showDOI{\tempurl}


\bibitem[Giray(2023)]%
        {giray2023prompt}
\bibfield{author}{\bibinfo{person}{Louie Giray}.} \bibinfo{year}{2023}\natexlab{}.
\newblock \showarticletitle{Prompt Engineering with ChatGPT: A Guide for Academic Writers}.
\newblock \bibinfo{journal}{\emph{Annals of Biomedical Engineering}} (\bibinfo{year}{2023}), \bibinfo{pages}{1--5}.
\newblock
\urldef\tempurl%
\url{https://doi.org/10.1007/s10439-023-03272-4}
\showDOI{\tempurl}


\bibitem[Golpayegani et~al\mbox{.}(2023)]%
        {Golpayegani2023Risk}
\bibfield{author}{\bibinfo{person}{Delaram Golpayegani}, \bibinfo{person}{Harshvardhan~J. Pandit}, {and} \bibinfo{person}{Dave Lewis}.} \bibinfo{year}{2023}\natexlab{}.
\newblock \showarticletitle{To Be High-Risk, or Not To Be--Semantic Specifications and Implications of the AI Act's High-Risk AI Applications and Harmonised Standards}. In \bibinfo{booktitle}{\emph{Proceedings of the 2023 ACM Conference on Fairness, Accountability, and Transparency}} (Chicago, IL, USA) \emph{(\bibinfo{series}{FAccT '23})}. \bibinfo{publisher}{ACM}, \bibinfo{address}{New York, NY, USA}, \bibinfo{pages}{905–915}.
\newblock
\urldef\tempurl%
\url{https://doi.org/10.1145/3593013.3594050}
\showDOI{\tempurl}


\bibitem[Gor et~al\mbox{.}(2017)]%
        {gor2017gata}
\bibfield{author}{\bibinfo{person}{M Gor}, \bibinfo{person}{Jayneel Vora}, \bibinfo{person}{Sudeep Tanwar}, \bibinfo{person}{Sudhanshu Tyagi}, \bibinfo{person}{Neeraj Kumar}, \bibinfo{person}{Mohammad~S Obaidat}, {and} \bibinfo{person}{Balqies Sadoun}.} \bibinfo{year}{2017}\natexlab{}.
\newblock \showarticletitle{GATA: GPS-Arduino based Tracking and Alarm system for protection of wildlife animals}. In \bibinfo{booktitle}{\emph{2017 international conference on computer, information and telecommunication systems (CITS)}}. IEEE, \bibinfo{pages}{166--170}.
\newblock


\bibitem[Hagendorff(2020)]%
        {hagendorff2020ethics}
\bibfield{author}{\bibinfo{person}{Thilo Hagendorff}.} \bibinfo{year}{2020}\natexlab{}.
\newblock \showarticletitle{The ethics of AI ethics: An evaluation of guidelines}.
\newblock \bibinfo{journal}{\emph{Minds and machines}} \bibinfo{volume}{30}, \bibinfo{number}{1} (\bibinfo{year}{2020}), \bibinfo{pages}{99--120}.
\newblock
\urldef\tempurl%
\url{https://doi.org/10.1007/s11023-020-09517-8}
\showDOI{\tempurl}


\bibitem[Harari(2021)]%
        {harari}
\bibfield{author}{\bibinfo{person}{Noah~Yuval Harari}.} \bibinfo{year}{2021}\natexlab{}.
\newblock \bibinfo{title}{The world after coronavirus}.
\newblock
\newblock


\bibitem[Haris et~al\mbox{.}(2014)]%
        {haris2014privacy}
\bibfield{author}{\bibinfo{person}{Muhammad Haris}, \bibinfo{person}{Hamed Haddadi}, {and} \bibinfo{person}{Pan Hui}.} \bibinfo{year}{2014}\natexlab{}.
\newblock \showarticletitle{Privacy leakage in mobile computing: Tools, methods, and characteristics}.
\newblock \bibinfo{journal}{\emph{arXiv preprint arXiv:1410.4978}} (\bibinfo{year}{2014}).
\newblock


\bibitem[Hassel and {\"O}zkiziltan(2023)]%
        {hassel2023governing}
\bibfield{author}{\bibinfo{person}{Anke Hassel} {and} \bibinfo{person}{Didem {\"O}zkiziltan}.} \bibinfo{year}{2023}\natexlab{}.
\newblock \showarticletitle{Governing the work-related risks of AI: implications for the German government and trade unions}.
\newblock \bibinfo{journal}{\emph{Transfer: European Review of Labour and Research}} \bibinfo{volume}{29}, \bibinfo{number}{1} (\bibinfo{year}{2023}), \bibinfo{pages}{71--86}.
\newblock
\urldef\tempurl%
\url{https://doi.org/10.1177/10242589221147228}
\showDOI{\tempurl}


\bibitem[Heidel et~al\mbox{.}(2020)]%
        {heidel2020potential}
\bibfield{author}{\bibinfo{person}{Alexandra Heidel}, \bibinfo{person}{Christian Hagist}, {et~al\mbox{.}}} \bibinfo{year}{2020}\natexlab{}.
\newblock \showarticletitle{Potential benefits and risks resulting from the introduction of health apps and wearables into the German statutory health care system: scoping review}.
\newblock \bibinfo{journal}{\emph{JMIR mHealth and uHealth}} \bibinfo{volume}{8}, \bibinfo{number}{9} (\bibinfo{year}{2020}), \bibinfo{pages}{e16444}.
\newblock


\bibitem[Helwatkar et~al\mbox{.}(2014)]%
        {helwatkar2014sensor}
\bibfield{author}{\bibinfo{person}{Amruta Helwatkar}, \bibinfo{person}{Daniel Riordan}, {and} \bibinfo{person}{Joseph Walsh}.} \bibinfo{year}{2014}\natexlab{}.
\newblock \showarticletitle{Sensor technology for animal health monitoring}.
\newblock \bibinfo{journal}{\emph{International Journal on Smart Sensing and Intelligent Systems}} \bibinfo{volume}{7}, \bibinfo{number}{5} (\bibinfo{year}{2014}), \bibinfo{pages}{1--6}.
\newblock


\bibitem[Hoth et~al\mbox{.}(2016)]%
        {hoth2016uncovering}
\bibfield{author}{\bibinfo{person}{Jessica Hoth}, \bibinfo{person}{Bj{\"o}rn Schwarz}, \bibinfo{person}{Gabriele Kaiser}, \bibinfo{person}{Andreas Busse}, \bibinfo{person}{Johannes K{\"o}nig}, {and} \bibinfo{person}{Sigrid Bl{\"o}meke}.} \bibinfo{year}{2016}\natexlab{}.
\newblock \showarticletitle{Uncovering predictors of disagreement: ensuring the quality of expert ratings}.
\newblock \bibinfo{journal}{\emph{ZDM}}  \bibinfo{volume}{48} (\bibinfo{year}{2016}), \bibinfo{pages}{83--95}.
\newblock


\bibitem[Ilyas and Ahmad(2020)]%
        {ilyas2020smart}
\bibfield{author}{\bibinfo{person}{Qazi~Mudassar Ilyas} {and} \bibinfo{person}{Muneer Ahmad}.} \bibinfo{year}{2020}\natexlab{}.
\newblock \showarticletitle{Smart farming: An enhanced pursuit of sustainable remote livestock tracking and geofencing using IoT and GPRS}.
\newblock \bibinfo{journal}{\emph{Wireless communications and mobile computing}}  \bibinfo{volume}{2020} (\bibinfo{year}{2020}), \bibinfo{pages}{1--12}.
\newblock


\bibitem[{International Federation of Accountants}(2018)]%
        {ifac}
\bibfield{author}{\bibinfo{person}{{International Federation of Accountants}}.} \bibinfo{year}{2018}\natexlab{}.
\newblock \bibinfo{booktitle}{\emph{Regulatory divergence: costs, risks, impacts.}}
\newblock IFAC.
\newblock
\urldef\tempurl%
\url{https://www.ifac.org/_flysystem/azure-private/publications/files/IFAC-OECD-Regulatory-Divergence.pdf}
\showURL{%
Retrieved April 25, 2024 from \tempurl}


\bibitem[Jobin et~al\mbox{.}(2019)]%
        {jobin2019global}
\bibfield{author}{\bibinfo{person}{Anna Jobin}, \bibinfo{person}{Marcello Ienca}, {and} \bibinfo{person}{Effy Vayena}.} \bibinfo{year}{2019}\natexlab{}.
\newblock \showarticletitle{The global landscape of AI ethics guidelines}.
\newblock \bibinfo{journal}{\emph{Nature machine intelligence}} \bibinfo{volume}{1}, \bibinfo{number}{9} (\bibinfo{year}{2019}), \bibinfo{pages}{389--399}.
\newblock
\urldef\tempurl%
\url{https://doi.org/10.1038/s42256-019-0088-2}
\showDOI{\tempurl}


\bibitem[Joshi et~al\mbox{.}(2008)]%
        {joshi2008gps}
\bibfield{author}{\bibinfo{person}{Apurva Joshi}, \bibinfo{person}{I~Naga VishnuKanth}, \bibinfo{person}{Navkar Samdaria}, \bibinfo{person}{Sumit Bagla}, {and} \bibinfo{person}{Prabhat Ranjan}.} \bibinfo{year}{2008}\natexlab{}.
\newblock \showarticletitle{GPS-less animal tracking system}. In \bibinfo{booktitle}{\emph{2008 Fourth International Conference on Wireless Communication and Sensor Networks}}. IEEE, \bibinfo{pages}{120--125}.
\newblock


\bibitem[Kasirzadeh(2021)]%
        {kasirzadeh2021reasons}
\bibfield{author}{\bibinfo{person}{Atoosa Kasirzadeh}.} \bibinfo{year}{2021}\natexlab{}.
\newblock \showarticletitle{Reasons, Values, Stakeholders: A Philosophical Framework for Explainable Artificial Intelligence}. In \bibinfo{booktitle}{\emph{Proceedings of the 2021 ACM Conference on Fairness, Accountability, and Transparency}}. \bibinfo{publisher}{Association for Computing Machinery}.
\newblock
\urldef\tempurl%
\url{https://doi.org/10.1145/3442188.3445866}
\showDOI{\tempurl}


\bibitem[Kelly(2014)]%
        {fitbit_irritations}
\bibfield{author}{\bibinfo{person}{Heather Kelly}.} \bibinfo{year}{2014}\natexlab{}.
\newblock \bibinfo{booktitle}{\emph{Fitbit recalls activity tracker due to skin rashes}}.
\newblock CNN.
\newblock
\urldef\tempurl%
\url{https://edition.cnn.com/2014/02/21/tech/gaming-gadgets/fitbit-force-recall/}
\showURL{%
\tempurl}


\bibitem[Laha et~al\mbox{.}(2022)]%
        {laha2022advancement}
\bibfield{author}{\bibinfo{person}{Suprava~Ranjan Laha}, \bibinfo{person}{Binod~Kumar Pattanayak}, {and} \bibinfo{person}{Saumendra Pattnaik}.} \bibinfo{year}{2022}\natexlab{}.
\newblock \showarticletitle{Advancement of Environmental Monitoring System Using IoT and Sensor: A Comprehensive Analysis}.
\newblock \bibinfo{journal}{\emph{AIMS Environmental Science}} \bibinfo{volume}{9}, \bibinfo{number}{6} (\bibinfo{year}{2022}), \bibinfo{pages}{771--800}.
\newblock


\bibitem[Lee et~al\mbox{.}(2022)]%
        {chi_llm_coauthor}
\bibfield{author}{\bibinfo{person}{Mina Lee}, \bibinfo{person}{Percy Liang}, {and} \bibinfo{person}{Qian Yang}.} \bibinfo{year}{2022}\natexlab{}.
\newblock \showarticletitle{CoAuthor: Designing a Human-AI Collaborative Writing Dataset for Exploring Language Model Capabilities}. In \bibinfo{booktitle}{\emph{Proceedings of the 2022 CHI Conference on Human Factors in Computing Systems}} (New Orleans, LA, USA) \emph{(\bibinfo{series}{CHI '22})}. \bibinfo{publisher}{Association for Computing Machinery}, \bibinfo{address}{New York, NY, USA}, Article \bibinfo{articleno}{388}, \bibinfo{numpages}{19}~pages.
\newblock
\showISBNx{9781450391573}
\urldef\tempurl%
\url{https://doi.org/10.1145/3491102.3502030}
\showDOI{\tempurl}


\bibitem[Li et~al\mbox{.}(2016)]%
        {li2016examining}
\bibfield{author}{\bibinfo{person}{He Li}, \bibinfo{person}{Jing Wu}, \bibinfo{person}{Yiwen Gao}, {and} \bibinfo{person}{Yao Shi}.} \bibinfo{year}{2016}\natexlab{}.
\newblock \showarticletitle{Examining individuals’ adoption of healthcare wearable devices: An empirical study from privacy calculus perspective}.
\newblock \bibinfo{journal}{\emph{International journal of medical informatics}}  \bibinfo{volume}{88} (\bibinfo{year}{2016}), \bibinfo{pages}{8--17}.
\newblock


\bibitem[Lin et~al\mbox{.}(2022)]%
        {lin2022iot}
\bibfield{author}{\bibinfo{person}{Hsin-Chang Lin}, \bibinfo{person}{Ming-Jen Chen}, {and} \bibinfo{person}{Jung-Tang Huang}.} \bibinfo{year}{2022}\natexlab{}.
\newblock \showarticletitle{An IoT-based smart healthcare system using location-based mesh network and big data analytics}.
\newblock \bibinfo{journal}{\emph{Journal of Ambient Intelligence and Smart Environments}} \bibinfo{number}{Preprint} (\bibinfo{year}{2022}), \bibinfo{pages}{1--27}.
\newblock


\bibitem[Liu et~al\mbox{.}(2023)]%
        {liu2023lost}
\bibfield{author}{\bibinfo{person}{Nelson~F. Liu}, \bibinfo{person}{Kevin Lin}, \bibinfo{person}{John Hewitt}, \bibinfo{person}{Ashwin Paranjape}, \bibinfo{person}{Michele Bevilacqua}, \bibinfo{person}{Fabio Petroni}, {and} \bibinfo{person}{Percy Liang}.} \bibinfo{year}{2023}\natexlab{}.
\newblock \showarticletitle{Lost in the middle: How language models use long contexts}.
\newblock \bibinfo{journal}{\emph{arXiv preprint arXiv:2307.03172}} (\bibinfo{year}{2023}).
\newblock
\urldef\tempurl%
\url{https://arxiv.org/abs/2307.03172}
\showURL{%
\tempurl}


\bibitem[Mann et~al\mbox{.}(2003)]%
        {mann2003sousveillance}
\bibfield{author}{\bibinfo{person}{Steve Mann}, \bibinfo{person}{Jason Nolan}, {and} \bibinfo{person}{Barry Wellman}.} \bibinfo{year}{2003}\natexlab{}.
\newblock \showarticletitle{Sousveillance: Inventing and using wearable computing devices for data collection in surveillance environments.}
\newblock \bibinfo{journal}{\emph{Surveillance \& society}} \bibinfo{volume}{1}, \bibinfo{number}{3} (\bibinfo{year}{2003}), \bibinfo{pages}{331--355}.
\newblock


\bibitem[McDonald et~al\mbox{.}(2019)]%
        {mcdonald2019reliability}
\bibfield{author}{\bibinfo{person}{Nora McDonald}, \bibinfo{person}{Sarita Schoenebeck}, {and} \bibinfo{person}{Andrea Forte}.} \bibinfo{year}{2019}\natexlab{}.
\newblock \showarticletitle{{Reliability and Inter-Rater Reliability in Qualitative Research: Norms and Guidelines for CSCW and HCI Practice}}.
\newblock \bibinfo{journal}{\emph{Proceedings of the ACM on Human-Computer Interaction}} \bibinfo{volume}{3}, \bibinfo{number}{CSCW} (\bibinfo{year}{2019}).
\newblock
\urldef\tempurl%
\url{https://doi.org/10.1145/3359174}
\showDOI{\tempurl}


\bibitem[McMahan et~al\mbox{.}(2017)]%
        {mcmahan2017communication}
\bibfield{author}{\bibinfo{person}{Brendan McMahan}, \bibinfo{person}{Eider Moore}, \bibinfo{person}{Daniel Ramage}, \bibinfo{person}{Seth Hampson}, {and} \bibinfo{person}{Blaise~Aguera y Arcas}.} \bibinfo{year}{2017}\natexlab{}.
\newblock \showarticletitle{Communication-efficient learning of deep networks from decentralized data}. In \bibinfo{booktitle}{\emph{Artificial intelligence and statistics}}. PMLR, \bibinfo{pages}{1273--1282}.
\newblock
\urldef\tempurl%
\url{https://proceedings.mlr.press/v54/mcmahan17a/mcmahan17a.pdf}
\showURL{%
\tempurl}


\bibitem[McNary and Hunter(2018)]%
        {mcnary2018using}
\bibfield{author}{\bibinfo{person}{Sarah McNary} {and} \bibinfo{person}{Aaron Hunter}.} \bibinfo{year}{2018}\natexlab{}.
\newblock \showarticletitle{Using Wearable Device Data in a Criminal Investigation: A Preliminary Study}. In \bibinfo{booktitle}{\emph{2018 European Intelligence and Security Informatics Conference (EISIC)}}. IEEE, \bibinfo{pages}{85--85}.
\newblock


\bibitem[Meegahapola et~al\mbox{.}(2023)]%
        {meegahapola2023quantified}
\bibfield{author}{\bibinfo{person}{Lakmal Meegahapola}, \bibinfo{person}{Marios Constantinides}, \bibinfo{person}{Zoran Radivojevic}, \bibinfo{person}{Hongwei Li}, \bibinfo{person}{Daniele Quercia}, {and} \bibinfo{person}{Michael~S Eggleston}.} \bibinfo{year}{2023}\natexlab{}.
\newblock \showarticletitle{Quantified Canine: Inferring Dog Personality From Wearables}. In \bibinfo{booktitle}{\emph{Proceedings of the 2023 CHI Conference on Human Factors in Computing Systems}}. \bibinfo{pages}{1--19}.
\newblock


\bibitem[Metcalf et~al\mbox{.}(2021)]%
        {metcalf2021algorithmic}
\bibfield{author}{\bibinfo{person}{Jacob Metcalf}, \bibinfo{person}{Emanuel Moss}, \bibinfo{person}{Elizabeth~Anne Watkins}, \bibinfo{person}{Ranjit Singh}, {and} \bibinfo{person}{Madeleine~Clare Elish}.} \bibinfo{year}{2021}\natexlab{}.
\newblock \showarticletitle{Algorithmic impact assessments and accountability: The co-construction of impacts}. In \bibinfo{booktitle}{\emph{Proceedings of the 2021 ACM Conference on Fairness, Accountability, and Transparency}}. \bibinfo{publisher}{Association for Computing Machinery}, \bibinfo{pages}{735--746}.
\newblock
\urldef\tempurl%
\url{https://doi.org/10.1145/3442188.3445935}
\showDOI{\tempurl}


\bibitem[{Microsoft}(2022)]%
        {microsoft2022Assessment}
\bibfield{author}{\bibinfo{person}{{Microsoft}}.} \bibinfo{year}{2022}\natexlab{}.
\newblock \bibinfo{booktitle}{\emph{{Microsoft’s framework for building AI systems responsibly}}}.
\newblock
\urldef\tempurl%
\url{https://blogs.microsoft.com/on-the-issues/2022/06/21/microsofts-framework-for-building-ai-systems-responsibly/}
\showURL{%
Retrieved January 22, 2024 from \tempurl}


\bibitem[Miles and Huberman(1994)]%
        {miles1994qualitative}
\bibfield{author}{\bibinfo{person}{Matthew Miles} {and} \bibinfo{person}{Michael Huberman}.} \bibinfo{year}{1994}\natexlab{}.
\newblock \bibinfo{booktitle}{\emph{{Qualitative Data Analysis: A Methods Sourcebook}}}.
\newblock \bibinfo{publisher}{Sage}.
\newblock


\bibitem[Mills et~al\mbox{.}(2016)]%
        {mills2016wearing}
\bibfield{author}{\bibinfo{person}{Adam~J Mills}, \bibinfo{person}{Richard~T Watson}, \bibinfo{person}{Leyland Pitt}, {and} \bibinfo{person}{Jan Kietzmann}.} \bibinfo{year}{2016}\natexlab{}.
\newblock \showarticletitle{Wearing safe: Physical and informational security in the age of the wearable device}.
\newblock \bibinfo{journal}{\emph{Business Horizons}} \bibinfo{volume}{59}, \bibinfo{number}{6} (\bibinfo{year}{2016}), \bibinfo{pages}{615--622}.
\newblock


\bibitem[Mirjafari et~al\mbox{.}(2019)]%
        {mirjafari2019differentiating}
\bibfield{author}{\bibinfo{person}{Shayan Mirjafari}, \bibinfo{person}{Kizito Masaba}, \bibinfo{person}{Ted Grover}, \bibinfo{person}{Weichen Wang}, \bibinfo{person}{Pino Audia}, \bibinfo{person}{Andrew~T Campbell}, \bibinfo{person}{Nitesh~V Chawla}, \bibinfo{person}{Vedant~Das Swain}, \bibinfo{person}{Munmun~De Choudhury}, \bibinfo{person}{Anind~K Dey}, {et~al\mbox{.}}} \bibinfo{year}{2019}\natexlab{}.
\newblock \showarticletitle{Differentiating higher and lower job performers in the workplace using mobile sensing}.
\newblock \bibinfo{journal}{\emph{Proceedings of the ACM on Interactive, Mobile, Wearable and Ubiquitous Technologies}} \bibinfo{volume}{3}, \bibinfo{number}{2} (\bibinfo{year}{2019}), \bibinfo{pages}{1--24}.
\newblock


\bibitem[Miro(2022)]%
        {miro2022}
\bibfield{author}{\bibinfo{person}{Miro}.} \bibinfo{year}{2022}\natexlab{}.
\newblock \bibinfo{booktitle}{\emph{{Miro | Online Whiteboard for Visual Collaboration}}}.
\newblock
\urldef\tempurl%
\url{https://miro.com/}
\showURL{%
Retrieved August 2022 from \tempurl}


\bibitem[Mitchell et~al\mbox{.}(2019)]%
        {mitchell2019model}
\bibfield{author}{\bibinfo{person}{Margaret Mitchell}, \bibinfo{person}{Simone Wu}, \bibinfo{person}{Andrew Zaldivar}, \bibinfo{person}{Parker Barnes}, \bibinfo{person}{Lucy Vasserman}, \bibinfo{person}{Ben Hutchinson}, \bibinfo{person}{Elena Spitzer}, \bibinfo{person}{Inioluwa~Deborah Raji}, {and} \bibinfo{person}{Timnit Gebru}.} \bibinfo{year}{2019}\natexlab{}.
\newblock \showarticletitle{Model cards for model reporting}. In \bibinfo{booktitle}{\emph{Proceedings of the conference on fairness, accountability, and transparency}}. \bibinfo{pages}{220--229}.
\newblock


\bibitem[Mittelstadt et~al\mbox{.}(2016)]%
        {mittelstadt2016ethics}
\bibfield{author}{\bibinfo{person}{Brent~Daniel Mittelstadt}, \bibinfo{person}{Patrick Allo}, \bibinfo{person}{Mariarosaria Taddeo}, \bibinfo{person}{Sandra Wachter}, {and} \bibinfo{person}{Luciano Floridi}.} \bibinfo{year}{2016}\natexlab{}.
\newblock \showarticletitle{The ethics of algorithms: Mapping the debate}.
\newblock \bibinfo{journal}{\emph{Big Data \& Society}} \bibinfo{volume}{3}, \bibinfo{number}{2} (\bibinfo{year}{2016}), \bibinfo{pages}{2053951716679679}.
\newblock
\urldef\tempurl%
\url{https://doi.org/10.1177/2053951716679679}
\showDOI{\tempurl}


\bibitem[Mohamed et~al\mbox{.}(2021)]%
        {mohamed2021smart}
\bibfield{author}{\bibinfo{person}{Elsayed~Said Mohamed}, \bibinfo{person}{AA Belal}, \bibinfo{person}{Sameh~Kotb Abd-Elmabod}, \bibinfo{person}{Mohammed~A El-Shirbeny}, \bibinfo{person}{A Gad}, {and} \bibinfo{person}{Mohamed~B Zahran}.} \bibinfo{year}{2021}\natexlab{}.
\newblock \showarticletitle{Smart farming for improving agricultural management}.
\newblock \bibinfo{journal}{\emph{The Egyptian Journal of Remote Sensing and Space Science}} \bibinfo{volume}{24}, \bibinfo{number}{3} (\bibinfo{year}{2021}), \bibinfo{pages}{971--981}.
\newblock


\bibitem[Moraes et~al\mbox{.}(2021)]%
        {moraes2021smile}
\bibfield{author}{\bibinfo{person}{Thiago~Guimar{\~a}es Moraes}, \bibinfo{person}{Eduarda~Costa Almeida}, {and} \bibinfo{person}{Jos{\'e} Renato~Laranjeira de Pereira}.} \bibinfo{year}{2021}\natexlab{}.
\newblock \showarticletitle{Smile, you are being identified! Risks and measures for the use of facial recognition in (semi-) public spaces}.
\newblock \bibinfo{journal}{\emph{AI and Ethics}} \bibinfo{volume}{1}, \bibinfo{number}{2} (\bibinfo{year}{2021}), \bibinfo{pages}{159--172}.
\newblock
\urldef\tempurl%
\url{https://doi.org/10.1007/s43681-020-00014-3}
\showDOI{\tempurl}


\bibitem[Morris et~al\mbox{.}(2015)]%
        {morris2015partially}
\bibfield{author}{\bibinfo{person}{Sinead~E Morris}, \bibinfo{person}{Jonathan~L Zelner}, \bibinfo{person}{Deborah~A Fauquier}, \bibinfo{person}{Teresa~K Rowles}, \bibinfo{person}{Patricia~E Rosel}, \bibinfo{person}{Frances Gulland}, {and} \bibinfo{person}{Bryan~T Grenfell}.} \bibinfo{year}{2015}\natexlab{}.
\newblock \showarticletitle{Partially observed epidemics in wildlife hosts: modelling an outbreak of dolphin morbillivirus in the northwestern Atlantic, June 2013--2014}.
\newblock \bibinfo{journal}{\emph{Journal of the Royal Society Interface}} \bibinfo{volume}{12}, \bibinfo{number}{112} (\bibinfo{year}{2015}), \bibinfo{pages}{20150676}.
\newblock


\bibitem[{National Institute of Standards and Technology}(2023a)]%
        {nist2023aiRisk}
\bibfield{author}{\bibinfo{person}{{National Institute of Standards and Technology}}.} \bibinfo{year}{2023}\natexlab{a}.
\newblock \bibinfo{booktitle}{}.
\newblock
\urldef\tempurl%
\url{https://www.nist.gov/itl/ai-risk-management-framework}
\showURL{%
Retrieved May 2023 from \tempurl}


\bibitem[{National Institute of Standards and Technology}(2023b)]%
        {equalAI2023nist}
\bibfield{author}{\bibinfo{person}{{National Institute of Standards and Technology}}.} \bibinfo{year}{2023}\natexlab{b}.
\newblock \bibinfo{booktitle}{\emph{{The EqualAI Algorithmic Impact Assessment Tool}}}.
\newblock
\urldef\tempurl%
\url{https://www.equalai.org/aia/}
\showURL{%
Retrieved January 2024 from \tempurl}


\bibitem[Neethirajan(2017)]%
        {animal-health}
\bibfield{author}{\bibinfo{person}{Suresh Neethirajan}.} \bibinfo{year}{2017}\natexlab{}.
\newblock \showarticletitle{Recent advances in wearable sensors for animal health management}.
\newblock \bibinfo{journal}{\emph{Sensing and Bio-Sensing Research}}  \bibinfo{volume}{12} (\bibinfo{year}{2017}), \bibinfo{pages}{15--29}.
\newblock
\showISSN{2214-1804}
\urldef\tempurl%
\url{https://doi.org/10.1016/j.sbsr.2016.11.004}
\showDOI{\tempurl}


\bibitem[Nweke et~al\mbox{.}(2019)]%
        {nweke2019data}
\bibfield{author}{\bibinfo{person}{Henry~Friday Nweke}, \bibinfo{person}{Ying~Wah Teh}, \bibinfo{person}{Ghulam Mujtaba}, {and} \bibinfo{person}{Mohammed~Ali Al-Garadi}.} \bibinfo{year}{2019}\natexlab{}.
\newblock \showarticletitle{Data fusion and multiple classifier systems for human activity detection and health monitoring: Review and open research directions}.
\newblock \bibinfo{journal}{\emph{Information Fusion}}  \bibinfo{volume}{46} (\bibinfo{year}{2019}), \bibinfo{pages}{147--170}.
\newblock


\bibitem[Obaideen et~al\mbox{.}(2022)]%
        {obaideen2022overview}
\bibfield{author}{\bibinfo{person}{Khaled Obaideen}, \bibinfo{person}{Bashria~AA Yousef}, \bibinfo{person}{Maryam~Nooman AlMallahi}, \bibinfo{person}{Yong~Chai Tan}, \bibinfo{person}{Montaser Mahmoud}, \bibinfo{person}{Hadi Jaber}, {and} \bibinfo{person}{Mohamad Ramadan}.} \bibinfo{year}{2022}\natexlab{}.
\newblock \showarticletitle{An overview of smart irrigation systems using IoT}.
\newblock \bibinfo{journal}{\emph{Energy Nexus}} (\bibinfo{year}{2022}), \bibinfo{pages}{100124}.
\newblock


\bibitem[Olteanu et~al\mbox{.}(2023)]%
        {olteanu2023responsible}
\bibfield{author}{\bibinfo{person}{Alexandra Olteanu}, \bibinfo{person}{Michael Ekstrand}, \bibinfo{person}{Carlos Castillo}, {and} \bibinfo{person}{Jina Suh}.} \bibinfo{year}{2023}\natexlab{}.
\newblock \showarticletitle{Responsible AI Research Needs Impact Statements Too}.
\newblock \bibinfo{journal}{\emph{arXiv preprint arXiv:2311.11776}} (\bibinfo{year}{2023}).
\newblock


\bibitem[\"{O}mer Ayd{\i}n and Karaarslan(2022)]%
        {aydin2022openai}
\bibfield{author}{\bibinfo{person}{\"{O}mer Ayd{\i}n} {and} \bibinfo{person}{Enis Karaarslan}.} \bibinfo{year}{2022}\natexlab{}.
\newblock \showarticletitle{{OpenAI} {ChatGPT} Generated Literature Review: Digital Twin in Healthcare}.
\newblock \bibinfo{journal}{\emph{{SSRN} Electronic Journal}} (\bibinfo{year}{2022}), \bibinfo{pages}{22--31}.
\newblock
\urldef\tempurl%
\url{https://doi.org/10.2139/ssrn.4308687}
\showDOI{\tempurl}


\bibitem[OpenAI(2023)]%
        {OpenAI}
\bibfield{author}{\bibinfo{person}{OpenAI}.} \bibinfo{year}{2023}\natexlab{}.
\newblock \bibinfo{booktitle}{\emph{Research on GPT-4 - latest updates.}}
\newblock
\urldef\tempurl%
\url{https://openai.com/research/gpt-4}
\showURL{%
Retrieved August 7, 2023 from \tempurl}


\bibitem[Park et~al\mbox{.}(2023)]%
        {park2023social}
\bibfield{author}{\bibinfo{person}{Sungkyu Park}, \bibinfo{person}{Assem Zhunis}, \bibinfo{person}{Marios Constantinides}, \bibinfo{person}{Luca~Maria Aiello}, \bibinfo{person}{Daniele Quercia}, {and} \bibinfo{person}{Meeyoung Cha}.} \bibinfo{year}{2023}\natexlab{}.
\newblock \showarticletitle{Social dimensions impact individual sleep quantity and quality}.
\newblock \bibinfo{journal}{\emph{Scientific Reports}} \bibinfo{volume}{13}, \bibinfo{number}{1} (\bibinfo{year}{2023}), \bibinfo{pages}{1--11}.
\newblock


\bibitem[Perez-Pozuelo et~al\mbox{.}(2021)]%
        {perez2021wearables}
\bibfield{author}{\bibinfo{person}{Ignacio Perez-Pozuelo}, \bibinfo{person}{Dimitris Spathis}, \bibinfo{person}{Emma~AD Clifton}, {and} \bibinfo{person}{Cecilia Mascolo}.} \bibinfo{year}{2021}\natexlab{}.
\newblock \showarticletitle{Wearables, smartphones, and artificial intelligence for digital phenotyping and health}.
\newblock In \bibinfo{booktitle}{\emph{Digital Health}}. \bibinfo{publisher}{Elsevier}, \bibinfo{pages}{33--54}.
\newblock


\bibitem[Puntoni et~al\mbox{.}(2021)]%
        {puntoni2021consumers}
\bibfield{author}{\bibinfo{person}{Stefano Puntoni}, \bibinfo{person}{Rebecca~Walker Reczek}, \bibinfo{person}{Markus Giesler}, {and} \bibinfo{person}{Simona Botti}.} \bibinfo{year}{2021}\natexlab{}.
\newblock \showarticletitle{Consumers and artificial intelligence: An experiential perspective}.
\newblock \bibinfo{journal}{\emph{Journal of Marketing}} \bibinfo{volume}{85}, \bibinfo{number}{1} (\bibinfo{year}{2021}), \bibinfo{pages}{131--151}.
\newblock


\bibitem[Read(2009)]%
        {read2009majority}
\bibfield{author}{\bibinfo{person}{James~H Read}.} \bibinfo{year}{2009}\natexlab{}.
\newblock \bibinfo{booktitle}{\emph{Majority rule versus consensus: the political thought of John C. Calhoun}}.
\newblock \bibinfo{publisher}{University Press of Kansas}.
\newblock


\bibitem[Rios-Navarro et~al\mbox{.}(2016)]%
        {rios2016sensor}
\bibfield{author}{\bibinfo{person}{Antonio Rios-Navarro}, \bibinfo{person}{Juan~Pedro Dominguez-Morales}, \bibinfo{person}{Ricardo Tapiador-Morales}, \bibinfo{person}{Manuel Dominguez-Morales}, \bibinfo{person}{Angel Jimenez-Fernandez}, {and} \bibinfo{person}{Alejandro Linares-Barranco}.} \bibinfo{year}{2016}\natexlab{}.
\newblock \showarticletitle{A sensor fusion horse gait classification by a spiking neural network on SpiNNaker}. In \bibinfo{booktitle}{\emph{Artificial Neural Networks and Machine Learning--ICANN 2016: 25th International Conference on Artificial Neural Networks, Barcelona, Spain, September 6-9, 2016, Proceedings, Part I 25}}. Springer, \bibinfo{pages}{36--44}.
\newblock


\bibitem[Roper et~al\mbox{.}(2021)]%
        {roper2021emerging}
\bibfield{author}{\bibinfo{person}{Jenna~M Roper}, \bibinfo{person}{Jose~F Garcia}, {and} \bibinfo{person}{Hideaki Tsutsui}.} \bibinfo{year}{2021}\natexlab{}.
\newblock \showarticletitle{Emerging technologies for monitoring plant health in vivo}.
\newblock \bibinfo{journal}{\emph{ACS omega}} \bibinfo{volume}{6}, \bibinfo{number}{8} (\bibinfo{year}{2021}), \bibinfo{pages}{5101--5107}.
\newblock


\bibitem[Salda{\~n}a(2015)]%
        {saldana2015coding}
\bibfield{author}{\bibinfo{person}{Johnny Salda{\~n}a}.} \bibinfo{year}{2015}\natexlab{}.
\newblock \bibinfo{booktitle}{\emph{{The Coding Manual for Qualitative Researchers}}}.
\newblock \bibinfo{publisher}{Sage}.
\newblock


\bibitem[Saraceni et~al\mbox{.}(2012)]%
        {saraceni2012active}
\bibfield{author}{\bibinfo{person}{Simone Saraceni}, \bibinfo{person}{Andrea Claudi}, {and} \bibinfo{person}{Aldo~Franco Dragoni}.} \bibinfo{year}{2012}\natexlab{}.
\newblock \showarticletitle{An active monitoring system for real-time face-tracking based on mobile sensors}. In \bibinfo{booktitle}{\emph{Proceedings ELMAR-2012}}. IEEE, \bibinfo{pages}{53--56}.
\newblock


\bibitem[Saraswat et~al\mbox{.}(2022)]%
        {saraswat2022explainable}
\bibfield{author}{\bibinfo{person}{Deepti Saraswat}, \bibinfo{person}{Pronaya Bhattacharya}, \bibinfo{person}{Ashwin Verma}, \bibinfo{person}{Vivek~Kumar Prasad}, \bibinfo{person}{Sudeep Tanwar}, \bibinfo{person}{Gulshan Sharma}, \bibinfo{person}{Pitshou~N Bokoro}, {and} \bibinfo{person}{Ravi Sharma}.} \bibinfo{year}{2022}\natexlab{}.
\newblock \showarticletitle{Explainable AI for healthcare 5.0: opportunities and challenges}.
\newblock \bibinfo{journal}{\emph{IEEE Access}} (\bibinfo{year}{2022}).
\newblock


\bibitem[Saravia(2022)]%
        {Saravia_Prompt_Engineering_Guide_2022}
\bibfield{author}{\bibinfo{person}{Elvis Saravia}.} \bibinfo{year}{2022}\natexlab{}.
\newblock \bibinfo{booktitle}{\emph{{Prompt Engineering Guide}}}.
\newblock DAIR.AI.
\newblock
\urldef\tempurl%
\url{https://github.com/dair-ai/Prompt-Engineering-Guide}
\showURL{%
Retrieved August 7, 2023 from \tempurl}


\bibitem[Schaekermann et~al\mbox{.}(2019)]%
        {schaekermann2019understanding}
\bibfield{author}{\bibinfo{person}{Mike Schaekermann}, \bibinfo{person}{Graeme Beaton}, \bibinfo{person}{Minahz Habib}, \bibinfo{person}{Andrew Lim}, \bibinfo{person}{Kate Larson}, {and} \bibinfo{person}{Edith Law}.} \bibinfo{year}{2019}\natexlab{}.
\newblock \showarticletitle{Understanding expert disagreement in medical data analysis through structured adjudication}.
\newblock \bibinfo{journal}{\emph{Proceedings of the ACM on Human-Computer Interaction}} \bibinfo{volume}{3}, \bibinfo{number}{CSCW} (\bibinfo{year}{2019}), \bibinfo{pages}{1--23}.
\newblock


\bibitem[Schriewer and Bulaj(2016)]%
        {schriewer2016music}
\bibfield{author}{\bibinfo{person}{Karl Schriewer} {and} \bibinfo{person}{Grzegorz Bulaj}.} \bibinfo{year}{2016}\natexlab{}.
\newblock \showarticletitle{Music streaming services as adjunct therapies for depression, anxiety, and bipolar symptoms: convergence of digital technologies, mobile apps, emotions, and global mental health}.
\newblock \bibinfo{journal}{\emph{Frontiers in public health}}  \bibinfo{volume}{4} (\bibinfo{year}{2016}), \bibinfo{pages}{217}.
\newblock


\bibitem[Seshadri et~al\mbox{.}(2020)]%
        {seshadri2020wearable}
\bibfield{author}{\bibinfo{person}{Dhruv~R Seshadri}, \bibinfo{person}{Evan~V Davies}, \bibinfo{person}{Ethan~R Harlow}, \bibinfo{person}{Jeffrey~J Hsu}, \bibinfo{person}{Shanina~C Knighton}, \bibinfo{person}{Timothy~A Walker}, \bibinfo{person}{James~E Voos}, {and} \bibinfo{person}{Colin~K Drummond}.} \bibinfo{year}{2020}\natexlab{}.
\newblock \showarticletitle{Wearable sensors for COVID-19: a call to action to harness our digital infrastructure for remote patient monitoring and virtual assessments}.
\newblock \bibinfo{journal}{\emph{Frontiers in Digital Health}} (\bibinfo{year}{2020}), \bibinfo{pages}{8}.
\newblock


\bibitem[Sharma et~al\mbox{.}(2023)]%
        {sharma2023cognitive}
\bibfield{author}{\bibinfo{person}{Ashish Sharma}, \bibinfo{person}{Kevin Rushton}, \bibinfo{person}{Inna~Wanyin Lin}, \bibinfo{person}{David Wadden}, \bibinfo{person}{Khendra~G. Lucas}, \bibinfo{person}{Adam~S Miner}, \bibinfo{person}{Theresa Nguyen}, {and} \bibinfo{person}{Tim Althoff}.} \bibinfo{year}{2023}\natexlab{}.
\newblock \showarticletitle{Cognitive Reframing of Negative Thoughts through Human-Language Model Interaction}.
\newblock \bibinfo{journal}{\emph{arXiv preprint arXiv:2305.02466}} (\bibinfo{year}{2023}).
\newblock
\urldef\tempurl%
\url{https://arxiv.org/abs/2305.02466}
\showURL{%
\tempurl}


\bibitem[Sherman and Eisenberg(2023)]%
        {sherman2023riskProfiles}
\bibfield{author}{\bibinfo{person}{Eli Sherman} {and} \bibinfo{person}{Ian~W. Eisenberg}.} \bibinfo{year}{2023}\natexlab{}.
\newblock \bibinfo{title}{AI Risk Profiles: A Standards Proposal for Pre-Deployment AI Risk Disclosures}.
\newblock
\newblock
\showeprint[arxiv]{2309.13176}~[cs.AI]


\bibitem[Shieh(2023)]%
        {OpenAIGuide}
\bibfield{author}{\bibinfo{person}{Jessica Shieh}.} \bibinfo{year}{2023}\natexlab{}.
\newblock \bibinfo{booktitle}{\emph{Best practices for prompt engineering with OpenAI API}}.
\newblock OpenAI.
\newblock
\urldef\tempurl%
\url{https://help.openai.com/en/articles/6654000-best-practices-for-prompt-engineering-with-openai-api}
\showURL{%
Retrieved August 7, 2023 from \tempurl}


\bibitem[Sjoding et~al\mbox{.}(2020)]%
        {sjoding2020racial}
\bibfield{author}{\bibinfo{person}{Michael~W Sjoding}, \bibinfo{person}{Robert~P Dickson}, \bibinfo{person}{Theodore~J Iwashyna}, \bibinfo{person}{Steven~E Gay}, {and} \bibinfo{person}{Thomas~S Valley}.} \bibinfo{year}{2020}\natexlab{}.
\newblock \showarticletitle{Racial bias in pulse oximetry measurement}.
\newblock \bibinfo{journal}{\emph{New England Journal of Medicine}} \bibinfo{volume}{383}, \bibinfo{number}{25} (\bibinfo{year}{2020}), \bibinfo{pages}{2477--2478}.
\newblock


\bibitem[Sumrall et~al\mbox{.}(2001)]%
        {sumrall2001global}
\bibfield{author}{\bibinfo{person}{Colin~D Sumrall}, \bibinfo{person}{Christopher~A Brochu}, {and} \bibinfo{person}{John~W Merck}.} \bibinfo{year}{2001}\natexlab{}.
\newblock \showarticletitle{Global lability, regional resolution, and majority-rule consensus bias}.
\newblock \bibinfo{journal}{\emph{Paleobiology}} \bibinfo{volume}{27}, \bibinfo{number}{2} (\bibinfo{year}{2001}), \bibinfo{pages}{254--261}.
\newblock


\bibitem[Sunehra et~al\mbox{.}(2020)]%
        {sunehra2020raspberry}
\bibfield{author}{\bibinfo{person}{Dhiraj Sunehra}, \bibinfo{person}{V~Sai Sreshta}, \bibinfo{person}{V Shashank}, {and} \bibinfo{person}{B~Uday~Kumar Goud}.} \bibinfo{year}{2020}\natexlab{}.
\newblock \showarticletitle{Raspberry pi based smart wearable device for women safety using gps and gsm technology}. In \bibinfo{booktitle}{\emph{2020 IEEE International Conference for Innovation in Technology (INOCON)}}. IEEE, \bibinfo{pages}{1--5}.
\newblock


\bibitem[Tahaei et~al\mbox{.}(2023)]%
        {tahaei2023human}
\bibfield{author}{\bibinfo{person}{Mohammad Tahaei}, \bibinfo{person}{Marios Constantinides}, \bibinfo{person}{Daniele Quercia}, \bibinfo{person}{Sean Kennedy}, \bibinfo{person}{Michael Muller}, \bibinfo{person}{Simone Stumpf}, \bibinfo{person}{Q~Vera Liao}, \bibinfo{person}{Ricardo Baeza-Yates}, \bibinfo{person}{Lora Aroyo}, \bibinfo{person}{Jess Holbrook}, {et~al\mbox{.}}} \bibinfo{year}{2023}\natexlab{}.
\newblock \showarticletitle{Human-Centered Responsible Artificial Intelligence: Current \& Future Trends}. In \bibinfo{booktitle}{\emph{Extended Abstracts of the 2023 CHI Conference on Human Factors in Computing Systems}}. \bibinfo{pages}{1--4}.
\newblock


\bibitem[Tolmeijer et~al\mbox{.}(2022)]%
        {chi_human_ai_colab}
\bibfield{author}{\bibinfo{person}{Suzanne Tolmeijer}, \bibinfo{person}{Markus Christen}, \bibinfo{person}{Serhiy Kandul}, \bibinfo{person}{Markus Kneer}, {and} \bibinfo{person}{Abraham Bernstein}.} \bibinfo{year}{2022}\natexlab{}.
\newblock \showarticletitle{Capable but Amoral? Comparing AI and Human Expert Collaboration in Ethical Decision Making}. In \bibinfo{booktitle}{\emph{Proceedings of the 2022 CHI Conference on Human Factors in Computing Systems}} (New Orleans, LA, USA) \emph{(\bibinfo{series}{CHI '22})}. \bibinfo{publisher}{Association for Computing Machinery}, \bibinfo{address}{New York, NY, USA}, Article \bibinfo{articleno}{160}, \bibinfo{numpages}{17}~pages.
\newblock
\showISBNx{9781450391573}
\urldef\tempurl%
\url{https://doi.org/10.1145/3491102.3517732}
\showDOI{\tempurl}


\bibitem[Torous et~al\mbox{.}(2017)]%
        {torous2017new}
\bibfield{author}{\bibinfo{person}{John Torous}, \bibinfo{person}{JP Onnela}, {and} \bibinfo{person}{Matcheri Keshavan}.} \bibinfo{year}{2017}\natexlab{}.
\newblock \showarticletitle{New dimensions and new tools to realize the potential of RDoC: digital phenotyping via smartphones and connected devices}.
\newblock \bibinfo{journal}{\emph{Translational psychiatry}} \bibinfo{volume}{7}, \bibinfo{number}{3} (\bibinfo{year}{2017}), \bibinfo{pages}{e1053--e1053}.
\newblock


\bibitem[Tyagi et~al\mbox{.}(2020)]%
        {tyagi2020crop}
\bibfield{author}{\bibinfo{person}{Kirti Tyagi}, \bibinfo{person}{Aabha Karmarkar}, \bibinfo{person}{Simran Kaur}, \bibinfo{person}{Sukanya Kulkarni}, {and} \bibinfo{person}{Rita Das}.} \bibinfo{year}{2020}\natexlab{}.
\newblock \showarticletitle{Crop health monitoring system}. In \bibinfo{booktitle}{\emph{2020 International Conference for Emerging Technology (INCET)}}. IEEE, \bibinfo{pages}{1--5}.
\newblock


\bibitem[{U}nited {N}ations: Department~of Economic and Development(2023)]%
        {sdgs}
\bibfield{author}{\bibinfo{person}{{U}nited {N}ations: Department~of Economic} {and} \bibinfo{person}{Social Affairs~Sustainable Development}.} \bibinfo{year}{2023}\natexlab{}.
\newblock \bibinfo{booktitle}{\emph{The 17 Goals}}.
\newblock
\urldef\tempurl%
\url{https://sdgs.un.org/goals}
\showURL{%
Retrieved November 8, 2023 from \tempurl}


\bibitem[Venkateswarlu et~al\mbox{.}(2006)]%
        {venkateswarlu2006wearable}
\bibfield{author}{\bibinfo{person}{Ronda Venkateswarlu}, \bibinfo{person}{Choo~Hui Wei}, \bibinfo{person}{Ngiam~Li Lian}, \bibinfo{person}{ET Lim}, \bibinfo{person}{Zijian Zhu}, {and} \bibinfo{person}{Mingjiang Yang}.} \bibinfo{year}{2006}\natexlab{}.
\newblock \showarticletitle{Wearable high-tech gear for homeland security personnel}. In \bibinfo{booktitle}{\emph{Sensors, and Command, Control, Communications, and Intelligence (C3I) Technologies for Homeland Security and Homeland Defense V}}, Vol.~\bibinfo{volume}{6201}. SPIE, \bibinfo{pages}{382--391}.
\newblock


\bibitem[Vurro et~al\mbox{.}(2023)]%
        {vurro2023vivo}
\bibfield{author}{\bibinfo{person}{Filippo Vurro}, \bibinfo{person}{Riccardo Manfredi}, \bibinfo{person}{Manuele Bettelli}, \bibinfo{person}{Gionata Bocci}, \bibinfo{person}{Alberto~Luigi Cologni}, \bibinfo{person}{Sandro Cornali}, \bibinfo{person}{Roberto Reggiani}, \bibinfo{person}{Edoardo Marchetti}, \bibinfo{person}{Nicola Copped{\`e}}, \bibinfo{person}{Stefano Caselli}, {et~al\mbox{.}}} \bibinfo{year}{2023}\natexlab{}.
\newblock \showarticletitle{In vivo sensing to monitor tomato plants in field conditions and optimize crop water management}.
\newblock \bibinfo{journal}{\emph{Precision Agriculture}} (\bibinfo{year}{2023}), \bibinfo{pages}{1--21}.
\newblock


\bibitem[Wall et~al\mbox{.}(2014)]%
        {wall2014novel}
\bibfield{author}{\bibinfo{person}{Jake Wall}, \bibinfo{person}{George Wittemyer}, \bibinfo{person}{Brian Klinkenberg}, {and} \bibinfo{person}{Iain Douglas-Hamilton}.} \bibinfo{year}{2014}\natexlab{}.
\newblock \showarticletitle{Novel opportunities for wildlife conservation and research with real-time monitoring}.
\newblock \bibinfo{journal}{\emph{Ecological Applications}} \bibinfo{volume}{24}, \bibinfo{number}{4} (\bibinfo{year}{2014}), \bibinfo{pages}{593--601}.
\newblock


\bibitem[Wei et~al\mbox{.}(2023)]%
        {Wei2023AnOO}
\bibfield{author}{\bibinfo{person}{Chen Wei}, \bibinfo{person}{Yun~Cheng Wang}, \bibinfo{person}{Bin Wang}, {and} \bibinfo{person}{C.-C.~Jay Kuo}.} \bibinfo{year}{2023}\natexlab{}.
\newblock \showarticletitle{An Overview on Language Models: Recent Developments and Outlook}.
\newblock \bibinfo{journal}{\emph{ArXiv}}  \bibinfo{volume}{abs/2303.05759} (\bibinfo{year}{2023}).
\newblock
\urldef\tempurl%
\url{https://api.semanticscholar.org/CorpusID:257482688}
\showURL{%
\tempurl}


\bibitem[Wei et~al\mbox{.}(2022a)]%
        {wei2021finetuned}
\bibfield{author}{\bibinfo{person}{Jason Wei}, \bibinfo{person}{Maarten Bosma}, \bibinfo{person}{Vincent~Y. Zhao}, \bibinfo{person}{Kelvin Guu}, \bibinfo{person}{Adams~Wei Yu}, \bibinfo{person}{Brian Lester}, \bibinfo{person}{Nan Du}, \bibinfo{person}{Andrew~M. Dai}, {and} \bibinfo{person}{Quoc~V. Le}.} \bibinfo{year}{2022}\natexlab{a}.
\newblock \showarticletitle{Finetuned language models are zero-shot learners}.
\newblock \bibinfo{journal}{\emph{The International Conference on Learning Representations (ICLR)}} (\bibinfo{year}{2022}).
\newblock


\bibitem[Wei et~al\mbox{.}(2022b)]%
        {NEURIPS2022_9d560961}
\bibfield{author}{\bibinfo{person}{Jason Wei}, \bibinfo{person}{Xuezhi Wang}, \bibinfo{person}{Dale Schuurmans}, \bibinfo{person}{Maarten Bosma}, \bibinfo{person}{Brian Ichter}, \bibinfo{person}{Fei Xia}, \bibinfo{person}{Ed Chi}, \bibinfo{person}{Quoc~V. Le}, {and} \bibinfo{person}{Denny Zhou}.} \bibinfo{year}{2022}\natexlab{b}.
\newblock \showarticletitle{Chain-of-Thought Prompting Elicits Reasoning in Large Language Models}. In \bibinfo{booktitle}{\emph{Advances in Neural Information Processing Systems}}, \bibfield{editor}{\bibinfo{person}{S.~Koyejo}, \bibinfo{person}{S.~Mohamed}, \bibinfo{person}{A.~Agarwal}, \bibinfo{person}{D.~Belgrave}, \bibinfo{person}{K.~Cho}, {and} \bibinfo{person}{A.~Oh}} (Eds.), Vol.~\bibinfo{volume}{35}. \bibinfo{publisher}{Curran Associates, Inc.}, \bibinfo{pages}{24824--24837}.
\newblock
\urldef\tempurl%
\url{https://proceedings.neurips.cc/paper_files/paper/2022/file/9d5609613524ecf4f15af0f7b31abca4-Paper-Conference.pdf}
\showURL{%
\tempurl}


\bibitem[Wong and Kim(2016)]%
        {wong2016enhanced}
\bibfield{author}{\bibinfo{person}{Kok-Seng Wong} {and} \bibinfo{person}{Myung~Ho Kim}.} \bibinfo{year}{2016}\natexlab{}.
\newblock \showarticletitle{An enhanced user authentication solution for mobile payment systems using wearables}.
\newblock \bibinfo{journal}{\emph{Security and Communication Networks}} \bibinfo{volume}{9}, \bibinfo{number}{17} (\bibinfo{year}{2016}), \bibinfo{pages}{4639--4649}.
\newblock


\bibitem[Wu et~al\mbox{.}(2022)]%
        {CHI_AI_CHAINS}
\bibfield{author}{\bibinfo{person}{Tongshuang Wu}, \bibinfo{person}{Michael Terry}, {and} \bibinfo{person}{Carrie~Jun Cai}.} \bibinfo{year}{2022}\natexlab{}.
\newblock \showarticletitle{AI Chains: Transparent and Controllable Human-AI Interaction by Chaining Large Language Model Prompts} \emph{(\bibinfo{series}{CHI '22})}. \bibinfo{publisher}{Association for Computing Machinery}, \bibinfo{address}{New York, NY, USA}, Article \bibinfo{articleno}{385}, \bibinfo{numpages}{22}~pages.
\newblock
\showISBNx{9781450391573}
\urldef\tempurl%
\url{https://doi.org/10.1145/3491102.3517582}
\showDOI{\tempurl}


\bibitem[Xu and Ma(2023)]%
        {xu2023regulation}
\bibfield{author}{\bibinfo{person}{Xiangyun Xu} {and} \bibinfo{person}{Wei Ma}.} \bibinfo{year}{2023}\natexlab{}.
\newblock \showarticletitle{The regulation of the body by smart wearable devices and their social risk progression}.
\newblock \bibinfo{journal}{\emph{Wearable Technology}} \bibinfo{volume}{2}, \bibinfo{number}{1} (\bibinfo{year}{2023}), \bibinfo{pages}{60--72}.
\newblock


\bibitem[Yang et~al\mbox{.}(2018b)]%
        {yang2018towards}
\bibfield{author}{\bibinfo{person}{Liyun Yang}, \bibinfo{person}{Ke Lu}, \bibinfo{person}{Jose~A Diaz-Olivares}, \bibinfo{person}{Fernando Seoane}, \bibinfo{person}{Kaj Lindecrantz}, \bibinfo{person}{Mikael Forsman}, \bibinfo{person}{Farhad Abtahi}, {and} \bibinfo{person}{J{\"o}Rgen~AE Eklund}.} \bibinfo{year}{2018}\natexlab{b}.
\newblock \showarticletitle{Towards smart work clothing for automatic risk assessment of physical workload}.
\newblock \bibinfo{journal}{\emph{Ieee Access}}  \bibinfo{volume}{6} (\bibinfo{year}{2018}), \bibinfo{pages}{40059--40072}.
\newblock


\bibitem[Yang et~al\mbox{.}(2018a)]%
        {yang2018backscatter}
\bibfield{author}{\bibinfo{person}{Zhice Yang}, \bibinfo{person}{Qianyi Huang}, {and} \bibinfo{person}{Qian Zhang}.} \bibinfo{year}{2018}\natexlab{a}.
\newblock \showarticletitle{Backscatter as a covert channel in mobile devices}.
\newblock \bibinfo{journal}{\emph{GetMobile: Mobile Computing and Communications}} \bibinfo{volume}{22}, \bibinfo{number}{1} (\bibinfo{year}{2018}), \bibinfo{pages}{31--34}.
\newblock


\bibitem[Zheng et~al\mbox{.}(2023)]%
        {zheng2023llm}
\bibfield{author}{\bibinfo{person}{Zhe Zheng}, \bibinfo{person}{Ke-Yin Chen}, \bibinfo{person}{Xin-Yu Cao}, \bibinfo{person}{Xin-Zheng Lu}, {and} \bibinfo{person}{Jia-Rui Lin}.} \bibinfo{year}{2023}\natexlab{}.
\newblock \showarticletitle{LLM-FuncMapper: Function Identification for Interpreting Complex Clauses in Building Codes via LLM}.
\newblock \bibinfo{journal}{\emph{arXiv preprint arXiv:2308.08728}} (\bibinfo{year}{2023}).
\newblock


\bibitem[Zhou et~al\mbox{.}(2023)]%
        {zhou2023circadian}
\bibfield{author}{\bibinfo{person}{Ke Zhou}, \bibinfo{person}{Marios Constantinides}, \bibinfo{person}{Daniele Quercia}, {and} \bibinfo{person}{Sanja {\v{S}}{\'c}epanovi{\'c}}.} \bibinfo{year}{2023}\natexlab{}.
\newblock \showarticletitle{How Circadian Rhythms Extracted From Social Media Relate to Physical Activity and Sleep}. In \bibinfo{booktitle}{\emph{Proceedings of the International AAAI Conference on Web and Social Media}}, Vol.~\bibinfo{volume}{17}. \bibinfo{pages}{948--959}.
\newblock


\bibitem[Zuboff(2023)]%
        {zuboff2023age}
\bibfield{author}{\bibinfo{person}{Shoshana Zuboff}.} \bibinfo{year}{2023}\natexlab{}.
\newblock \showarticletitle{The age of surveillance capitalism}.
\newblock In \bibinfo{booktitle}{\emph{Social Theory Re-Wired}}. \bibinfo{publisher}{Routledge}, \bibinfo{pages}{203--213}.
\newblock


\end{thebibliography}


\clearpage
\appendix
\section*{Appendix}

\renewcommand{\arraystretch}{0.8} 


 \tiny
\setlength{\tabcolsep}{1.3pt}
\begin{longtable}{rllll}
\caption{List of LLM-generated AI uses for mobile and wearable technologies. Each entry includes an ID, description, an indication of whether it is high risk, the goals it supports, and the specific clauses from the EU AI Act that categorize the use as high risk or not. Use \#64 is classified as unacceptable, and was marked with a (*).}\label{tab:all_uses}\\
\toprule
  ID &                                        Description & High Risk &                                 SDG Goals &                                          AIActText \\
\midrule
\endhead
\midrule
\multicolumn{5}{r}{{Continued on next page}} \\
\midrule
\endfoot

\bottomrule
\endlastfoot
   {\fontsize{4}{6}\selectfont 1} &  {\fontsize{4}{6}\selectfont Enhancing security through facial recognition using high-performance cameras in smartphones} &    {\fontsize{4}{6}\selectfont   Yes} & {\fontsize{4}{6}\selectfont 16, 5, 8} &  {\fontsize{4}{6}\selectfont Article 6, Amendment 711 - biometric identification/categorization of natural persons}  \\ 
   \arrayrulecolor{black!20}\midrule
  {\fontsize{4}{6}\selectfont 2} & {\fontsize{4}{6}\selectfont Monitoring and analyzing crowd behavior using mobile cameras} &    {\fontsize{4}{6}\selectfont   Yes} &                 {\fontsize{4}{6}\selectfont  16, 11, 5, 8, 3} & {\fontsize{4}{6}\selectfont Article 6, Amendment 711 - biometric identification/categorization of natural persons } \\ \arrayrulecolor{black!20}\midrule
   {\fontsize{4}{6}\selectfont 3} & {\fontsize{4}{6}\selectfont Identifying individuals in criminal investigations using smartphones and wearable watches } &      {\fontsize{4}{6}\selectfont Yes} &                {\fontsize{4}{6}\selectfont 16, 10, 5, 8, 3, 11} & {\fontsize{4}{6}\selectfont Article 6, Amendment 711  - biometric identification/categorization of natural persons}  \\ \arrayrulecolor{black!20}\midrule
   {\fontsize{4}{6}\selectfont 4} & {\fontsize{4}{6}\selectfont Monitoring children's location for safety using wristbands} &     {\fontsize{4}{6}\selectfont  Yes} &                  {\fontsize{4}{6}\selectfont 3, 11, 16, 5, 8} & {\fontsize{4}{6}\selectfont Amendment 38 -  perceive or exploit vulnerabilities of children and people} \\ \arrayrulecolor{black!20}\midrule
   {\fontsize{4}{6}\selectfont 5} & {\fontsize{4}{6}\selectfont Monitoring elderly health and activity using wristbands} &       {\fontsize{4}{6}\selectfont Yes} &                          {\fontsize{4}{6}\selectfont 3, 11} & {\fontsize{4}{6}\selectfont Amendment 38 - perceive or exploit vulnerabilities of children and people}  \\ \arrayrulecolor{black!20}\midrule
   {\fontsize{4}{6}\selectfont 6} &           {\fontsize{4}{6}\selectfont   Coordinating family calendar and tasks on mobile phones} &    {\fontsize{4}{6}\selectfont    No} &            {\fontsize{4}{6}\selectfont   5, 8, 12} &                                                {\fontsize{4}{6}\selectfont N/A} \\ \arrayrulecolor{black!20}\midrule
   {\fontsize{4}{6}\selectfont 7} &           {\fontsize{4}{6}\selectfont    Sharing calendars and reminders on mobile phones} &      {\fontsize{4}{6}\selectfont  No} &  {\fontsize{4}{6}\selectfont 5, 8, 12} &                                               {\fontsize{4}{6}\selectfont N/A } \\ \arrayrulecolor{black!20}\midrule
   {\fontsize{4}{6}\selectfont 8} &       {\fontsize{4}{6}\selectfont  Sharing location for coordination through smartwatches} &       {\fontsize{4}{6}\selectfont No} &                   {\fontsize{4}{6}\selectfont 11, 5, 3} &                                             {\fontsize{4}{6}\selectfont   N/A} \\ \arrayrulecolor{black!20}\midrule
   {\fontsize{4}{6}\selectfont 9} &         {\fontsize{4}{6}\selectfont  Tracking emotions using mobile app for improving emotional well-being} &     {\fontsize{4}{6}\selectfont  Yes} &                     {\fontsize{4}{6}\selectfont  5, 3} &  {\fontsize{4}{6}\selectfont Article 6, Amendment 712 - biometric identification/categorization of natural persons}  \\ \arrayrulecolor{black!20}\midrule
  {\fontsize{4}{6}\selectfont 10} &            {\fontsize{4}{6}\selectfont  Tracking health status using smart rings for chronic conditions}  &     {\fontsize{4}{6}\selectfont  Yes} &                               {\fontsize{4}{6}\selectfont 8, 3} &  {\fontsize{4}{6}\selectfont Article 6 - access to essential private services}  \\ \arrayrulecolor{black!20}\midrule
  {\fontsize{4}{6}\selectfont 11} &                {\fontsize{4}{6}\selectfont  Tracking movements using accelerometers for improving physical therapy outcomes}  &     {\fontsize{4}{6}\selectfont  Yes} &                         {\fontsize{4}{6}\selectfont 8, 3} &  {\fontsize{4}{6}\selectfont Article 6 - access to essential private services }\\ \arrayrulecolor{black!20}\midrule
  {\fontsize{4}{6}\selectfont 12} & {\fontsize{4}{6}\selectfont Enhancing mental health treatment using smartwatches}  &      {\fontsize{4}{6}\selectfont Yes} &                        {\fontsize{4}{6}\selectfont 8, 3} &  {\fontsize{4}{6}\selectfont Article 6 - biometric identification/categorization of natural persons }\\ \arrayrulecolor{black!20}\midrule
  {\fontsize{4}{6}\selectfont 13} &    {\fontsize{4}{6}\selectfont                       Promoting physical fitness through activity tracking on smartwatches} &      {\fontsize{4}{6}\selectfont  No} &                   {\fontsize{4}{6}\selectfont       5, 8, 3} &                                               {\fontsize{4}{6}\selectfont N/A} \\ \arrayrulecolor{black!20}\midrule
  {\fontsize{4}{6}\selectfont 14} &                           {\fontsize{4}{6}\selectfont   Tracking sleep quality} &        {\fontsize{4}{6}\selectfont No} &                            {\fontsize{4}{6}\selectfont  8, 3} &                                               {\fontsize{4}{6}\selectfont N/A} \\ \arrayrulecolor{black!20}\midrule
 {\fontsize{4}{6}\selectfont 15} &         {\fontsize{4}{6}\selectfont  Providing meditation guidance on a mobile app}  &      {\fontsize{4}{6}\selectfont  No} &                         {\fontsize{4}{6}\selectfont 8, 3 }&                                              {\fontsize{4}{6}\selectfont  N/A} \\ \arrayrulecolor{black!20}\midrule
  {\fontsize{4}{6}\selectfont 16} &            {\fontsize{4}{6}\selectfont   Performing gesture recognition using accelerometers for UI accessibility } &    {\fontsize{4}{6}\selectfont    No} &  {\fontsize{4}{6}\selectfont 4, 8, 3, 9} &                                               {\fontsize{4}{6}\selectfont N/A} \\ \arrayrulecolor{black!20}\midrule
  {\fontsize{4}{6}\selectfont 17} &              {\fontsize{4}{6}\selectfont  Tracking motion and environmental interaction using VR headsets} &      {\fontsize{4}{6}\selectfont  No} &                   {\fontsize{4}{6}\selectfont 8, 4, 3} &                                             {\fontsize{4}{6}\selectfont   N/A} \\ \arrayrulecolor{black!20}\midrule
  {\fontsize{4}{6}\selectfont 18} &                {\fontsize{4}{6}\selectfont   Enabling hands-free device control through voice recognition and command execution} &      {\fontsize{4}{6}\selectfont  No} & {\fontsize{4}{6}\selectfont 8, 4, 3} &                                               {\fontsize{4}{6}\selectfont N/A} \\ \arrayrulecolor{black!20}\midrule
  {\fontsize{4}{6}\selectfont 19} &           {\fontsize{4}{6}\selectfont   Tracking expenses and monitoring personal finances using a mobile app} &      {\fontsize{4}{6}\selectfont  No} &              {\fontsize{4}{6}\selectfont 12, 5, 8, 4, 1} &                                              {\fontsize{4}{6}\selectfont  N/A }\\ \arrayrulecolor{black!20}\midrule
  {\fontsize{4}{6}\selectfont 20} &              {\fontsize{4}{6}\selectfont Enabling secure mobile payments and transfers with a smartwatch} &      {\fontsize{4}{6}\selectfont Yes} &                     {\fontsize{4}{6}\selectfont  17, 10, 8, 1} & {\fontsize{4}{6}\selectfont Article 6, Amendment 64 - AI systems as safety component}  \\ \arrayrulecolor{black!20}\midrule
  {\fontsize{4}{6}\selectfont 21} &             {\fontsize{4}{6}\selectfont  Providing investment and financial advice using a mobile app} &      {\fontsize{4}{6}\selectfont Yes} &                {\fontsize{4}{6}\selectfont 10, 5, 8, 4, 1} &  {\fontsize{4}{6}\selectfont Article 6, Amendment 64 - AI systems as safety component} \\ \arrayrulecolor{black!20}\midrule
  {\fontsize{4}{6}\selectfont 22} &                     {\fontsize{4}{6}\selectfont  Delivering interactive educational content using a mobile app} &     {\fontsize{4}{6}\selectfont  Yes} &                       {\fontsize{4}{6}\selectfont 4, 8} & {\fontsize{4}{6}\selectfont Article 6 - Education and vocational training} \\ \arrayrulecolor{black!20}\midrule
  {\fontsize{4}{6}\selectfont 23} &     {\fontsize{4}{6}\selectfont Providing personalized learning recommendations using a tablet}  &     {\fontsize{4}{6}\selectfont  Yes} &                   {\fontsize{4}{6}\selectfont  4, 8} & {\fontsize{4}{6}\selectfont Amendment 715, 716 - assign natural persons to educational and vocational training} \\ \arrayrulecolor{black!20}\midrule
  {\fontsize{4}{6}\selectfont 24} &        {\fontsize{4}{6}\selectfont   Enabling online exams using a 360 degree camera} &     {\fontsize{4}{6}\selectfont  Yes} &  {\fontsize{4}{6}\selectfont 4, 17, 10, 8} & {\fontsize{4}{6}\selectfont Article 6 - Education and vocational training} \\ \arrayrulecolor{black!20}\midrule
  {\fontsize{4}{6}\selectfont 25} &                  {\fontsize{4}{6}\selectfont    Monitoring employee productivity using smart badges} &      {\fontsize{4}{6}\selectfont Yes} &                          {\fontsize{4}{6}\selectfont 8, 4, 5, 16} & {\fontsize{4}{6}\selectfont Article 6 - AI systems as safety component} \\ \arrayrulecolor{black!20}\midrule
  {\fontsize{4}{6}\selectfont 26} &                        {\fontsize{4}{6}\selectfont   Monitoring physical movements for workplace safety using smartwatches} &     {\fontsize{4}{6}\selectfont  Yes} &              {\fontsize{4}{6}\selectfont    3, 8, 5, 16, 11} &  
 {\fontsize{4}{6}\selectfont Article 6 - AI systems as safety component and for monitoring/evaluating performance}
    \\ \arrayrulecolor{black!20}\midrule
  {\fontsize{4}{6}\selectfont 27} &                           {\fontsize{4}{6}\selectfont  Tracking work hours and tasks during remote work using a mobile app} &     {\fontsize{4}{6}\selectfont  Yes} &     {\fontsize{4}{6}\selectfont             12, 5, 8, 4, 3} &   {\fontsize{4}{6}\selectfont Article 6 - AI systems as safety component and for monitoring/evaluating performance} \\ \arrayrulecolor{black!20}\midrule
  {\fontsize{4}{6}\selectfont 28} &          {\fontsize{4}{6}\selectfont Tracking travel patterns for public transportation  efficiency using smart cards} &     {\fontsize{4}{6}\selectfont  Yes} &  {\fontsize{4}{6}\selectfont 17, 13, 12, 11, 5, 9, 8} &  {\fontsize{4}{6}\selectfont  Article 6 - AI systems as safety component} \\ \arrayrulecolor{black!20}\midrule
  {\fontsize{4}{6}\selectfont 29} &                   {\fontsize{4}{6}\selectfont     Monitoring health status and symptoms using headbands} &      {\fontsize{4}{6}\selectfont Yes} &  {\fontsize{4}{6}\selectfont 10, 9, 8, 3} & {\fontsize{4}{6}\selectfont Amendment 721 - evaluate natural persons' eligibility for public benefits and services} \\ \arrayrulecolor{black!20}\midrule
  {\fontsize{4}{6}\selectfont 30 } &             {\fontsize{4}{6}\selectfont  Tracking eligibility criteria and application status for social benefits using a mobile app} &       {\fontsize{4}{6}\selectfont Yes} &  {\fontsize{4}{6}\selectfont 10, 8, 5, 3, 1} &  {\fontsize{4}{6}\selectfont Amendment 721 - evaluate natural persons' eligibility for public benefits and services} \\ \arrayrulecolor{black!20}\midrule
  {\fontsize{4}{6}\selectfont 31} &         {\fontsize{4}{6}\selectfont    Tracking user preferences for personalized online shopping experience using a mobile app} &      {\fontsize{4}{6}\selectfont  No} &                 {\fontsize{4}{6}\selectfont      12, 8} & {\fontsize{4}{6}\selectfont N/A} \\ \arrayrulecolor{black!20}\midrule
  {\fontsize{4}{6}\selectfont 32} &        {\fontsize{4}{6}\selectfont  Customizing music and video recommendations using earbuds} &     {\fontsize{4}{6}\selectfont  Yes} &                            {\fontsize{4}{6}\selectfont 12, 8} &  {\fontsize{4}{6}\selectfont Amendment 740 - recommend user-generated content available on the platform} \\ \arrayrulecolor{black!20}\midrule
  {\fontsize{4}{6}\selectfont 33} &                {\fontsize{4}{6}\selectfont  Tailoring news and information feeds using a mobile app} &  {\fontsize{4}{6}\selectfont     Yes} &               {\fontsize{4}{6}\selectfont 5, 4, 8, 12} & {\fontsize{4}{6}\selectfont Amendment 740 - recommend user-generated content available on the platform}  \\ 
  {\fontsize{4}{6}\selectfont 34} &                       {\fontsize{4}{6}\selectfont     Tracking user interactions for enhancing user engagement using a mobile app} &     {\fontsize{4}{6}\selectfont   No} &              {\fontsize{4}{6}\selectfont         8, 12} &  {\fontsize{4}{6}\selectfont N/A}  \\ \arrayrulecolor{black!20}\midrule
  {\fontsize{4}{6}\selectfont 35 } &                 {\fontsize{4}{6}\selectfont     Facilitating social connections using a mobile app} &      {\fontsize{4}{6}\selectfont Yes} &  {\fontsize{4}{6}\selectfont 3, 8} & {\fontsize{4}{6}\selectfont Amendment 740 - recommend user-generated content available on the platform}  \\ \arrayrulecolor{black!20}\midrule
  {\fontsize{4}{6}\selectfont 36} &                       {\fontsize{4}{6}\selectfont         Monitoring harmful content using a mobile app} &     {\fontsize{4}{6}\selectfont  Yes} &  {\fontsize{4}{6}\selectfont 10, 12, 5, 4, 3 } & {\fontsize{4}{6}\selectfont Amendment 740 - recommend user-generated content available on the platform}  \\ \arrayrulecolor{black!20}\midrule
  {\fontsize{4}{6}\selectfont 37} &            {\fontsize{4}{6}\selectfont           Monitoring physical activity and vital signs for improving athletic performance using armbands} &      {\fontsize{4}{6}\selectfont  No} &             {\fontsize{4}{6}\selectfont              5, 3, 4} &                                               {\fontsize{4}{6}\selectfont N/A }\\ \arrayrulecolor{black!20}\midrule
  {\fontsize{4}{6}\selectfont 38} &                   {\fontsize{4}{6}\selectfont          Tracking environmental conditions in outdoor spaces using a mobile app} &     {\fontsize{4}{6}\selectfont   No} &                  {\fontsize{4}{6}\selectfont    5, 3, 4, 8, 11, 16} &                                     {\fontsize{4}{6}\selectfont           N/A }\\ \arrayrulecolor{black!20}\midrule
 {\fontsize{4}{6}\selectfont 39} &                  {\fontsize{4}{6}\selectfont       Monitoring player health for team management using armbands} &       {\fontsize{4}{6}\selectfont No} &                     {\fontsize{4}{6}\selectfont       5, 3, 4} &                                              {\fontsize{4}{6}\selectfont  N/A} \\ \arrayrulecolor{black!20}\midrule
  {\fontsize{4}{6}\selectfont 40} &       {\fontsize{4}{6}\selectfont Tracking user motion for enhancing user experience using VR headsets} &    {\fontsize{4}{6}\selectfont    No} &   {\fontsize{4}{6}\selectfont 5, 8, 3, 4} &                                               {\fontsize{4}{6}\selectfont N/A} \\ \arrayrulecolor{black!20}\midrule
  {\fontsize{4}{6}\selectfont 41} &            {\fontsize{4}{6}\selectfont   Monitoring audience reactions for facilitating interactive performances using mobile cameras} &       {\fontsize{4}{6}\selectfont Yes} &                  {\fontsize{4}{6}\selectfont  5, 8, 4} &                                             {\fontsize{4}{6}\selectfont   Amendment 52, 191 - infer emotions, thoughts, states of mind individuals or groups} \\ \arrayrulecolor{black!20}\midrule
  {\fontsize{4}{6}\selectfont 42} &             {\fontsize{4}{6}\selectfont   Analyzing user preferences for personalized content recommendations using a mobile app} &   {\fontsize{4}{6}\selectfont    No} &                   {\fontsize{4}{6}\selectfont        5, 8, 4}
 &                                             {\fontsize{4}{6}\selectfont   N/A} \\ \arrayrulecolor{black!20}\midrule
 {\fontsize{4}{6}\selectfont 43} &                 {\fontsize{4}{6}\selectfont       Preventing unauthorized access through biometric authentication using mobile scanners} &       {\fontsize{4}{6}\selectfont Yes} &  {\fontsize{4}{6}\selectfont 16, 5, 8} & {\fontsize{4}{6}\selectfont Article 6, Amendment 711 - biometric identification/categorization of natural persons}  \\ \arrayrulecolor{black!20}\midrule
 {\fontsize{4}{6}\selectfont 44} &               {\fontsize{4}{6}\selectfont        Analyzing network traffic for security threats using mobile dashboards} &    {\fontsize{4}{6}\selectfont   Yes} &                     {\fontsize{4}{6}\selectfont   16, 5, 8, 9} &  {\fontsize{4}{6}\selectfont AI Article 6, Amendment 713 -  AI systems as safety component }\\ \arrayrulecolor{black!20}\midrule
 {\fontsize{4}{6}\selectfont 45} &                    {\fontsize{4}{6}\selectfont        Tracking user location and providing emergency alerts for personal safety using smartwatches} &    {\fontsize{4}{6}\selectfont   Yes} &                {\fontsize{4}{6}\selectfont   5, 8, 3, 16, 11} &                                     {\fontsize{4}{6}\selectfont           N/A} \\ 

 {\fontsize{4}{6}\selectfont 46} &                    {\fontsize{4}{6}\selectfont        Analyzing user behavior for personalized advertising using a mobile app} &      {\fontsize{4}{6}\selectfont  No} &                        {\fontsize{4}{6}\selectfont   8} &                                           {\fontsize{4}{6}\selectfont     N/A} \\ \arrayrulecolor{black!20}\midrule
  {\fontsize{4}{6}\selectfont 47} &                       {\fontsize{4}{6}\selectfont Tracking user interactions with ads for customer engagement using a mobile app} &    {\fontsize{4}{6}\selectfont    No} &      {\fontsize{4}{6}\selectfont 8} &                                               {\fontsize{4}{6}\selectfont N/A} \\ \arrayrulecolor{black!20}\midrule
  {\fontsize{4}{6}\selectfont 48} &                     {\fontsize{4}{6}\selectfont  Analyzing marketing campaign performance data using a mobile app } &       {\fontsize{4}{6}\selectfont No} &                  {\fontsize{4}{6}\selectfont  8, 12} &                                              {\fontsize{4}{6}\selectfont  N/A} \\ \arrayrulecolor{black!20}\midrule
 {\fontsize{4}{6}\selectfont 49} &                  {\fontsize{4}{6}\selectfont   Monitoring crop health and growth using IoT sensors} &       {\fontsize{4}{6}\selectfont No} &   {\fontsize{4}{6}\selectfont 17, 11, 13, 9, 12, 2, 15, 8} &                                              {\fontsize{4}{6}\selectfont  N/A} \\ \arrayrulecolor{black!20}\midrule
  {\fontsize{4}{6}\selectfont 50} &              {\fontsize{4}{6}\selectfont  Tracking livestock health and location using GPS collars} &      {\fontsize{4}{6}\selectfont  No} &             {\fontsize{4}{6}\selectfont  13, 12, 17, 11, 9, 2, 15, 8} &                                               {\fontsize{4}{6}\selectfont N/A} \\ \arrayrulecolor{black!20}\midrule
  {\fontsize{4}{6}\selectfont 51} &                       {\fontsize{4}{6}\selectfont Optimizing irrigation through soil moisture detection using IoT sensors} &      {\fontsize{4}{6}\selectfont  No} &         {\fontsize{4}{6}\selectfont 17, 13, 9, 12, 2, 11, 15, 8, 6} &                                             {\fontsize{4}{6}\selectfont   N/A} \\ \arrayrulecolor{black!20}\midrule
  {\fontsize{4}{6}\selectfont 52} &                      {\fontsize{4}{6}\selectfont  Improving customer engagement through personalized recommendations using a mobile app} &   {\fontsize{4}{6}\selectfont No} &              {\fontsize{4}{6}\selectfont  12, 8} &                                         {\fontsize{4}{6}\selectfont   N/A }\\ \arrayrulecolor{black!20}\midrule
  {\fontsize{4}{6}\selectfont 53} &                 {\fontsize{4}{6}\selectfont   Automating task management in business operations using a tablet} &     {\fontsize{4}{6}\selectfont   No } & {\fontsize{4}{6}\selectfont 12, 17, 9, 8 } &                                                {\fontsize{4}{6}\selectfont N/A }\\ \arrayrulecolor{black!20}\midrule
  {\fontsize{4}{6}\selectfont 54} &                     {\fontsize{4}{6}\selectfont   Enhancing product development using customer feedback from mobile apps} &     {\fontsize{4}{6}\selectfont   No } &                 {\fontsize{4}{6}\selectfont  12, 8} &                                               {\fontsize{4}{6}\selectfont N/A} \\ \arrayrulecolor{black!20}\midrule
  {\fontsize{4}{6}\selectfont 55} &                 {\fontsize{4}{6}\selectfont  Facilitating autonomous navigation using mobile cameras} &      {\fontsize{4}{6}\selectfont  No} &                {\fontsize{4}{6}\selectfont  11, 9, 8, 15} &                                               {\fontsize{4}{6}\selectfont N/A} \\ \arrayrulecolor{black!20}\midrule
  {\fontsize{4}{6}\selectfont 56} &                   {\fontsize{4}{6}\selectfont  Enabling human-robot interaction using mobile cameras} &      {\fontsize{4}{6}\selectfont Yes} &               {\fontsize{4}{6}\selectfont  9, 3, 8} &                                               {\fontsize{4}{6}\selectfont Article 6 - AI as safety component} \\ \arrayrulecolor{black!20}\midrule
  {\fontsize{4}{6}\selectfont 57} & {\fontsize{4}{6}\selectfont Improving the quality of multi-sensor data fusion in fall detection smartwatches} &      {\fontsize{4}{6}\selectfont Yes} &  {\fontsize{4}{6}\selectfont 9, 3, 8} & {\fontsize{4}{6}\selectfont  AI Article 6 - AI systems as safety component } \\ \arrayrulecolor{black!20}\midrule
  {\fontsize{4}{6}\selectfont 58} &                   {\fontsize{4}{6}\selectfont Accelerating scientific discovery through analyzing data from sensor-equipped backpack} &        {\fontsize{4}{6}\selectfont No} &   {\fontsize{4}{6}\selectfont 17, 12, 9, 8, 4} &                                               {\fontsize{4}{6}\selectfont N/A} \\ \arrayrulecolor{black!20}\midrule
  {\fontsize{4}{6}\selectfont 59} &                  {\fontsize{4}{6}\selectfont  Facilitating remote collaboration using mobile cameras}  &      {\fontsize{4}{6}\selectfont  No} &                  {\fontsize{4}{6}\selectfont  17, 13, 9, 8, 4} &                                              {\fontsize{4}{6}\selectfont  N/A} \\ \arrayrulecolor{black!20}\midrule
 {\fontsize{4}{6}\selectfont 60} &                 {\fontsize{4}{6}\selectfont  Improving research data management using a mobile app} &       {\fontsize{4}{6}\selectfont No} &                       {\fontsize{4}{6}\selectfont 17, 9, 8, 4} &                                               {\fontsize{4}{6}\selectfont N/A} \\ \arrayrulecolor{black!20}\midrule
  {\fontsize{4}{6}\selectfont 61} &                     {\fontsize{4}{6}\selectfont Monitoring infrastructure health using pressure sensors} &      {\fontsize{4}{6}\selectfont Yes} &         {\fontsize{4}{6}\selectfont  17, 11, 3, 15, 13, 8, 16, 9, 12, 7} &  {\fontsize{4}{6}\selectfont Article 6 - AI systems for management and operation of critical infrastructure} \\ \arrayrulecolor{black!20}\midrule
  {\fontsize{4}{6}\selectfont 62} &                  {\fontsize{4}{6}\selectfont  Predicting infrastructure failures using wearable sensors} &      {\fontsize{4}{6}\selectfont Yes} &       {\fontsize{4}{6}\selectfont 17, 11, 3, 15, 13, 8, 16, 9, 12, 7} &  {\fontsize{4}{6}\selectfont Article 6 - AI systems for management and operation of critical infrastructure} \\ \arrayrulecolor{black!20}\midrule
  {\fontsize{4}{6}\selectfont 63} &               {\fontsize{4}{6}\selectfont  Optimizing infrastructure operations using tablets} &     {\fontsize{4}{6}\selectfont  Yes} & {\fontsize{4}{6}\selectfont 17, 11, 3, 15, 13, 8, 16, 9, 6, 12, 7 } & {\fontsize{4}{6}\selectfont Article 6 - AI systems for management and operation of critical infrastructure } \\ \arrayrulecolor{black!20}\midrule
 {\fontsize{4}{6}\selectfont 64} &              {\fontsize{4}{6}\selectfont    Enhancing surveillance capabilities using mobile cameras} &    {\fontsize{4}{6}\selectfont   *} &   {\fontsize{4}{6}\selectfont                 16, 5, 8, 11} & {\fontsize{4}{6}\selectfont Article 5, 6 - remote biometric identification in public spaces used by law enforcement} \\ \arrayrulecolor{black!20}\midrule
 {\fontsize{4}{6}\selectfont 65} &                  {\fontsize{4}{6}\selectfont         Improving crime prediction using mobile cameras} &      {\fontsize{4}{6}\selectfont Yes} & {\fontsize{4}{6}\selectfont 16, 5, 8} & {\fontsize{4}{6}\selectfont Article 6 - predict criminal offence based on profiling of natural persons} \\ \arrayrulecolor{black!20}\midrule
  {\fontsize{4}{6}\selectfont 66} &           {\fontsize{4}{6}\selectfont          Facilitating evidence collection using a mobile app} & {\fontsize{4}{6}\selectfont      Yes} & {\fontsize{4}{6}\selectfont 16, 5, 8} & {\fontsize{4}{6}\selectfont Amendment 728 - AI Systems for investigation or prosecution of criminal offences} \\ 
 {\fontsize{4}{6}\selectfont 67} &                           {\fontsize{4}{6}\selectfont Enhancing border security using mobile fingerpint scanners} &     {\fontsize{4}{6}\selectfont  Yes} &                       {\fontsize{4}{6}\selectfont 16, 5, 8, 17, 10, 11} & {\fontsize{4}{6}\selectfont Amendment 736 - AI Systems for border control management}  \\ \arrayrulecolor{black!20}\midrule
 {\fontsize{4}{6}\selectfont 68} &           {\fontsize{4}{6}\selectfont   Improving asylum application processing using mobile document scanners} &     {\fontsize{4}{6}\selectfont  Yes} &                 {\fontsize{4}{6}\selectfont      16, 1, 3, 5, 8, 17, 10, 11} & {\fontsize{4}{6}\selectfont Amendment 735 - AI Systems for assessment in relation to applications for asylum} \\ \arrayrulecolor{black!20}\midrule
  {\fontsize{4}{6}\selectfont 69} &                      {\fontsize{4}{6}\selectfont  Tracking migrant location using GPS} &    {\fontsize{4}{6}\selectfont   Yes} &                           {\fontsize{4}{6}\selectfont     16, 5, 17, 10} & {\fontsize{4}{6}\selectfont Amendment 736 - AI Systems for border control management} \\ 
  {\fontsize{4}{6}\selectfont 70} &                      {\fontsize{4}{6}\selectfont  Enhancing voter participation using a mobile app} &      {\fontsize{4}{6}\selectfont Yes} &             {\fontsize{4}{6}\selectfont   16, 4} & {\fontsize{4}{6}\selectfont Amendment 739 - influence the outcome of election/referendum/voting behaviour} \\ \arrayrulecolor{black!20}\midrule
  {\fontsize{4}{6}\selectfont 71} &                {\fontsize{4}{6}\selectfont      Improving election transparency using a mobile app} &      {\fontsize{4}{6}\selectfont Yes} &                   {\fontsize{4}{6}\selectfont    16, 10, 8, 9} & {\fontsize{4}{6}\selectfont Amendment 739 - influence the outcome of election/referendum/voting behaviour}\\ \arrayrulecolor{black!20}\midrule
  {\fontsize{4}{6}\selectfont 72} &                   {\fontsize{4}{6}\selectfont   Facilitating citizen engagement using digital kiosks} &      {\fontsize{4}{6}\selectfont  No} &             {\fontsize{4}{6}\selectfont   16, 11, 5, 17, 8, 4, 9} & {\fontsize{4}{6}\selectfont N/A} \\ \arrayrulecolor{black!20}\midrule
  {\fontsize{4}{6}\selectfont 73} &       {\fontsize{4}{6}\selectfont  Enhancing immersive storytelling experiences using smart glasses} &       {\fontsize{4}{6}\selectfont No} &              {\fontsize{4}{6}\selectfont  4} &                                          {\fontsize{4}{6}\selectfont      N/A} \\ \arrayrulecolor{black!20}\midrule
 {\fontsize{4}{6}\selectfont 74} &         {\fontsize{4}{6}\selectfont Facilitating real-time language translation using earbuds} &    {\fontsize{4}{6}\selectfont    No} &         {\fontsize{4}{6}\selectfont  4, 8, 10} &                                      {\fontsize{4}{6}\selectfont          N/A} \\ \arrayrulecolor{black!20}\midrule
 {\fontsize{4}{6}\selectfont 75} &                   {\fontsize{4}{6}\selectfont   Improving content accessibility using mobile screen readers} &      {\fontsize{4}{6}\selectfont  No} &            {\fontsize{4}{6}\selectfont    16, 8, 4, 9} &                                              {\fontsize{4}{6}\selectfont  N/A} \\ \arrayrulecolor{black!20}\midrule
  {\fontsize{4}{6}\selectfont 76} &             {\fontsize{4}{6}\selectfont  Assisting visually impaired individuals using smart canes} &      {\fontsize{4}{6}\selectfont Yes} &           {\fontsize{4}{6}\selectfont  11, 16, 4, 10, 8, 3} & {\fontsize{4}{6}\selectfont Amendment 38 - perceive or exploit vulnerabilities of children and people} \\ \arrayrulecolor{black!20}\midrule
  {\fontsize{4}{6}\selectfont 77} &      {\fontsize{4}{6}\selectfont  Supporting individuals with hearing impairment using earbuds} &       {\fontsize{4}{6}\selectfont Yes} & {\fontsize{4}{6}\selectfont 11, 16, 4, 10, 8, 3} & {\fontsize{4}{6}\selectfont Amendment 38 - perceive or exploit vulnerabilities of children and people}\\ \arrayrulecolor{black!20}\midrule
  {\fontsize{4}{6}\selectfont 78} &               {\fontsize{4}{6}\selectfont Tracking movement for assisting individuals with mobility issues using accelerometers} & {\fontsize{4}{6}\selectfont      Yes} &                 {\fontsize{4}{6}\selectfont   11, 16, 4, 10, 8, 3} &  {\fontsize{4}{6}\selectfont Article 6 - AI system as a safety component}  \\ \arrayrulecolor{black!20}\midrule
 {\fontsize{4}{6}\selectfont 79} &                     {\fontsize{4}{6}\selectfont    Monitoring energy consumption using infrared sensors} &     {\fontsize{4}{6}\selectfont   No} & {\fontsize{4}{6}\selectfont  13, 12, 17, 11, 15, 9, 7} &                                                {\fontsize{4}{6}\selectfont N/A} \\ \arrayrulecolor{black!20}\midrule
  {\fontsize{4}{6}\selectfont 80} &             {\fontsize{4}{6}\selectfont  Optimizing renewable energy production using heat sensors } &      {\fontsize{4}{6}\selectfont  No} &                         {\fontsize{4}{6}\selectfont 13, 12, 17, 11, 15, 9, 7} &                                               {\fontsize{4}{6}\selectfont N/A} \\ \arrayrulecolor{black!20}\midrule
  {\fontsize{4}{6}\selectfont 81} &                        {\fontsize{4}{6}\selectfont Improving energy efficiency using infrared sensors} &       {\fontsize{4}{6}\selectfont Yes} &                 {\fontsize{4}{6}\selectfont 13, 12, 11, 15, 9, 7} &    {\fontsize{4}{6}\selectfont Article 6 - AI system as a safety component } \\ \arrayrulecolor{black!20}\midrule
  {\fontsize{4}{6}\selectfont 82} &                      {\fontsize{4}{6}\selectfont     Enhancing soldier training using VR headsets} &       {\fontsize{4}{6}\selectfont Yes} &                         {\fontsize{4}{6}\selectfont  16, 17, 9, 8, 4} & {\fontsize{4}{6}\selectfont Article 6 - AI system as a safety component} \\ \arrayrulecolor{black!20}\midrule
 {\fontsize{4}{6}\selectfont 83} &      {\fontsize{4}{6}\selectfont    Improving battlefield situational awareness using smart glassess} &       {\fontsize{4}{6}\selectfont Yes} &                 {\fontsize{4}{6}\selectfont      16, 17, 9, 8, 4 } &  {\fontsize{4}{6}\selectfont Article 6 - AI system as a safety component} \\ \arrayrulecolor{black!20}\midrule
 {\fontsize{4}{6}\selectfont 84} &             {\fontsize{4}{6}\selectfont     Enhancing soldier health monitoring using smart belts} &      {\fontsize{4}{6}\selectfont Yes} &  {\fontsize{4}{6}\selectfont   16, 17, 9, 8, 4, 3} &  {\fontsize{4}{6}\selectfont Article 6 - AI system as a safety component} \\ \arrayrulecolor{black!20}\midrule
   {\fontsize{4}{6}\selectfont 85} &               {\fontsize{4}{6}\selectfont Facilitating remote court proceedings using a mobile app} &       {\fontsize{4}{6}\selectfont Yes} &              {\fontsize{4}{6}\selectfont  17, 10, 16, 11, 8, 9} & {\fontsize{4}{6}\selectfont Amendment 738 – assist in judicial authority in researching and interpreting facts} \\ \arrayrulecolor{black!20}\midrule
  {\fontsize{4}{6}\selectfont 86} &       {\fontsize{4}{6}\selectfont Improving voter registration and verification using a mobile app} &     {\fontsize{4}{6}\selectfont  Yes} & {\fontsize{4}{6}\selectfont 17, 16, 10, 11, 9} & {\fontsize{4}{6}\selectfont Article 6 - AI Systems for administration of justice and democratic processes} \\ \arrayrulecolor{black!20}\midrule
  {\fontsize{4}{6}\selectfont 87} & {\fontsize{4}{6}\selectfont Enhancing public participation in democratic processes using smart kiosks} &      {\fontsize{4}{6}\selectfont Yes} &               {\fontsize{4}{6}\selectfont 17, 16, 10, 11, 5, 8, 4, 9} & {\fontsize{4}{6}\selectfont Amendment 738 – assist in judicial authority in researching and interpreting facts} \\ \arrayrulecolor{black!20}\midrule
  {\fontsize{4}{6}\selectfont 88} &                   {\fontsize{4}{6}\selectfont Improving public service delivery using a mobile app} &     {\fontsize{4}{6}\selectfont  Yes} &     {\fontsize{4}{6}\selectfont 16, 11, 5, 8, 4, 3, 9} &                                               {\fontsize{4}{6}\selectfont Amendment 67 - apply for or receive public assistance benefits and services} \\ \arrayrulecolor{black!20}\midrule
  {\fontsize{4}{6}\selectfont 89} &                  {\fontsize{4}{6}\selectfont      Enhancing disaster management using accelerometers} &     {\fontsize{4}{6}\selectfont  Yes} &   {\fontsize{4}{6}\selectfont  17, 13, 9, 8, 3, 11, 16, 14, 15} & {\fontsize{4}{6}\selectfont Amendment 724 - AI systems for dispatching of emergency first response services} \\ \arrayrulecolor{black!20}\midrule
  {\fontsize{4}{6}\selectfont 90} &                    {\fontsize{4}{6}\selectfont     Improving traffic management through cameras} &     {\fontsize{4}{6}\selectfont  Yes} &          {\fontsize{4}{6}\selectfont    13, 12, 11, 16, 9} & {\fontsize{4}{6}\selectfont Article 6 - AI systems as safety component }\\ \arrayrulecolor{black!20}\midrule
 {\fontsize{4}{6}\selectfont 91} &             {\fontsize{4}{6}\selectfont   Facilitating diplomatic communication using earbuds} &      {\fontsize{4}{6}\selectfont  No} &    {\fontsize{4}{6}\selectfont  16, 17, 10, 8, 4} &                                              {\fontsize{4}{6}\selectfont  N/A} \\ \arrayrulecolor{black!20}\midrule
  {\fontsize{4}{6}\selectfont 92} &                  {\fontsize{4}{6}\selectfont Monitoring global protest actions using a mobile app} &     {\fontsize{4}{6}\selectfont   No} &    {\fontsize{4}{6}\selectfont 16, 17, 5, 13, 9, 11} &                                        {\fontsize{4}{6}\selectfont        N/A} \\ \arrayrulecolor{black!20}\midrule
  {\fontsize{4}{6}\selectfont 93} &     {\fontsize{4}{6}\selectfont   Improving international negotiation processes using a mobile app} &       {\fontsize{4}{6}\selectfont No} &                             {\fontsize{4}{6}\selectfont 16, 17, 10, 4 } &                                               {\fontsize{4}{6}\selectfont N/A} \\ \arrayrulecolor{black!20}\midrule
 {\fontsize{4}{6}\selectfont 94} &                {\fontsize{4}{6}\selectfont    Enhancing food quality inspection using mobile cameras} &      {\fontsize{4}{6}\selectfont Yes} &                   {\fontsize{4}{6}\selectfont 2, 12, 3, 17, 11, 9, 8} &                                               {\fontsize{4}{6}\selectfont Amendment 207 - AI Systems for critical infrastructure} \\ \arrayrulecolor{black!20}\midrule
 {\fontsize{4}{6}\selectfont 95} &                       {\fontsize{4}{6}\selectfont   Improving food traceability using RFID tags} &      {\fontsize{4}{6}\selectfont Yes} &               {\fontsize{4}{6}\selectfont     2, 12, 3, 17, 11, 9, 8} &                                                {\fontsize{4}{6}\selectfont Amendment 207 - AI Systems for critical infrastructure} \\ \arrayrulecolor{black!20}\midrule
 {\fontsize{4}{6}\selectfont 96} &       {\fontsize{4}{6}\selectfont   Enhancing food safety compliance using mobile cameras} &  {\fontsize{4}{6}\selectfont     Yes} &            {\fontsize{4}{6}\selectfont   2, 12, 3, 17, 11, 9, 8} &  {\fontsize{4}{6}\selectfont Amendment 207 - AI Systems for critical infrastructure}\\ \arrayrulecolor{black!20}\midrule
  {\fontsize{4}{6}\selectfont 97} &           {\fontsize{4}{6}\selectfont    Enhancing disaster response using GPS} &     {\fontsize{4}{6}\selectfont  Yes} &    {\fontsize{4}{6}\selectfont  17, 13, 8, 3, 16, 9, 11} & {\fontsize{4}{6}\selectfont Amendment 724 - AI systems for dispatching of emergency first response services} \\ \arrayrulecolor{black!20}\midrule
 {\fontsize{4}{6}\selectfont 98} &                {\fontsize{4}{6}\selectfont   Monitoring health status of victims using wristbands} &      {\fontsize{4}{6}\selectfont Yes} &             {\fontsize{4}{6}\selectfont    3, 9, 8, 11} & {\fontsize{4}{6}\selectfont Amendment 724 - AI systems for dispatching of emergency first response services} \\ \arrayrulecolor{black!20}\midrule
 {\fontsize{4}{6}\selectfont 99} &                      {\fontsize{4}{6}\selectfont Coordinating rescue operations using a mobile app} &     {\fontsize{4}{6}\selectfont  Yes} &           {\fontsize{4}{6}\selectfont     17, 13, 8, 9, 3, 11, 16} & {\fontsize{4}{6}\selectfont Amendment 724 - AI systems for dispatching of emergency first response services} \\ \arrayrulecolor{black!20}\midrule
 {\fontsize{4}{6}\selectfont 100} &             {\fontsize{4}{6}\selectfont   Improving aid distribution efficiency using motion sensors} &      {\fontsize{4}{6}\selectfont Yes} & {\fontsize{4}{6}\selectfont 12, 17, 16, 11, 1, 2, 3, 13, 8, 9} & {\fontsize{4}{6}\selectfont Amendment 721 - evaluate natural persons' eligibility for public benefits and services} \\ \arrayrulecolor{black!20}\midrule
{\fontsize{4}{6}\selectfont 101} &        {\fontsize{4}{6}\selectfont    Monitoring health status of aid recipients using wristbands} &     {\fontsize{4}{6}\selectfont  Yes} &                 {\fontsize{4}{6}\selectfont      17, 3, 9, 8, 11} & {\fontsize{4}{6}\selectfont Amendment 721 - evaluate natural persons' eligibility for public benefits and services}  \\ \arrayrulecolor{black!20}\midrule
{\fontsize{4}{6}\selectfont 102} &        {\fontsize{4}{6}\selectfont Facilitating communication in disaster zones using a mobile app} &      {\fontsize{4}{6}\selectfont  No} &  {\fontsize{4}{6}\selectfont 16, 17, 13, 8, 9, 11} &                                           {\fontsize{4}{6}\selectfont     N/A} \\ \arrayrulecolor{black!20}\midrule
 {\fontsize{4}{6}\selectfont 103} &              {\fontsize{4}{6}\selectfont  Improving fleet management efficiency using GPS} &     {\fontsize{4}{6}\selectfont   No} &                {\fontsize{4}{6}\selectfont 12, 11, 17, 9, 8} &                                 {\fontsize{4}{6}\selectfont               N/A} \\ \arrayrulecolor{black!20}\midrule
 {\fontsize{4}{6}\selectfont 104} &                          {\fontsize{4}{6}\selectfont    Enhancing driver safety using motion sensors} &      {\fontsize{4}{6}\selectfont Yes} &    {\fontsize{4}{6}\selectfont  3, 12, 11, 8, 9, 16} & {\fontsize{4}{6}\selectfont Amendment 713 - AI Systems for management and operation of road traffic} \\ \arrayrulecolor{black!20}\midrule
 {\fontsize{4}{6}\selectfont 105} &                   {\fontsize{4}{6}\selectfont       Optimizing fuel consumption using pressure sensors} &     {\fontsize{4}{6}\selectfont   No} &  {\fontsize{4}{6}\selectfont 13, 12, 11, 9, 7, 15} &                                          {\fontsize{4}{6}\selectfont      N/A} \\ \arrayrulecolor{black!20}\midrule
 {\fontsize{4}{6}\selectfont 106} &                 {\fontsize{4}{6}\selectfont   Monitoring urban mobility patterns using GPS} &   {\fontsize{4}{6}\selectfont    Yes} &         {\fontsize{4}{6}\selectfont   13, 12, 11, 16, 9} & {\fontsize{4}{6}\selectfont Article 6 - AI Systems as safety component} \\ \arrayrulecolor{black!20}\midrule
 {\fontsize{4}{6}\selectfont 107} &                      {\fontsize{4}{6}\selectfont    Assessing public space usage using mobile cameras} &     {\fontsize{4}{6}\selectfont  Yes} &                       {\fontsize{4}{6}\selectfont  16, 11, 12, 9} &  {\fontsize{4}{6}\selectfont Article 6 - Biometric identification and categorisation of natural persons} \\ \arrayrulecolor{black!20}\midrule
 {\fontsize{4}{6}\selectfont 108} &     {\fontsize{4}{6}\selectfont      Improving public transportation using mobile cameras} &       {\fontsize{4}{6}\selectfont Yes} &                     {\fontsize{4}{6}\selectfont 12, 11, 9} & {\fontsize{4}{6}\selectfont Article 6 - AI Systems as safety component}\\ \arrayrulecolor{black!20}\midrule
 {\fontsize{4}{6}\selectfont 109} &                   {\fontsize{4}{6}\selectfont      Detecting suspicious behavior using motion sensors} &      {\fontsize{4}{6}\selectfont Yes} &                       {\fontsize{4}{6}\selectfont   16, 10, 17, 8, 3, 11} & {\fontsize{4}{6}\selectfont Article 6 - Biometric identification and categorisation of natural persons} \\ \arrayrulecolor{black!20}\midrule
 {\fontsize{4}{6}\selectfont 110} &            {\fontsize{4}{6}\selectfont      Enhancing surveillance capabilities for counter-terrorism using GPS} &    {\fontsize{4}{6}\selectfont   Yes} &     {\fontsize{4}{6}\selectfont   16, 10, 17, 8, 3, 11, 9} &  {\fontsize{4}{6}\selectfont Article 6 - Biometric identification and categorisation of natural persons} \\ \arrayrulecolor{black!20}\midrule
 {\fontsize{4}{6}\selectfont 111} &      {\fontsize{4}{6}\selectfont    Improving communication in field operations using earbuds} &    {\fontsize{4}{6}\selectfont   Yes} &  {\fontsize{4}{6}\selectfont 17, 16, 8, 9, 4} &  {\fontsize{4}{6}\selectfont Article 6 - AI Systems as safety component}  \\ \arrayrulecolor{black!20}\midrule
 {\fontsize{4}{6}\selectfont 112} &            {\fontsize{4}{6}\selectfont       Monitoring environmental conditions using heat sensors} &      {\fontsize{4}{6}\selectfont  No} &   {\fontsize{4}{6}\selectfont 13, 17, 12, 6, 3, 14, 9, 15, 8, 11, 7} &                                              {\fontsize{4}{6}\selectfont  N/A} \\ \arrayrulecolor{black!20}\midrule
 {\fontsize{4}{6}\selectfont 113} &                   {\fontsize{4}{6}\selectfont        Tracking wildlife movements using GPS collars} &      {\fontsize{4}{6}\selectfont  No} &      {\fontsize{4}{6}\selectfont     17, 14, 15, 13, 9, 8} &                                          {\fontsize{4}{6}\selectfont      N/A} \\ \arrayrulecolor{black!20}\midrule
 {\fontsize{4}{6}\selectfont 114} &                 {\fontsize{4}{6}\selectfont  Assessing resource usage efficiency using motion sensors} &    {\fontsize{4}{6}\selectfont    No} &         {\fontsize{4}{6}\selectfont   13, 12, 17, 11, 9, 6, 7, 15}&                                              {\fontsize{4}{6}\selectfont  N/A} \\ \arrayrulecolor{black!20}\midrule
 {\fontsize{4}{6}\selectfont 115} &                  {\fontsize{4}{6}\selectfont Enhancing cross-border surveillance using mobile fingerprint scanners} &      {\fontsize{4}{6}\selectfont Yes} &                  {\fontsize{4}{6}\selectfont 16, 17, 10, 8, 11, 9} & {\fontsize{4}{6}\selectfont Amendment 736 - AI Systems for border control management} \\ \arrayrulecolor{black!20}\midrule
 {\fontsize{4}{6}\selectfont 116} &        {\fontsize{4}{6}\selectfont  Improving communication in joint operations using smart glasses} &    {\fontsize{4}{6}\selectfont   Yes} &                       {\fontsize{4}{6}\selectfont 16, 17, 8, 9, 4 }& {\fontsize{4}{6}\selectfont Article 6 - AI Systems as safety component} \\ \arrayrulecolor{black!20}\midrule
{\fontsize{4}{6}\selectfont 117} &                {\fontsize{4}{6}\selectfont          Detecting illegal activities using motion sensors} &   {\fontsize{4}{6}\selectfont    Yes} &                     {\fontsize{4}{6}\selectfont  16, 10, 11, 8, 5} &  {\fontsize{4}{6}\selectfont Amendment 209 - AI Systems used for determining social behavior} \\ \arrayrulecolor{black!20}\midrule
 {\fontsize{4}{6}\selectfont 118} &       {\fontsize{4}{6}\selectfont           Monitoring climate change using heat sensors} &  {\fontsize{4}{6}\selectfont      No} &                {\fontsize{4}{6}\selectfont      13, 17, 11, 6, 9, 15, 10, 7} &                                                {\fontsize{4}{6}\selectfont N/A} \\ \arrayrulecolor{black!20}\midrule
 {\fontsize{4}{6}\selectfont 119} &      {\fontsize{4}{6}\selectfont   Tracking impact of climate change on wildlife using GPS collars and heat sensors} &      {\fontsize{4}{6}\selectfont  No} &             {\fontsize{4}{6}\selectfont   17, 11, 15, 14, 13, 9, 8} &                                               {\fontsize{4}{6}\selectfont N/A} \\ \arrayrulecolor{black!20}\midrule
 {\fontsize{4}{6}\selectfont 120} &      {\fontsize{4}{6}\selectfont  Assessing effectiveness of energy-saving interventions using smart electricity meters} &      {\fontsize{4}{6}\selectfont  No} &   {\fontsize{4}{6}\selectfont 13, 12, 17, 6, 11, 7, 9, 8, 14, 15} &                                              {\fontsize{4}{6}\selectfont  N/A} \\ \arrayrulecolor{black!20}\midrule
{\fontsize{4}{6}\selectfont 121} &             {\fontsize{4}{6}\selectfont   Enhancing immersive gaming experience using VR headsets}  &     {\fontsize{4}{6}\selectfont   No} &                           {\fontsize{4}{6}\selectfont 9 } &                                     {\fontsize{4}{6}\selectfont           N/A} \\ \arrayrulecolor{black!20}\midrule
{\fontsize{4}{6}\selectfont 122} &           {\fontsize{4}{6}\selectfont Tracking physical activity for improving physical fitness using wristbands} &     {\fontsize{4}{6}\selectfont   No} &    {\fontsize{4}{6}\selectfont 8, 3, 4} &                                                {\fontsize{4}{6}\selectfont N/A} \\ \arrayrulecolor{black!20}\midrule
{\fontsize{4}{6}\selectfont 123} &   {\fontsize{4}{6}\selectfont Facilitating virtual social interaction using VR headsets} &     {\fontsize{4}{6}\selectfont   No} & {\fontsize{4}{6}\selectfont  4, 8} &                                      {\fontsize{4}{6}\selectfont          N/A} \\ \arrayrulecolor{black!20}\midrule
 {\fontsize{4}{6}\selectfont 124} &          {\fontsize{4}{6}\selectfont    Tracking and improving musical performance using a mobile app} &     {\fontsize{4}{6}\selectfont   No} &                      {\fontsize{4}{6}\selectfont 4, 8} &                                    {\fontsize{4}{6}\selectfont            N/A} \\ \arrayrulecolor{black!20}\midrule
 {\fontsize{4}{6}\selectfont 125} &         {\fontsize{4}{6}\selectfont  Monitoring gardening activities using motion sensors} &      {\fontsize{4}{6}\selectfont  No} &               {\fontsize{4}{6}\selectfont    11, 12, 2, 15, 8} &                                           {\fontsize{4}{6}\selectfont     N/A} \\ \arrayrulecolor{black!20}\midrule
 {\fontsize{4}{6}\selectfont 126} &                    {\fontsize{4}{6}\selectfont     Enhancing photography skills using mobile cameras} &      {\fontsize{4}{6}\selectfont  No} &                       {\fontsize{4}{6}\selectfont 4, 8} &                                             {\fontsize{4}{6}\selectfont  N/A} \\ \arrayrulecolor{black!20}\midrule
 {\fontsize{4}{6}\selectfont 127} &                          {\fontsize{4}{6}\selectfont    Improving home security using mobile cameras} &      {\fontsize{4}{6}\selectfont Yes} &                                 {\fontsize{4}{6}\selectfont 5, 11} & {\fontsize{4}{6}\selectfont Article 6 - AI Systems as safety component} \\ \arrayrulecolor{black!20}\midrule
 {\fontsize{4}{6}\selectfont 128} &                       {\fontsize{4}{6}\selectfont   Enhancing energy efficiency using heat sensors} &     {\fontsize{4}{6}\selectfont  Yes} &                      {\fontsize{4}{6}\selectfont 13, 11, 12, 9, 7} &                                                {\fontsize{4}{6}\selectfont Article 6 - AI system as a safety component} \\ \arrayrulecolor{black!20}\midrule
 {\fontsize{4}{6}\selectfont 129} & {\fontsize{4}{6}\selectfont Facilitating home automation using smart thermostats} &       {\fontsize{4}{6}\selectfont Yes} &      {\fontsize{4}{6}\selectfont 13, 12, 9} &                                {\fontsize{4}{6}\selectfont               Article 6 - AI system as a safety component} \\ \arrayrulecolor{black!20}\midrule
 {\fontsize{4}{6}\selectfont 130} & {\fontsize{4}{6}\selectfont Improving public safety using mobile cameras} &    {\fontsize{4}{6}\selectfont   Yes} &           {\fontsize{4}{6}\selectfont        11, 5, 8, 3, 16} & {\fontsize{4}{6}\selectfont Article 6 - AI Systems as safety component} \\ \arrayrulecolor{black!20}\midrule
 {\fontsize{4}{6}\selectfont 131 } &     {\fontsize{4}{6}\selectfont Enhancing accessibility for disabled individuals using smart glasses} &     {\fontsize{4}{6}\selectfont  Yes } &     {\fontsize{4}{6}\selectfont  11, 10, 8, 3} & {\fontsize{4}{6}\selectfont Article 6 - access to essential private services and public services and benefits }\\ \arrayrulecolor{black!20}\midrule
{\fontsize{4}{6}\selectfont 132} &              {\fontsize{4}{6}\selectfont         Facilitating disaster response using a mobile app} &    {\fontsize{4}{6}\selectfont   Yes} &            {\fontsize{4}{6}\selectfont    17, 13, 9, 8, 3, 11, 16, 10} & {\fontsize{4}{6}\selectfont Amendment 724 - AI systems for dispatching of emergency first response services} \\ \arrayrulecolor{black!20}\midrule
{\fontsize{4}{6}\selectfont 133} &                     {\fontsize{4}{6}\selectfont           Improving road safety using mobile cameras} &    {\fontsize{4}{6}\selectfont   Yes} &               {\fontsize{4}{6}\selectfont3, 11, 9, 8, 16} &  {\fontsize{4}{6}\selectfont Article 6 - AI Systems as safety component} \\ \arrayrulecolor{black!20}\midrule
 {\fontsize{4}{6}\selectfont 134} &         {\fontsize{4}{6}\selectfont     Enhancing navigation and route planning using GPS} &     {\fontsize{4}{6}\selectfont   No} &     {\fontsize{4}{6}\selectfont       12, 11, 8, 9} &  {\fontsize{4}{6}\selectfont   Article 6 - AI systems as safety component } \\
 \arrayrulecolor{black!20}\midrule
{\fontsize{4}{6}\selectfont 135} &           {\fontsize{4}{6}\selectfont           Facilitating autonomous driving using motion sensors} &     {\fontsize{4}{6}\selectfont  Yes} &    {\fontsize{4}{6}\selectfont   11, 9, 8} & {\fontsize{4}{6}\selectfont Article 6 - AI systems as safety component }  \\ \arrayrulecolor{black!20}\midrule
 {\fontsize{4}{6}\selectfont 136} &              {\fontsize{4}{6}\selectfont            Enhancing language learning using a mobile app} &    {\fontsize{4}{6}\selectfont    No} &       {\fontsize{4}{6}\selectfont         4, 8} &                                               {\fontsize{4}{6}\selectfont N/A }\\ \arrayrulecolor{black!20}\midrule
 {\fontsize{4}{6}\selectfont 137} &           {\fontsize{4}{6}\selectfont            Improving remote communication using mobile cameras} &      {\fontsize{4}{6}\selectfont  No} &  {\fontsize{4}{6}\selectfont          4, 8, 9 } &                                            {\fontsize{4}{6}\selectfont    N/A} \\ \arrayrulecolor{black!20}\midrule
 {\fontsize{4}{6}\selectfont 138} &                     {\fontsize{4}{6}\selectfont  Facilitating social networking using a mobile app} &     {\fontsize{4}{6}\selectfont   No} &     {\fontsize{4}{6}\selectfont   4, 8, 9 }&                                             {\fontsize{4}{6}\selectfont   N/A} \\
\end{longtable}



\end{document}